\documentclass{ldbc}

\usepackage{multirow}
\usepackage{float}
\usepackage{amsfonts}
\usepackage{rotating}
\usepackage{pifont}
\usepackage{rotating}
\usepackage{subcaption}
\usepackage{amsmath}
\usepackage{longtable}
\usepackage{tabularx}
\usepackage{varwidth}
\usepackage{hyperref}
\usepackage{xspace}
\usepackage[table]{xcolor}
\usepackage{listings}
\usepackage[export]{adjustbox}
\usepackage{array}
\usepackage[inline]{enumitem}
\usepackage{makecell}

\usepackage{todonotes}
\presetkeys{todonotes}{inline}{}

\usepackage[normalem]{ulem}
\usepackage{cellspace}
\usepackage{rotate}
\usepackage{MnSymbol}
\usepackage{xifthen}
\usepackage{numprint}
\usepackage[scale=0.78,ttdefault=true]{AnonymousPro}

\usepackage{bm}

\usepackage{caption}
\usepackage{tikz}
\usetikzlibrary{arrows}
\usetikzlibrary{arrows.meta}
\usetikzlibrary{patterns}

\usepackage[
	sorting=none,
	eprint=false
]{biblatex}
\renewbibmacro*{doi+eprint+url}{%
	\printfield{doi}%
	\newunit\newblock%
	\iftoggle{bbx:eprint}{%
		\usebibmacro{eprint}%
	}{}%
	\newunit\newblock%
	\iffieldundef{doi}{%
		\usebibmacro{url+urldate}}%
	{}%
}

\providecommand{\tightlist}{%
  \setlength{\itemsep}{0pt}\setlength{\parskip}{0pt}}

\setlist[]{noitemsep, topsep=5pt}

\newcommand{\asc}{\uparrow}
\newcommand{\desc}{\downarrow}

\newcommand{\yedscale}{0.43}

\newcommand{\DataGen}{DataGen\xspace}
\newcommand{\ldbcfinbench}{LDBC FinBench\xspace}
\newcommand{\ldbcsnb}{LDBC SNB~\cite{DBLP:journals/corr/abs-2001-02299,ldbc_snb_docs}\xspace}

\newcommand{\type}[1]{\textsf{#1}}

\def\shadedBox(#1,#2,#3){

  \fill[pattern=north west lines,pattern color=grey] (#1,#2) --  (#1,#2 - #3) -- (#1 + 0.3,#2 - #3) --  (#1 + 0.3,#2);
  \draw [grey,thin,dashed] (#1,#2)  -- (#1,#2 - #3);
  \draw [grey,thin,dashed] (#1 + 0.3,#2) -- (#1 + 0.3,#2 - #3);
  \draw [grey,line width=0.6mm] (#1,#2 - #3) -- node[midway,below,grey] {$\Delta$} (#1 + 0.3,#2 - #3);

}

\definecolor{Person}{HTML}{fdb462}
\definecolor{Message}{HTML}{bebada}
\definecolor{Forum}{HTML}{b3de69}
\definecolor{Comment}{HTML}{80b1d3}
\definecolor{Post}{HTML}{fb8072}
\definecolor{Company}{HTML}{ccebc5}
\definecolor{University}{HTML}{ffed6f}
\definecolor{City}{HTML}{8dd3c7}
\definecolor{Tag}{HTML}{fccde5}
\definecolor{Country}{HTML}{ffffb3}

\definecolor{grey}{rgb}{0.52, 0.52, 0.51}
\definecolor{red}{rgb}{0.7, 0.11, 0.11}
\definecolor{blue}{rgb}{0.0, 0.0, 0.55}
\definecolor{green}{rgb}{0.0, 0.42, 0.24}

\definecolor{parameter}{HTML}{e41a1c}
\definecolor{result}{HTML}{377eb8}
\definecolor{sort}{HTML}{4daf4a}

\definecolor{TCR}{HTML}{ffffcc}
\definecolor{TSR}{HTML}{ffffcc}
\definecolor{TW}{HTML}{fed9a6}
\definecolor{TRW}{HTML}{e5d8bd}

\newboolean{standalone}

\reversemarginpar

\newcommand{\currentQueryCard}{0}
\newcommand{\queryRefCard}[3]{
	\ifthenelse{
		\equal{\currentQueryCard}{#3}
	}{%
		\colorbox{white}{\tt #2 #3}%
	}{%
		\hyperref[sec:#1]{\colorbox{#2}{\tt #2 #3}}%
	}%
}

\newcommand{\attributeNumberWidth}{0.33cm}
\newcommand{\attributeColumnWidth}{2.5cm}
\newcommand{\typeColumnWidth}{2.7cm}
\newcommand{\descriptionColumnWidth}{10.3cm}
\newcommand{\largeDescriptionColumnWidth}{13cm}

\newcommand{\tableHeaderFirst}[1]{\multicolumn{1}{|c|}{\bf #1}}
\newcommand{\tableHeader}[1]{\multicolumn{1}{c|}{\bf #1}}

\newcommand{\queryCardWidth}{17cm}
\newcommand{\queryPropertyCell}{\small \sf \centering}
\newcommand{\queryPropertyCellWidth}{1.48cm}

\newcommand{\attributeCardWidth}{14.66cm}
\newcommand{\typeWidth}{2.04cm}

\newcommand{\paramNumberCell}{\cellcolor{parameter}\color{white}\footnotesize}
\newcommand{\resultNumberCell}{\cellcolor{result}\color{white}\footnotesize}
\newcommand{\sortNumberCell}{\cellcolor{sort}\color{white}\footnotesize}

\newcommand{\directionCell}{\cellcolor{gray!20}}
\newcommand{\resultOriginCell}{\tt}
\newcommand{\edgeDirectionCell}{\tt}
\newcommand{\cardinalCell}{\tt}

\newcommand{\varNameText}{\tt}
\newcommand{\varNameCell}{\varNameText\raggedright}

\newcommand{\typeText}{\footnotesize\sf}
\newcommand{\typeCellBase}{\cellcolor{gray!20}\typeText}
\newcommand{\typeCell}{\typeCellBase\raggedright}

\newcommand{\chokePoint}[1]{\hyperref[choke_point_#1]{#1}}

\newcommand{\innerCardVSpace}{\vspace{1.1ex}}
\newcommand{\queryCardVSpace}{\vspace{2ex}}

\newcolumntype{Y}{>{\raggedright\arraybackslash}X}

\newcolumntype{C}[1]{>{\centering\let\newline\\\arraybackslash\hspace{0pt}}m{#1}}

\newcolumntype{M}{>{\begin{varwidth}{3.8cm}}Sl<{\end{varwidth}}}

\definecolor{lightgray}{RGB}{242,242,242}
\definecolor{keywordcolor}{RGB}{0,0,160}
\definecolor{commentcolor}{RGB}{0,128,64}
\definecolor{instruction}{HTML}{107762}

\lstdefinelanguage{cypher}
{
	morekeywords={
			MATCH, OPTIONAL, WHERE, NOT, AND, OR, XOR, RETURN, DISTINCT, ORDER, BY, ASC, ASCENDING, DESC, DESCENDING, UNWIND, AS, UNION, WITH, ALL, CREATE, DELETE, DETACH, REMOVE, SET, MERGE, SET, SKIP, LIMIT, IN, CALL, CASE, WHEN,
			INDEX, DROP, UNIQUE, CONSTRAINT, EXPLAIN, PROFILE, START, FOREACH, 
			GROUP, HAVING,
		},
	sensitive=true,
	morecomment=[l]{//},
	morecomment=[s]{/*}{*/},
	morestring=[b]{"},
	literate=*
		{<<}{\color{instruction}\guillemotleft{}}{1}
		{>>}{\textcolor{instruction}{\guillemotright{}}\color{black}}{1}
}

\lstdefinelanguage{sparql}{
	morekeywords={SELECT, DISTINCT, WHERE, OPTIONAL, FILTER, NOT, EXISTS, MINUS, sameTerm, bound},
}

\WPTitle{Financial Benchmark Task Force}

\renewcommand{\delIDText}{}

\dissPU 

\natR 

\author{LDBC Financial Benchmark Task Force}


\bibliography{bib/references}

\setboolean{standalone}{false}

\abstract{
    Motivated by \ldbcsnb, LDBC FinBench (Financial Benchmark) intends to define
    a benchmark characterized by special data and query patterns in financial
    industry to test graph database systems to make the evaluation of graph
    databases representative, reliable and comparable, especially in financial
    scenarios.

    Similar to \ldbcsnb, LDBC FinBench consists of two workloads that focus on
    different functionalities: the Transaction workload and the Analytics
    workload (future work for now). This document contains the definition of
    workloads including a detailed description of the datasets and queries, and
    also an explanation about the workflow to use the benchmark.
}

\execSummary{
    Inspired by \ldbcsnb (LDBC's Social Network Benchmark), a task force is
    organized by AntGroup(Ant Group Co., Ltd.) and formed by the principal actors
    in the field of financial graph-like data management under the guidance and
    help from LDBC to design \ldbcfinbench (LDBC's Financial Benchmark) which is
    more applicable to financial scenarios. The task force is committed to
    define a framework that can fairly test and compare different graph-based
    technologies where the dataset and workload are carefully designed with the
    rich practical experience of members in the financial industry by hosting
    the financial business itself or serving other financial entities. \ldbcfinbench
    is an industrial and academic initiative that is distinguished and
    characterized by the special features and patterns in the financial industry.

    In this version, the task force has finished the design of the benchmark
    framework without the analytics workload which is future work. Meantime, the
    task force has also organized a developer group to develop the benchmark
    suite for \ldbcfinbench. The benchmark suite is currently under development
    according to the benchmark framework design. In the future, \ldbcfinbench
    will be improved continuously with more feedback and more workloads
    including analytics workload will be designed and added. Please feel free to
    contact us if you have some suggestions, or if you are interested in joining
    in \ldbcfinbench.

    This document contains:
    \begin{itemize}
        \item A detailed specification of the data and workloads in the whole
        \ldbcfinbench.
        \item A detailed specification of the workflow and instructions about
        how to use the benchmark suite.
        \item A detailed specification of the auditing rules and the full
        disclosure report's required contents.
    \end{itemize}
}

\begin{document}

\maketitle

\chapter*{Acknowledgments}
\label{sec:acknowledgments}

\renewcommand{\labelitemii}{\textbullet}

Special thanks to the people who have actively contributed to the development of the benchmark suite:

\begin{itemize}
  \item Zhihui Guo, the chair of the FinBench Task Force (Ant Group)
  \item Shipeng Qi, the open-source projects leader of FinBench (Ant Group)
  \item Heng Lin (Ant Group)
  \item Bing Tong (CreateLink)
  \item Yan Zhou (CreateLink)
  \item Bin Yang (Ultipa)
  \item Jiansong Zhang (Ultipa)
  \item Youren Shen (StarGraph)
  \item Zheng Wang (StarGraph)
  \item Changyuan Wang (Vesoft)
  \item Parviz Peiravi (Intel)
  \item Gábor Szárnyas, the lead of LDBC SNB Task Force (CWI)
\end{itemize}

\chapter*{Definitions}

{\flushleft \textbf{\DataGen:}} The data generator provided by the
\ldbcfinbench, which is responsible for generating the data needed to run the
benchmark.

{\flushleft \textbf{DBMS:}} A DataBase Management System. 

{\flushleft \textbf{\ldbcfinbench:}} Linked Data Benchmark Council Financial
Benchmark. 

{\flushleft \textbf{Query Mix:}} Refers to the ratio between read and update
queries of a workload, and the frequency at which they are issued.

{\flushleft \textbf{SF (Scale Factor):}} The \ldbcfinbench is designed to target
systems of different sizes and scales. The scale factor determines the size of the
data used to run the benchmark, measured in Gigabytes.

{\flushleft \textbf{SUT:}} The System Under Test  is defined to be the database
system where the benchmark is executed.

{\flushleft \textbf{Test Driver:}}  A program provided by the \ldbcfinbench,
which is responsible for executing the different workloads and gathering the
results.

{\flushleft \textbf{Full Disclosure Report (FDR):}} The FDR is a document that
allows the reproduction of any benchmark result by a third-party. This contains
a complete description of the SUT and the circumstances of the benchmark run, \eg
the configuration of SUT, dataset and test driver, \etc

{\flushleft \textbf{Test Sponsor:}} The Test Sponsor is the company officially
submitting the Result with the FDR and will be charged the filing fee. Although
multiple companies may sponsor a Result together, for the purposes of the LDBC
processes the Test Sponsor must be a single company. A Test Sponsor need not be
a LDBC member. The Test Sponsor is responsible for maintaining the FDR with any
necessary updates or corrections. The Test Sponsor is also the name used to
identify the Result.

%


{\flushleft \textbf{Workload:}} A workload refers to a set of queries of a given nature
(\ie interactive, analytical, business), how they are issued and at which rate.

\chapter{Introduction}
\label{sec:introduction}


\section{Motivation}

Inspired by \ldbcsnb, a task force proposed by AntGroup~\cite{antgroup} is
formed by the principal actors in the field of financial graph-like data
management with help from LDBC to design a new benchmark, \ldbcfinbench (LDBC's
Financial Benchmark). The task force intends to define a framework that is more
applicable to financial scenarios to fairly test and compare different graph-based
technologies. To this end, they carefully design the dataset and workload using
their rich practical experience as members of the financial industry. \ldbcfinbench
is distinguished and characterized by the special features and patterns in
the financial industry.


\section{Relevance to the Industry}

\ldbcfinbench is intended to provide the following value to these relevant
stakeholders:

\begin{itemize}
      \item For \textbf{users} facing graph processing tasks in the financial industry,
            \ldbcfinbench provides a recognizable scenario against which it is possible
            to compare the merits of different products and technologies. By covering
            a wide variety of scales and price points, \ldbcfinbench can serve as an
            aid to technology selection.
      \item For \textbf{vendors} of graph database technology, \ldbcfinbench provides a
            checklist of features and performance characteristics that helps in product
            positioning and can serve to guide new development.
      \item For \textbf{researchers}, both industrial and academic, the \ldbcfinbench
            dataset and workload provide interesting challenges in multiple choke point
            areas, and help compare the efficiency of existing technology in these
            areas.
\end{itemize}

The technological scope of \ldbcfinbench comprises all systems that one might
conceivably use to perform financial data management tasks including
\textbf{Graph database management systems} (\eg Neo4j, TuGraph, Galaxybase, etc.), \textbf{
      Graph processing frameworks} (\eg Giraph, Ligra, etc.), \textbf{RDF database
      systems} (\eg Virtuoso, AWS Neptune, etc.), \textbf{Relational database systems}
(\eg MySQL, Oracle, etc.), \textbf{NoSQL database systems} (\eg key-value stores
such as HBase, Redis, MongoDB, CouchDB, or even MapReduce systems like Hadoop
and Pig).


\section{Participation of Industry and Academia}

Initially, the \ldbcfinbench task force is formed by relevant actors mainly from
industry. In the process of design and development, we also received supports and
suggestions from fellows in academia. All the participants have contributed with
their experience and expertise to make this benchmark a credible effort. The list
of participants is as follows.

\begin{itemize}
      \item AntGroup (entity)
      \item CreateLink (entity)
      \item Ultipa (entity)
      \item StarGraph (entity)
      \item Vesoft (entity)
      \item Pometry (entity)
      \item Katana (entity)
      \item Intel (entity)
      \item TigerGraph (entity)
      \item Koji Annoura (individual)
\end{itemize}


\section{Software Components}
\label{sec:software-components}

The source code of this specification and the benchmark suite is available
open-source:
\begin{itemize}
      \item \ldbcfinbench Specification: \url{https://github.com/ldbc/ldbc_finbench_docs}
      \item \ldbcfinbench Data Generator: \url{https://github.com/ldbc/ldbc_finbench_DataGen}
      \item \ldbcfinbench Driver: \url{https://github.com/ldbc/ldbc_finbench_driver}
      \item Transaction Workload Implementation: \url{https://github.com/ldbc/ldbc_finbench_transaction_impls}
      \item Analytics Workload: future work
\end{itemize}

Note that the \texttt{main} branch for these repositories is under development
by default. Please refer to the releases and branch started with \texttt{v} and
named \texttt{vX.X.X} for stable versions.


\section{Related Projects}

Along with \ldbcfinbench, LDBC~\cite{DBLP:journals/sigmod/AnglesBLF0ENMKT14}
provides other benchmarks as well:

\begin{itemize}
      \item \ldbcsnb measures the performance of \emph{all systems relevant to
                  linked data} operating a social network.
      \item The Semantic Publishing Benchmark
            (SPB)~\cite{DBLP:conf/semweb/SpasicJP16} measures the performance of
            \emph{semantic databases} operating on RDF datasets.
      \item The Graphalytics
            benchmark~\cite{DBLP:journals/pvldb/IosupHNHPMCCSAT16} measures the
            performance of \emph{graph analysis} operations (\eg PageRank, local
            clustering coefficient).
\end{itemize}

\chapter{Benchmark Overview}
\label{sec:benchmark-overview}

\section{Practice basis}

The task force members design \ldbcfinbench with their rich practical experience in
financial industry based on a comprehensive survey of financial scenarios including
Risk Control, AML (Anti-Money Laundering), KYC (Know Your Customer), Stock Recommendation
and so on.


\section{Design Concepts}

\ldbcfinbench is intended to be a credible, fair and representative benchmark.
It's designed with the following concepts:

\begin{itemize}
      \item \textbf{Based on real systems}. The task force members gathering
            together from industry and academia intend to design \ldbcfinbench
            to express and emphasize the special patterns of data and workload
            distinguished from other popular benchmarks. To do that,
            \ldbcfinbench is designed based on the rich practical experience of
            members and additional surveys.
      \item \textbf{Comprehensive and complete.} \ldbcfinbench is intended to
            cover most demands encountered in the management of complexly
            structured data in financial scenarios.
      \item \textbf{Challenging and instructive.} Benchmarks are known to direct
            product development in certain directions. \ldbcfinbench is informed
            by state-of-the-art in database research and industry practice
            to offer optimization challenges.
      \item \textbf{Easy to use and extendable.} As a benchmark offering value
            to many relevant stakeholders, \ldbcfinbench is designed to be easy
            to use. The effort for obtaining test results with it should be
            small.
      \item \textbf{Modularized.} \ldbcfinbench is broken into parts both in
            design and benchmark suite that can be individually addressed to
            stimulate innovation without imposing an overly high threshold for
            participation.
      \item \textbf{Reproducible and documented.} \ldbcfinbench is intended to
            specify the auditing rules and provide full disclosure reports of
            auditing of benchmark runs in accordance with the LDBC
            Bylaws~\cite{ldbc_byelaws}.
\end{itemize}


\section{New features in FinBench}

\ldbcsnb, one of the earlier LDBC benchmarks, is modeled around the operation of a real social network site. It defines a data schema that represents a realistic social network including people and their activities during a period of time and also the workloads mimic the different usage scenarios found in operating a real social network site. Compared with \ldbcsnb, \ldbcfinbench is characterized by the special features and patterns of the data schema and queries that represent the characteristics of financial scenarios.

\subsection{Data Schema}

The data schema for \ldbcfinbench is designed to reflect the real data in the financial systems. Frequent
financial entities in real systems include accounts, medium, persons, companies, loans, etc. The
entities are vertices in the data schema while the edges reflect financial activities, \eg fund transferred
from one account to another. In their data schema, financial scenarios have these distinguished characteristics
compared to regular social networks.
\begin{itemize}
      \item Multiple edges can exist between two vertices, \eg Many transfer records exist between two accounts
      \item Dynamic attribute exists in vertex to mark entities status, \eg an account is marked as blocked
      \item Quantity attribute exists in edge, \eg Transfer edge has quantity attribute amount
\end{itemize}

The designed data schema is specified in \autoref{sec:data-definition}.

\subsection{Workloads}

In workloads and queries, financial scenarios have these distinguished characteristics.
\begin{itemize}
      \item More tight latency, \eg some queries need to return in less than 100ms.
      \item Write operations updating attributes, \eg marking an account as blocked.
      \item Recursive Path Filtering. Some queries filter data with backward dependency
            in variable-length paths, \eg finding all transfer paths A-[${e_1}$]->..-[${e_k}$]->B
            where the timestamp of each transfer edge ${e_i}$ in the path is larger than that of
            the previous ${e_{i-1}}$. In this pattern, the variable length path is qualified by
            the edge quantity attributes or the aggregation in the path, either along one path
            or a set of paths.
      \item Read-write Query, which is a query sequence with a mix of reads and writes reflecting the
            complexity of financial systems. Read-write query describes a desired pattern that risk control
            policies are checked, and corresponding actions are taken before financial activities like
            transfers are written down to storage. See \autoref{sec:read-write-query} for details.
      \item Truncation. In financial scenarios, the degree of hub vertex may reach million and even
            billion scales, especially when traversing on a graph. To handle the discordance between the tight
            latency requirements and power-law distribution of data in the system, truncation is introduced
            to reduce the complexity of queries. See \autoref{sec:truncation-on-hub-vertices} for details.
\end{itemize}

In \ldbcfinbench, there are two kinds of workloads:
\begin{itemize}
      \item Transaction Workload. It includes queries with a tight latency bound, which are usually
            queries hopping a few steps from a start vertice. There are complex reads, simple reads, write
            operations, and read-write queries in transaction workload. The Transaction Workload is specified
            in \autoref{sec:transaction-workload}.
      \item Analytics Workload. It is supposed to include more complicated queries, \eg triggers and pre-computed
            values in online systems. This part is future work that will be designed and discussed in the
            following versions. The Analytics Workload is specified in \autoref{sec:analytics-workload}.
\end{itemize}


\section{Benchmark Workflow}

See \autoref{sec:auditing-rules} for the execution workflow of \ldbcfinbench.


\chapter{Data Definition}
\label{sec:data-definition}

This chapter describes the dataset used by \ldbcfinbench, including the data
schema design and the data generation process. Generally, we design
\ldbcfinbench balancing reality and abstraction. There are some annotations
about the compromises in data design,
\begin{itemize}
    \item Although multiple persons/companies may own the same account in
          reality, in the schema, an account is owned by only a single person
          or company for simplicity.
    \item Although rejected transactions may be recorded to support future loan
          decisions, only approved transactions/transfers are recorded in the
          benchmark dataset.
    \item Considering the number of daily active users (DAU) of financial systems
          in reality, there will be many signIn edges between medium and account
          vertices. However, we do not generate so many signIn edges aligning to
          reality with a limit in the simulation of the data generation process
          since systems usually circumvent the problem by adding a medium attribute
          to edges like transfer and withdraw to record the medium users used.
\end{itemize}

\section{Data Types}

\autoref{table:types} describes the different data types used in the benchmark.
Compared with \ldbcsnb, there is a new compound type, \textbf{Path}, which is
widely applied in financial scenarios reflecting traces, \eg fund transfer
traces.

\begin{table}[h]
    \centering
    \begin{tabular}{|>{\typeCell}p{\attributeColumnWidth}|p{\largeDescriptionColumnWidth}|}
        \hline
        \tableHeaderFirst{Type} & \tableHeader{Description}                      \\
        \hline
        ID                      & Integer type with 64-bit precision. All IDs
        within a single entity type (\eg Person) are unique, but different
        entity types (\eg a Person and an Account) might have the same ID.       \\
        \hline
        32-bit Integer          & Integer type with 32-bit precision             \\
        \hline
        64-bit Integer          & Integer type with 64-bit precision             \\
        \hline
        32-bit Float            & Floating type with 32-bit precision            \\
        \hline
        64-bit Float            & Floating type with 64-bit precision            \\
        \hline
        String                  & Variable length text of size 40 Unicode
        characters                                                               \\
        \hline
        Long String             & Variable length text of size 256 Unicode
        characters                                                               \\
        \hline
        Text                    & Variable length text of size 2000 Unicode
        characters                                                               \\
        \hline
        Date                    & Date with a precision of a day, encoded as a
        string with the following format: \textit{yyyy-mm-dd}, where
        \textit{yyyy} is a four-digit integer representing the year, the year,
        \textit{mm} is a two-digit integer representing the month and
        \textit{dd} is a two-digit integer representing the day.                 \\
        \hline
        DateTime                & Date with a precision of milliseconds, encoded
        as a string with the following format:
        \textit{yyyy-mm-ddTHH:MM:ss.sss+0000}, where \textit{yyyy} is a
        four-digit integer representing the year, the year, \textit{mm} is a
        two-digit integer representing the month and \textit{dd} is a two-digit
        integer representing the day, \textit{HH} is a two-digit integer
        representing the hour, \textit{MM} is a two-digit integer representing
        the minute and \textit{ss.sss} is a five-digit fixed point real number
        representing the seconds up to millisecond precision. Finally, the
        \textit{+0000} of the end represents the timezone, which in this case is
        always GMT.                                                              \\
        \hline
        Boolean                 & A logical type taking the value of either True
        of False.                                                                \\
        \hline
        Enum                    & Enumeration type                               \\
        \hline
        Path                    & A compound type representing a trace which is
        expressed in an ordered sequence of vertices' IDs in the trace. For
        example, [1,3,4,8] expresses a trace 1->3->4->8.                         \\
        \hline
    \end{tabular}
    \caption{Description of the data types.}
    \label{table:types}
\end{table}

\subsection{Enumerations}
{\flushleft \textbf{TRUNCATION\_ORDER:}} The enumeration describes the sort
order before truncation. \textbf{TIMESTAMP\_ASCENDING} means truncation on
ascending order of timestamp.

\section{Data Schema}

\begin{figure}[htbp]
    \centering
    \includegraphics[width=\linewidth]{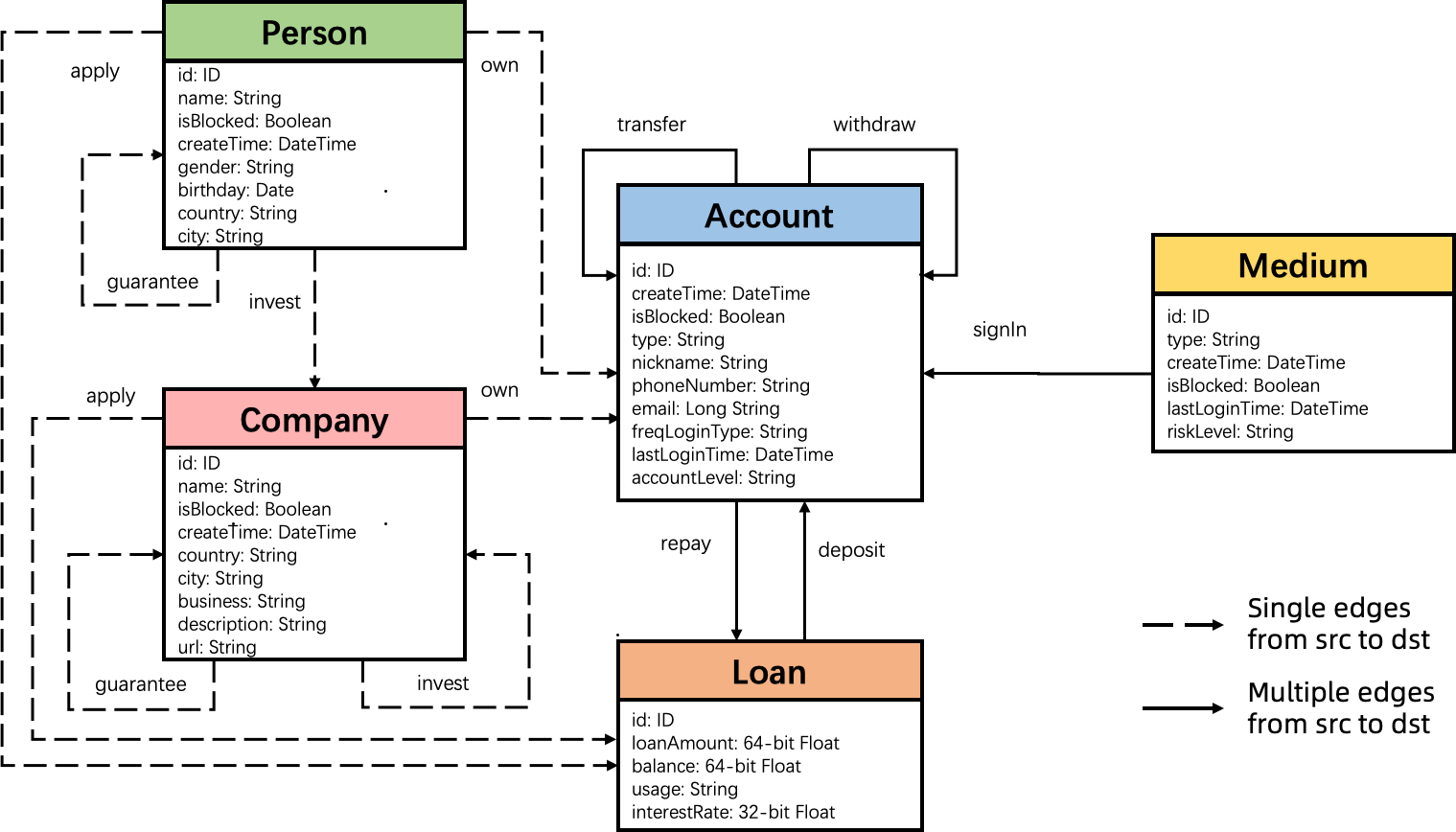}
    \caption{The \ldbcfinbench data schema}
    \label{figure:schema}
\end{figure}

\autoref{figure:schema} shows the data schema in UML. The schema defines the
structure of the data used in the benchmark in terms of entities and their
relations. The data represents a snapshot of the activity in several financial
scenarios during a period of time. The schema specifies different entities,
their attributes, and their relations. All of them are described in the
following sections.

\subsection{Entities}

{\flushleft \textbf{Person:}} A person of the real world. \autoref{table:person}
shows the attributes.
\begin{table}[H]
    \begin{tabular}{|>{\varNameCell}p{\attributeColumnWidth}|>{\typeCell}p{\typeColumnWidth}|p{\descriptionColumnWidth}|}
        \hline
        \tableHeaderFirst{Attribute} & \tableHeader{Type} &
        \tableHeader{Description}                                                                             \\
        \hline
        id                           & ID                 & The identifier of the person.                     \\
        \hline
        name                         & String             & The name of the person.                           \\
        \hline
        isBlocked                    & Boolean            & If the person is blocked or concerned in systems. \\
        \hline
        createTime                   & DateTime           & The time when the person created.                 \\
        \hline
        gender                       & String             & Gender of the person                              \\
        \hline
        birthday                     & Date               & Birthday of the person                            \\
        \hline
        country                      & String             & Country of the person                             \\
        \hline
        city                         & String             & City of the person                                \\
        \hline
    \end{tabular}
    \caption{Attributes of Person entity.}
    \label{table:person}
\end{table}

{\flushleft \textbf{Company:}} A company of the real world, which persons or other companies invest in. \autoref{table:company} shows the attributes.
\begin{table}[H]
    \begin{tabular}{|>{\varNameCell}p{\attributeColumnWidth}|>{\typeCell}p{\typeColumnWidth}|p{\descriptionColumnWidth}|}
        \hline
        \tableHeaderFirst{Attribute} & \tableHeader{Type} &
        \tableHeader{Description}                                                                              \\
        \hline
        id                           & ID                 & The identifier of the company.                     \\
        \hline
        name                         & String             & The name of the company.                           \\
        \hline
        isBlocked                    & Boolean            & If the company is blocked or concerned in systems. \\
        \hline
        createTime                   & DateTime           & The time when the company is created.              \\
        \hline
        country                      & String             & Country of the company                             \\
        \hline
        city                         & String             & City of the company                                \\
        \hline
        business                     & String             & The main business of the company                   \\
        \hline
        description                  & Long String        & The description of the company                     \\
        \hline
        url                          & String             & The url of the company's official site             \\
        \hline
    \end{tabular}
    \caption{Attributes of Company entity.}
    \label{table:company}
\end{table}

{\flushleft \textbf{Account:}} An account in real-world financial systems, which
is registered and owned by persons and companies. It includes many types such as
personalDeposit, personalCredit, etc. It can deal with other accounts.
\autoref{table:account} shows the attributes.
\begin{table}[H]
    \begin{tabular}{|>{\varNameCell}p{\attributeColumnWidth}|>{\typeCell}p{\typeColumnWidth}|p{\descriptionColumnWidth}|}
        \hline
        \tableHeaderFirst{Attribute} & \tableHeader{Type} &
        \tableHeader{Description}                                                                              \\
        \hline
        id                           & ID                 & The identifier of the account.                     \\
        \hline
        createTime                   & DateTime           & The time when the account is created.              \\
        \hline
        isBlocked                    & Boolean            & If the account is blocked or concerned in systems. \\
        \hline
        type                         & String             & The type of the account.                           \\
        \hline
        nickname                     & String             & The nickname of the account.                       \\
        \hline
        phoneNumber                  & String             & The phone number of the account.                   \\
        \hline
        email                        & String             & The email of the account.                          \\
        \hline
        freqLoginType                & String             & The frequent login type of the account.            \\
        \hline
        lastLoginTime                & DateTime           & The last login time of the account.                \\
        \hline
        accountLevel                 & String             & The level of the account.                          \\
        \hline
    \end{tabular}
    \caption{Attributes of Account entity.}
    \label{table:account}
\end{table}

{\flushleft \textbf{Loan:}} A loan for persons and companies to apply in real
world. \autoref{table:loan} shows the attributes.
\begin{table}[H]
    \begin{tabular}{|>{\varNameCell}p{\attributeColumnWidth}|>{\typeCell}p{\typeColumnWidth}|p{\descriptionColumnWidth}|}
        \hline
        \tableHeaderFirst{Attribute} & \tableHeader{Type} &
        \tableHeader{Description}                                                        \\
        \hline
        id                           & ID                 & The identifier of the loan.  \\
        \hline
        loanAmount                   & 64-bit Float       & The amount of a loan.        \\
        \hline
        balance                      & 64-bit Float       & The balance of a loan.       \\
        \hline
        usage                        & String             & The usage of a loan.         \\
        \hline
        interestRate                 & 32-bit Float       & The interest rate of a loan. \\
        \hline
    \end{tabular}
    \caption{Attributes of Loan entity.}
    \label{table:loan}
\end{table}

{\flushleft \textbf{Medium:}} An abstract standing for things that users use to
sign in account in the real world, such as IP address, MAC address, phone numbers.
\autoref{table:medium} shows the attributes.
\begin{table}[H]
    \begin{tabular}{|>{\varNameCell}p{\attributeColumnWidth}|>{\typeCell}p{\typeColumnWidth}|p{\descriptionColumnWidth}|}
        \hline
        \tableHeaderFirst{Attribute} & \tableHeader{Type} & \tableHeader{Description}                         \\
        \hline
        id                           & ID                 & The identifier of the medium.                     \\
        \hline
        type                         & String             & The medium type, \eg POS, IP.                     \\
        \hline
        createTime                   & DateTime           & The time when the medium is created.              \\
        \hline
        isBlocked                    & Boolean            & If the medium is blocked or concerned in systems. \\
        \hline
        lastLoginTime                & DateTime           & The last login time of the medium.                \\
        \hline
        riskLevel                    & String             & The risk level of the medium.                     \\
        \hline
    \end{tabular}
    \caption{Attributes of Medium entity.}
    \label{table:medium}
\end{table}

\subsection{Relations}
Relations connect entities of different types showed in
\autoref{table:relations}. Except that own has no attributes, the
attributes of other relations are shown in the following tables. Note that the
Cardinality means the cardinal relationship from the tail to the head of the edge
type and the Multiplicity means how many edges exist from the same tail to the
same head. For example, the 1 : N cardinality of own means an account can
only be owned by a person or a company.

\begin{longtable}{|>{\centering\varNameCell}p{1.5cm}|>{\typeCell}p{1.5cm}|>{\centering\cardinalCell}p{2cm}|>{\typeCell}p{1.5cm}|>{\centering\edgeDirectionCell}p{2cm}|p{5.5cm}|}
    \hline
    \tableHeaderFirst{Name} & \tableHeader{Tail}       & \tableHeader{Cardinality} & \tableHeader{Head}       & \tableHeader{Multiplicity} & \tableHeader{Description}                                            \\
    \hline
    signIn                  & Medium                   & N:N                       & Account                  & N                          & An account signed in with a media.                                   \\
    \hline
    own                     & Person/ \newline Company & 1:N                       & Account                  & 1                          & An account owned by a person or a company.                           \\
    \hline
    transfer                & Account                  & N:N                       & Account                  & N                          & Fund transferred between two accounts.                               \\
    \hline
    withdraw                & Account                  & N:N                       & Account                  & N                          & Fund transferred from an account to another account of type card.    \\
    \hline
    apply                   & Person/ \newline Company & 1:N                       & Loan                     & 1                          & A person or a company applies a Loan.                                \\
    \hline
    deposit                 & Loan                     & N:N                       & Account                  & N                          & Loan fund deposited to an account.                                   \\
    \hline
    repay                   & Account                  & N:N                       & Loan                     & N                          & Loan repaid from an account.                                         \\
    \hline
    invest                  & Person/ \newline Company & N:N                       & Company                  & 1                          & A person or a company invests into a company.                        \\
    \hline
    guarantee               & Person/ \newline Company & N:N                       & Person/ \newline Company & 1                          & A person or a company guarantees another for some reason like loans. \\
    \hline
    \caption{Description of the data relations.}
    \label{table:relations}
\end{longtable}

{\flushleft \textbf{transfer:}} Fund transfers between accounts. \autoref{table:transfer} shows the attributes.
\begin{table}[H]
    \begin{tabular}{|>{\varNameCell}p{\attributeColumnWidth}|>{\typeCell}p{\typeColumnWidth}|p{\descriptionColumnWidth}|}
        \hline
        \tableHeaderFirst{Attribute} & \tableHeader{Type} & \tableHeader{Description}         \\
        \hline
        timestamp                    & DateTime           & The time when transfer issues.    \\
        \hline
        amount                       & 64-bit Float       & The amount of the transfer.       \\
        \hline
        ordernumber                  & String             & The order number of the transfer. \\
        \hline
        comment                      & String             & The comment of the transfer.      \\
        \hline
        payType                      & String             & The pay type of the transfer.     \\
        \hline
        goodsType                    & String             & The goods type of the transfer.   \\
        \hline
    \end{tabular}
    \caption{Attributes of transfer relation.}
    \label{table:transfer}
\end{table}

{\flushleft \textbf{withdraw:}} Fund is transferred from one account to another of type card. \autoref{table:withdraw} shows the attributes.
\begin{table}[H]
    \begin{tabular}{|>{\varNameCell}p{\attributeColumnWidth}|>{\typeCell}p{\typeColumnWidth}|p{\descriptionColumnWidth}|}
        \hline
        \tableHeaderFirst{Attribute} & \tableHeader{Type} & \tableHeader{Description}      \\
        \hline
        timestamp                    & DateTime           & The time when withdraw issues. \\
        \hline
        amount                       & 64-bit Float       & The amount of the withdraw.    \\
        \hline
    \end{tabular}
    \caption{Attributes of withdraw relation.}
    \label{table:withdraw}
\end{table}

{\flushleft \textbf{repay:}} Loan is repaid from an account. \autoref{table:repay} shows the attributes.
\begin{table}[H]
    \begin{tabular}{|>{\varNameCell}p{\attributeColumnWidth}|>{\typeCell}p{\typeColumnWidth}|p{\descriptionColumnWidth}|}
        \hline
        \tableHeaderFirst{Attribute} & \tableHeader{Type} & \tableHeader{Description}   \\
        \hline
        timestamp                    & DateTime           & The time when repay issues. \\
        \hline
        amount                       & 64-bit Float       & The amount of the repay.    \\
        \hline
    \end{tabular}
    \caption{Attributes of repay relation.}
    \label{table:repay}
\end{table}

{\flushleft \textbf{deposit:}} Loan fund is deposited to an account. \autoref{table:deposit} shows the attributes.
\begin{table}[H]
    \begin{tabular}{|>{\varNameCell}p{\attributeColumnWidth}|>{\typeCell}p{\typeColumnWidth}|p{\descriptionColumnWidth}|}
        \hline
        \tableHeaderFirst{Attribute} & \tableHeader{Type} & \tableHeader{Description}     \\
        \hline
        timestamp                    & DateTime           & The time when deposit issues. \\
        \hline
        amount                       & 64-bit Float       & The amount of the deposit.    \\
        \hline
    \end{tabular}
    \caption{Attributes of deposit relation.}
    \label{table:deposit}
\end{table}

{\flushleft \textbf{signIn:}} An account is signed in with a Media. \autoref{table:signIn} shows the attributes.
\begin{table}[H]
    \begin{tabular}{|>{\varNameCell}p{\attributeColumnWidth}|>{\typeCell}p{\typeColumnWidth}|p{\descriptionColumnWidth}|}
        \hline
        \tableHeaderFirst{Attribute} & \tableHeader{Type} & \tableHeader{Description}     \\
        \hline
        timestamp                    & DateTime           & The time when signIn happens. \\
        \hline
        location                     & String             & The location of the signIn.   \\
        \hline
    \end{tabular}
    \caption{Attributes of signIn relation.}
    \label{table:signIn}
\end{table}

{\flushleft \textbf{invest:}} A person or a company invests in a company. \autoref{table:invest} shows the attributes.
\begin{table}[H]
    \begin{tabular}{|>{\varNameCell}p{\attributeColumnWidth}|>{\typeCell}p{\typeColumnWidth}|p{\descriptionColumnWidth}|}
        \hline
        \tableHeaderFirst{Attribute} & \tableHeader{Type} & \tableHeader{Description}             \\
        \hline
        timestamp                    & DateTime           & The time when the investment happens. \\
        \hline
        ratio                        & 32-bit Float       & The ratio of the investment.          \\
        \hline
    \end{tabular}
    \caption{Attributes of invest relation.}
    \label{table:invest}
\end{table}

{\flushleft \textbf{apply:}} A person or a company applies for a Loan. \autoref{table:apply} shows the attributes.
\begin{table}[H]
    \begin{tabular}{|>{\varNameCell}p{\attributeColumnWidth}|>{\typeCell}p{\typeColumnWidth}|p{\descriptionColumnWidth}|}
        \hline
        \tableHeaderFirst{Attribute} & \tableHeader{Type} & \tableHeader{Description}      \\
        \hline
        timestamp                    & DateTime           & The time when apply happens.   \\
        \hline
        organization                 & String             & The organization for the loan. \\
        \hline
    \end{tabular}
    \caption{Attributes of apply relation.}
    \label{table:apply}
\end{table}

{\flushleft \textbf{guarantee:}} A person or a company guarantees another for some reason like Loans. \autoref{table:guarantee} shows the attributes.
\begin{table}[H]
    \begin{tabular}{|>{\varNameCell}p{\attributeColumnWidth}|>{\typeCell}p{\typeColumnWidth}|p{\descriptionColumnWidth}|}
        \hline
        \tableHeaderFirst{Attribute} & \tableHeader{Type} & \tableHeader{Description}                       \\
        \hline
        timestamp                    & DateTime           & The time when guarantee happens.                \\
        \hline
        relationship                 & String             & The relationship between guarantor and applier. \\
        \hline
    \end{tabular}
    \caption{Attributes of guarantee relation.}
    \label{table:guarantee}
\end{table}

{\flushleft \textbf{own:}} A person or a company owns an account. This relation has no attributes.
\begin{table}[H]
    \begin{tabular}{|>{\varNameCell}p{\attributeColumnWidth}|>{\typeCell}p{\typeColumnWidth}|p{\descriptionColumnWidth}|}
        \hline
        \tableHeaderFirst{Attribute} & \tableHeader{Type} & \tableHeader{Description}        \\
        \hline
        timestamp                    & DateTime           & The time when guarantee happens. \\
        \hline
    \end{tabular}
    \caption{Attributes of guarantee relation.}
    \label{table:guarantee}
\end{table}

\section{Data Generation}

The data generation process is designed to produce a dataset that is as close as possible to the real-world data. The
data generator stimulates real-world financial activities in systems and generates the data according to the data
schema. See the data generator for more details at \url{https://github.com/ldbc/ldbc_finbench_DataGen}.

\section{Output Data}

\subsection{Data Precision}

The datasets are designed and created closely resembling real-world scenarios. {\DataGen} produces
financial data having the precision as follows:
\begin{itemize}
    \item The generated 64-bit Float numbers will have precision up to two decimal places for both
          the amount and balance values.
    \item The timestamps are generated with millisecond precision.
\end{itemize}

\subsection{Scale Factors}
\label{sec:scale-factors}

\ldbcfinbench defines a set of scale factors (SFs), targeting systems of different sizes and budgets. Namely, the SF1
dataset is 1 GiB, the SF10 is 10 GiB. In the initial version, CSV serializer is provided. We use the default settings
to split the data into an initial (bulk-loaded) dataset and incremental data, 97\% for initial data and 3\% for
incremental data. The currently available SFs are the following: 0.01, 0.1, 0.3, 1, 3, 10. By default, all SFs are
defined over three years,  starting from 2020, and SFs are computed by scaling the number of Persons and Companies in
the network. Please refer to \autoref{sec:sf-statistics} for the metrics of datasets of different scales.

\chapter{Workloads}
\label{sec:workloads}

\section{Query Annotations}

This section describes how to read the query cards in the following sections.
\subsection{Query Description Format}
\label{subsec:query-description-format}

Queries are described in natural language using a well-defined structure that consists of three sections:
\textit{description}, a concise textual description of the query,
\textit{parameters}, a list of input parameters and their types;
\textit{results}, a list of expected results and their types.
Additionally, queries returning multiple results specify \emph{sorting criteria} and a \emph{limit} (to return top-$k$ results).

We use the following notation:

\begin{itemize}
	\item \textbf{Vertex type}: vertice type in the dataset.
		One word, possibly constructed by appending multiple words together, starting with an uppercase character and following the camel case notation,
        \eg \textsf{TagClass} represents an entity of type ``TagClass''.
    \item \textbf{Edge type}: edge type in the dataset.
        One word, possibly constructed by appending multiple words together, starting with a lowercase character and following the camel case notation
        \eg \mbox{\textsf{workAt}} represents an edge of type ``workAt''.
    \item \textbf{Attribute}: attribute of a vertice or an edge in the dataset.
        One word, possibly constructed by appending multiple words together, starting with a lowercase character and following the camel case notation,
        and prefixed by a ``.'' to dereference the vertice/edge,
        \eg \textsf{person.firstName} refers to ``firstName'' attribute on the ``person'' entity,
        and \mbox{\textsf{studyAt.classYear}} refers to ``classYear'' attribute on the ``studyAt'' edge.
    \item \textbf{Unordered Set}: an unordered collection of distinct elements.
        Surrounded by \{ and \} braces, with the element type between them,
        \eg \textsf{\{String\}} refers to a set of strings.
    \item \textbf{Ordered List}: an ordered collection where duplicate elements are allowed.
        Surrounded by [ and ] braces, with the element type between them,
        \eg \textsf{[String]} refers to a list of strings.
    \item \textbf{Ordered Tuple}: a fixed-length, fixed-order list of elements, where elements at each position of the tuple have predefined, possibly different, types.
        Surrounded by < and > braces, with the element types between them in a specific order
        \eg \textsf{<String, Boolean>} refers to a 2-tuple containing a string value in the first element and a boolean value in the second,
        and \textsf{[<String, Boolean>]} is an ordered list of those 2-tuples.
\end{itemize}

\paragraph{Categorization of results.}

Results are categorized according to their source of origin:

\begin{itemize}
	\item \textbf{Raw} (\texttt{R}), if the result attribute is returned with an unmodified value and type.
	\item \textbf{Calculated} (\texttt{C}), if the result is calculated from attributes using arithmetic operators, functions, boolean conditions, etc.
	\item \textbf{Aggregated} (\texttt{A}), if the result is an aggregated value, \eg a count or a sum of another value. If a result is both calculated and aggregated (\eg \lstinline{count(x) + count(y)} or \lstinline{avg(x + y)}), it is considered an aggregated result.
	\item \textbf{Meta} (\texttt{M}), if the result is based on type information, \eg the type of a vertice.
\end{itemize}


\subsection{Returned Values}
\label{subsec:returned-values}

Return values are subject to the following rules:
\begin{itemize}
    \item Path type. The Path type is a sequence of vertices and edges. The
          Path type is returned as a sequence of vertex and edge identifiers
          ignoring the multiple edges between the same src and dst vertex.
    \item Precision of results. In order to maintain consistency of the 
          benchmark results, all floating-point results are rounded to 3
          decimal places using standard rounding rules (i.e., round half up).
\end{itemize}


\subsection{Other Annotations}
\label{subsec:other-annotations}

To express the patterns better, the pattern diagrams are drawn from the perspective
of data rather than the matching pattern in the graph. Here are some annotations to each
query card in this section.
\begin{itemize}
    \item Each row in the result cell represents an attribute to be returned.
    \item The second column means the data type of returned attribute.
          If the type is surrounded by \type{\{\}}, it means that the result is a
          set, \eg \type{\{String\}} means a string set is returned.
    \item For each row in the result cell, the third column annotates the
          category of type of result attribute returned, including \texttt{R} short
          for Raw, \texttt{A} short for Aggregated, \texttt{C} short for Calculated,
          \texttt{S} short for Structural. Among them, structural type means types
          such as \type{Path} while raw type means basic types in contrast.
   \item In the pattern of each query, the gray dashed box encapsulates the results
         to return. And the black solid arrows represent the multiple edges from src
         to dst while the black dashed arrows represent the single edges from src to dst.
\end{itemize}


\section{Truncation on Hub Vertices}
\label{sec:truncation-on-hub-vertices}

The high degree of hub vertex is a common feature not only in financial scenarios but also in other scenarios, which is
an inevitable challenge that systems face. To solve the problem, systems can either improve the performance to satisfy
the computation or just reduce the complexity to meet the latency requirements.

The mechanism is to do truncation on the edges when traversing out from the current vertex, which complies with the
discordance. Truncating less-important edges is a useful and practical mechanism to handle the discordance between the
tight latency requirements and hub vertices in the system, where the degree of hub vertex may reach a million and even
billion scales, especially when traversing the graph. To maintain the consistency of the results, a sort order has
to be specified when truncating. Since in financial graphs, users prefer newer data in business. It is reasonable that
attribute, \emph{timestamp}, in the edges is used as the sort order in truncation. With the sort order, truncation is
namely a deterministic sampling in traversing.

In the following queries, some parameters are added to describe the behavior of truncation reducing the complexity
including the \emph{TRUNCATION\_LIMIT} and \emph{TRUNCATION\_ORDER}. \emph{TRUNCATION\_ORDER} can be
\emph{TIMESTAMP\_ASCENDING, TIMESTAMP\_DESCENDING, AMOUNT\_ASCENDING, AMOUNT\_DESCENDING}. At most time,
\emph{TRUNCATION\_ORDER} is set to \emph{TIMESTAMP\_DESCENDING} by default.


\section{Read Write Query}
\label{sec:read-write-query}

In financial scenarios, risk control is a kind of hot and significant application.
Such applications usually detect a specific pattern in the form of linked data before
new records like transfers are written to systems. Read-write query, which can also
be seen as transaction-wrapped strategies, fits these applications very well since
users do not need to worry about translating the patterns to prevent malicious records.
A read-write query is composed of read queries and write queries in the previous sections.
In most cases, whether to commit the write query depends on the detection result of the
read queries. In the initial version, just 3 read-write queries are presented.

\chapter{Transaction Workload}
\label{sec:transaction-workload}

This workload consists of a set of relatively simple read queries, write queries
and read-write operations that touch a significant amount of data. These
queries and operations are usually considered online data processing and
analysis in online financial systems. The LDBC FinBench transaction workload
consists of four query types:
\begin{itemize}
    \item Complex-read queries. See \autoref{sec:complex-read-queries}. This
          section contains many basic read queries that are typical in financial
          scenarios.
    \item Simple-read queries. See \autoref{sec:simple-read-queries}. This
          section contains many basic read queries that are typical in financial
          scenarios.
    \item Write queries. See \autoref{sec:write-queries}. This section contains
          many basic write queries that are typical in financial scenarios.
    \item Read-write queries. See \autoref{sec:rw-queries}. This section
          contains many read-write operations composed of basic reads and writes.
\end{itemize}

\section{Complex Read Queries}
\label{sec:complex-read-queries}

\renewcommand*{\arraystretch}{1.1}

\subsection*{Transaction / complex-read / 1}
\label{sec:transaction-complex-read-01}

\let\oldemph\emph
\renewcommand{\emph}[1]{{\footnotesize \sf #1}}
\let\oldtextbf\textbf
\renewcommand{\textbf}[1]{{\it #1}}\renewcommand{\currentQueryCard}{1}
\marginpar{
	\raggedleft
	\scriptsize

	\queryRefCard{transaction-complex-read-01}{TCR}{1}\\
	\queryRefCard{transaction-complex-read-02}{TCR}{2}\\
	\queryRefCard{transaction-complex-read-03}{TCR}{3}\\
	\queryRefCard{transaction-complex-read-04}{TCR}{4}\\
	\queryRefCard{transaction-complex-read-05}{TCR}{5}\\
	\queryRefCard{transaction-complex-read-06}{TCR}{6}\\
	\queryRefCard{transaction-complex-read-07}{TCR}{7}\\
	\queryRefCard{transaction-complex-read-08}{TCR}{8}\\
	\queryRefCard{transaction-complex-read-09}{TCR}{9}\\
	\queryRefCard{transaction-complex-read-10}{TCR}{10}\\
	\queryRefCard{transaction-complex-read-11}{TCR}{11}\\
	\queryRefCard{transaction-complex-read-12}{TCR}{12}\\
}

\noindent\begin{tabularx}{\queryCardWidth}{|>{\queryPropertyCell}p{\queryPropertyCellWidth}|X|}
	\hline
	query & Transaction / complex-read / 1 \\ \hline
	title & Blocked medium related accounts \\ \hline

		pattern & \centering \includegraphics[scale=\yedscale,margin=0cm .2cm]{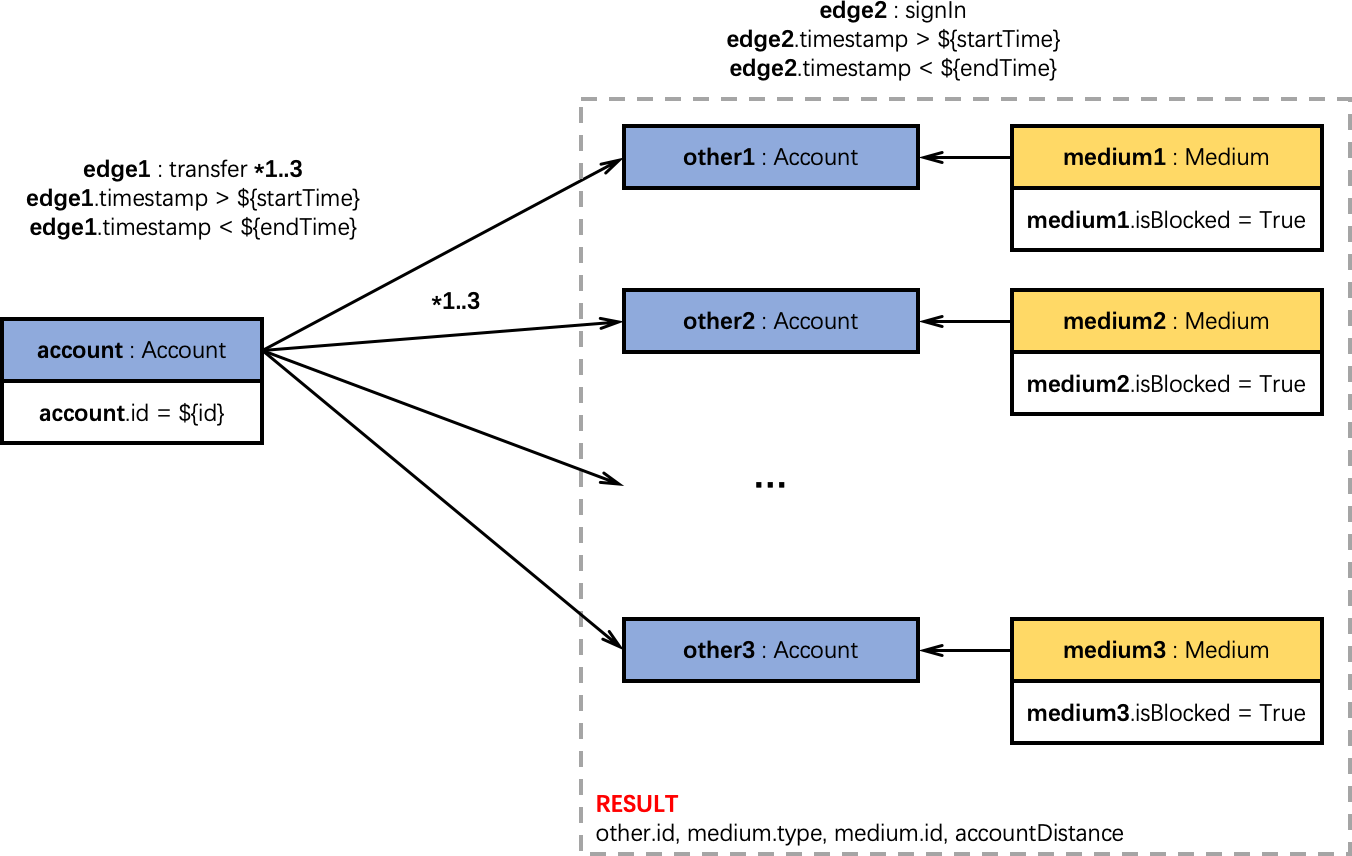} \tabularnewline \hline

	desc. & Given an \emph{Account} and a specified time window between
\emph{startTime} and \emph{endTime}, find all the \emph{Account} that is
signed in by a blocked \emph{Medium} and has fund transferred via
\emph{edge1} by at most 3 steps. Note that all timestamps in the
transfer trace must be in ascending order(only greater than). Return the
id of the account, the distance from the account to given one, the id
and type of the related medium.

Note: The returned accounts may exist in different distance from the
given one. \\ \hline

		params &
		\innerCardVSpace{\begin{tabularx}{\attributeCardWidth}{|>{\paramNumberCell}C{\attributeNumberWidth}|>{\varNameCell}M|>{\typeCell}m{\typeWidth}|Y|} \hline
		$\mathsf{1}$ & id & ID & id of the start \emph{Account} \\ \hline
		$\mathsf{2}$ & startTime & DateTime & begin of the time window \\ \hline
		$\mathsf{3}$ & endTime & DateTime & end of the time window \\ \hline
		$\mathsf{4}$ & truncationLimit & 32-bit Integer & maximum edges traversed at each step \\ \hline
		$\mathsf{5}$ & truncationOrder & Enum & the sort order before truncation at each step \\ \hline
		\end{tabularx}}\innerCardVSpace \\ \hline

		result &
		\innerCardVSpace{\begin{tabularx}{\attributeCardWidth}{|>{\resultNumberCell}C{\attributeNumberWidth}|>{\varNameCell}M|>{\typeCell}m{\typeWidth}|>{\resultOriginCell}c|Y|} \hline
		$\mathsf{1}$ & otherId & ID & R &
				the id of the account \\ \hline
		$\mathsf{2}$ & accountDistance & 32-bit Integer & C &
				the distance from the account to the given one \\ \hline
		$\mathsf{3}$ & mediumId & ID & R &
				the id of medium related to the account \\ \hline
		$\mathsf{4}$ & mediumType & String & R &
				the type of medium related to the account \\ \hline
		\end{tabularx}}\innerCardVSpace \\ \hline

		sort		&
		\innerCardVSpace{\begin{tabularx}{\attributeCardWidth}{|>{\sortNumberCell}C{\attributeNumberWidth}|>{\varNameCell}X|>{\directionCell}c|Y|} \hline
		$\mathsf{1}$ & accountDistance & $\asc$ &  \\ \hline
		$\mathsf{2}$ & otherId & $\asc$ &  \\ \hline
		$\mathsf{3}$ & mediumId & $\asc$ &  \\ \hline
		\end{tabularx}}\innerCardVSpace \\ \hline
	CPs &
	\multicolumn{1}{>{\raggedright}l|}{
		\chokePoint{3.2}, 
		\chokePoint{3.4}, 
		\chokePoint{6.2}, 
		\chokePoint{7.1}, 
		\chokePoint{7.4}, 
		\chokePoint{8.7}, 
		\chokePoint{8.8}
		} \\ \hline
	%
\end{tabularx}
\queryCardVSpace

\let\emph\oldemph
\let\textbf\oldtextbf

\renewcommand{\currentQueryCard}{0}
\renewcommand*{\arraystretch}{1.1}

\subsection*{Transaction / complex-read / 2}
\label{sec:transaction-complex-read-02}

\let\oldemph\emph
\renewcommand{\emph}[1]{{\footnotesize \sf #1}}
\let\oldtextbf\textbf
\renewcommand{\textbf}[1]{{\it #1}}\renewcommand{\currentQueryCard}{2}
\marginpar{
	\raggedleft
	\scriptsize

	\queryRefCard{transaction-complex-read-01}{TCR}{1}\\
	\queryRefCard{transaction-complex-read-02}{TCR}{2}\\
	\queryRefCard{transaction-complex-read-03}{TCR}{3}\\
	\queryRefCard{transaction-complex-read-04}{TCR}{4}\\
	\queryRefCard{transaction-complex-read-05}{TCR}{5}\\
	\queryRefCard{transaction-complex-read-06}{TCR}{6}\\
	\queryRefCard{transaction-complex-read-07}{TCR}{7}\\
	\queryRefCard{transaction-complex-read-08}{TCR}{8}\\
	\queryRefCard{transaction-complex-read-09}{TCR}{9}\\
	\queryRefCard{transaction-complex-read-10}{TCR}{10}\\
	\queryRefCard{transaction-complex-read-11}{TCR}{11}\\
	\queryRefCard{transaction-complex-read-12}{TCR}{12}\\
}

\noindent\begin{tabularx}{\queryCardWidth}{|>{\queryPropertyCell}p{\queryPropertyCellWidth}|X|}
	\hline
	query & Transaction / complex-read / 2 \\ \hline
	title & Fund gathered from the accounts applying loans \\ \hline

		pattern & \centering \includegraphics[scale=\yedscale,margin=0cm .2cm]{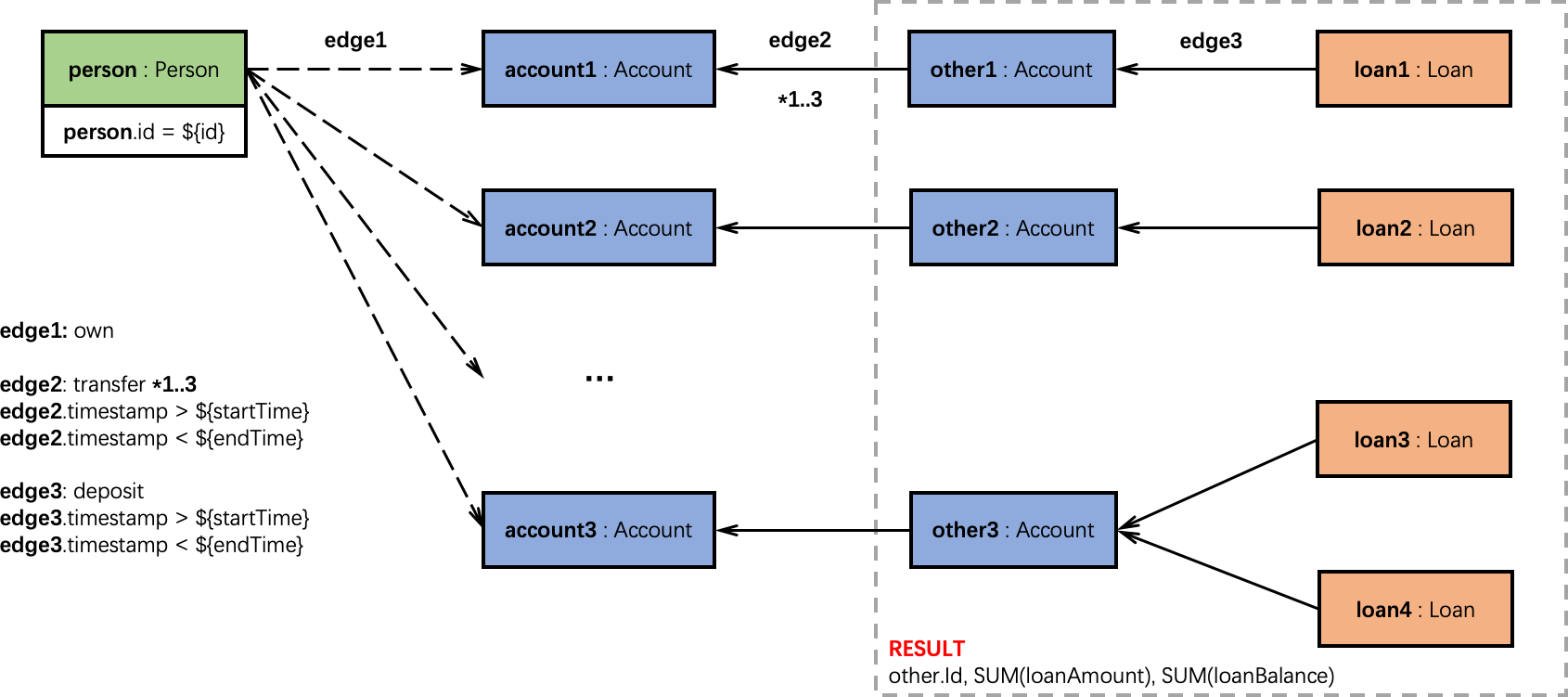} \tabularnewline \hline

	desc. & Given a \emph{Person} and a specified time window between
\emph{startTime} and \emph{endTime}, find an \emph{Account} \emph{owned}
by the \emph{Person} which has fund \emph{transferred} from other
\emph{Accounts} by at most 3 steps (\emph{edge2}) which has fund
\emph{deposited} from a \emph{loan}. The timestamps of in transfer trace
(\emph{edge2}) must be in ascending order(only greater than) from the
upstream to downstream. Return the sum of distinct \emph{loan} amount,
the sum of distinct \emph{loan} balance and the count of distinct
\emph{loans}. \\ \hline

		params &
		\innerCardVSpace{\begin{tabularx}{\attributeCardWidth}{|>{\paramNumberCell}C{\attributeNumberWidth}|>{\varNameCell}M|>{\typeCell}m{\typeWidth}|Y|} \hline
		$\mathsf{1}$ & id & ID & id of the start Person \\ \hline
		$\mathsf{2}$ & startTime & DateTime & begin of the time window \\ \hline
		$\mathsf{3}$ & endTime & DateTime & end of the time window \\ \hline
		$\mathsf{4}$ & truncationLimit & 32-bit Integer & maximum edges traversed at each step \\ \hline
		$\mathsf{5}$ & truncationOrder & Enum & the sort order before truncation at each step \\ \hline
		\end{tabularx}}\innerCardVSpace \\ \hline

		result &
		\innerCardVSpace{\begin{tabularx}{\attributeCardWidth}{|>{\resultNumberCell}C{\attributeNumberWidth}|>{\varNameCell}M|>{\typeCell}m{\typeWidth}|>{\resultOriginCell}c|Y|} \hline
		$\mathsf{1}$ & otherId & ID & R &
				id of the account related to loan \\ \hline
		$\mathsf{2}$ & sumLoanAmount & 64-bit Float & A &
				sum of all loans' amount of the account (rounded to 3 decimal places) \\ \hline
		$\mathsf{3}$ & sumLoanBalance & 64-bit Float & A &
				sum of all loans' balance of the account (rounded to 3 decimal places) \\ \hline
		\end{tabularx}}\innerCardVSpace \\ \hline

		sort		&
		\innerCardVSpace{\begin{tabularx}{\attributeCardWidth}{|>{\sortNumberCell}C{\attributeNumberWidth}|>{\varNameCell}X|>{\directionCell}c|Y|} \hline
		$\mathsf{1}$ & sumLoanAmount & $\desc$ &  \\ \hline
		$\mathsf{2}$ & otherId & $\asc$ &  \\ \hline
		\end{tabularx}}\innerCardVSpace \\ \hline
	CPs &
	\multicolumn{1}{>{\raggedright}l|}{
		\chokePoint{3.2}, 
		\chokePoint{3.4}, 
		\chokePoint{6.2}, 
		\chokePoint{7.1}, 
		\chokePoint{7.4}, 
		\chokePoint{8.7}, 
		\chokePoint{8.8}
		} \\ \hline
	%
\end{tabularx}
\queryCardVSpace

\let\emph\oldemph
\let\textbf\oldtextbf

\renewcommand{\currentQueryCard}{0}
\renewcommand*{\arraystretch}{1.1}

\subsection*{Transaction / complex-read / 3}
\label{sec:transaction-complex-read-03}

\let\oldemph\emph
\renewcommand{\emph}[1]{{\footnotesize \sf #1}}
\let\oldtextbf\textbf
\renewcommand{\textbf}[1]{{\it #1}}\renewcommand{\currentQueryCard}{3}
\marginpar{
	\raggedleft
	\scriptsize

	\queryRefCard{transaction-complex-read-01}{TCR}{1}\\
	\queryRefCard{transaction-complex-read-02}{TCR}{2}\\
	\queryRefCard{transaction-complex-read-03}{TCR}{3}\\
	\queryRefCard{transaction-complex-read-04}{TCR}{4}\\
	\queryRefCard{transaction-complex-read-05}{TCR}{5}\\
	\queryRefCard{transaction-complex-read-06}{TCR}{6}\\
	\queryRefCard{transaction-complex-read-07}{TCR}{7}\\
	\queryRefCard{transaction-complex-read-08}{TCR}{8}\\
	\queryRefCard{transaction-complex-read-09}{TCR}{9}\\
	\queryRefCard{transaction-complex-read-10}{TCR}{10}\\
	\queryRefCard{transaction-complex-read-11}{TCR}{11}\\
	\queryRefCard{transaction-complex-read-12}{TCR}{12}\\
}

\noindent\begin{tabularx}{\queryCardWidth}{|>{\queryPropertyCell}p{\queryPropertyCellWidth}|X|}
	\hline
	query & Transaction / complex-read / 3 \\ \hline
	title & Shortest transfer path \\ \hline

		pattern & \centering \includegraphics[scale=\yedscale,margin=0cm .2cm]{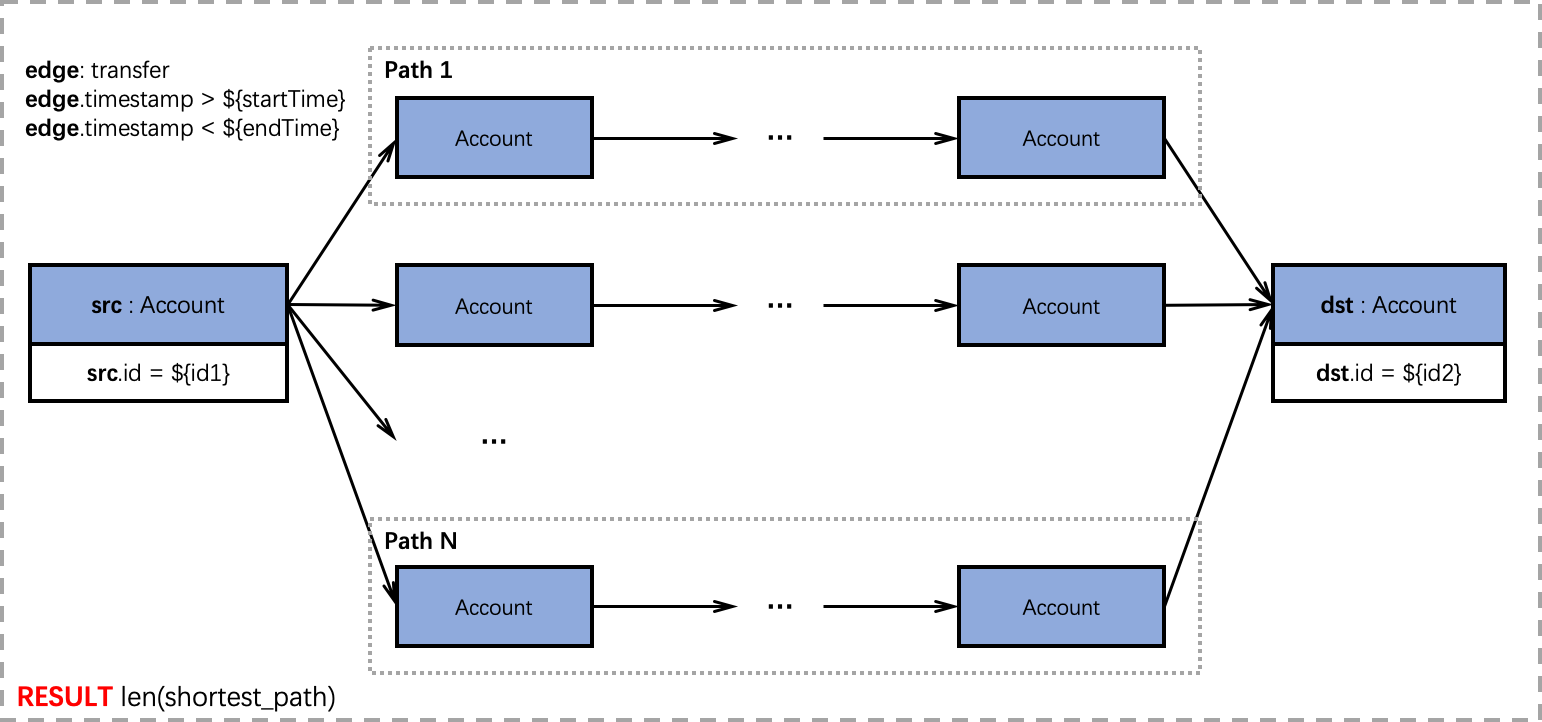} \tabularnewline \hline

	desc. & Given two \emph{accounts} and a specified time window between
\emph{startTime} and \emph{endTime}, find the length of shortest path
between these two \emph{accounts} by the \emph{transfer} relationships.
Note that all the edges in the path should be in the time window and of
type \emph{transfer}. Return 1 if src and dst are directly connected.
Return -1 if there is no path found. \\ \hline

		params &
		\innerCardVSpace{\begin{tabularx}{\attributeCardWidth}{|>{\paramNumberCell}C{\attributeNumberWidth}|>{\varNameCell}M|>{\typeCell}m{\typeWidth}|Y|} \hline
		$\mathsf{1}$ & id1 & ID & id of src Account \\ \hline
		$\mathsf{2}$ & id2 & ID & id of dst Account \\ \hline
		$\mathsf{3}$ & startTime & DateTime & begin of the time window \\ \hline
		$\mathsf{4}$ & endTime & DateTime & end of the time window \\ \hline
		\end{tabularx}}\innerCardVSpace \\ \hline

		result &
		\innerCardVSpace{\begin{tabularx}{\attributeCardWidth}{|>{\resultNumberCell}C{\attributeNumberWidth}|>{\varNameCell}M|>{\typeCell}m{\typeWidth}|>{\resultOriginCell}c|Y|} \hline
		$\mathsf{1}$ & shortestPathLength & 64-bit Integer & C &
				the length of shortest path \\ \hline
		\end{tabularx}}\innerCardVSpace \\ \hline

	CPs &
	\multicolumn{1}{>{\raggedright}l|}{
		\chokePoint{3.2}, 
		\chokePoint{3.4}, 
		\chokePoint{6.2}, 
		\chokePoint{8.7}
		} \\ \hline
	%
\end{tabularx}
\queryCardVSpace

\let\emph\oldemph
\let\textbf\oldtextbf

\renewcommand{\currentQueryCard}{0}
\renewcommand*{\arraystretch}{1.1}

\subsection*{Transaction / complex-read / 4}
\label{sec:transaction-complex-read-04}

\let\oldemph\emph
\renewcommand{\emph}[1]{{\footnotesize \sf #1}}
\let\oldtextbf\textbf
\renewcommand{\textbf}[1]{{\it #1}}\renewcommand{\currentQueryCard}{4}
\marginpar{
	\raggedleft
	\scriptsize

	\queryRefCard{transaction-complex-read-01}{TCR}{1}\\
	\queryRefCard{transaction-complex-read-02}{TCR}{2}\\
	\queryRefCard{transaction-complex-read-03}{TCR}{3}\\
	\queryRefCard{transaction-complex-read-04}{TCR}{4}\\
	\queryRefCard{transaction-complex-read-05}{TCR}{5}\\
	\queryRefCard{transaction-complex-read-06}{TCR}{6}\\
	\queryRefCard{transaction-complex-read-07}{TCR}{7}\\
	\queryRefCard{transaction-complex-read-08}{TCR}{8}\\
	\queryRefCard{transaction-complex-read-09}{TCR}{9}\\
	\queryRefCard{transaction-complex-read-10}{TCR}{10}\\
	\queryRefCard{transaction-complex-read-11}{TCR}{11}\\
	\queryRefCard{transaction-complex-read-12}{TCR}{12}\\
}

\noindent\begin{tabularx}{\queryCardWidth}{|>{\queryPropertyCell}p{\queryPropertyCellWidth}|X|}
	\hline
	query & Transaction / complex-read / 4 \\ \hline
	title & Three accounts in a transfer cycle \\ \hline

		pattern & \centering \includegraphics[scale=\yedscale,margin=0cm .2cm]{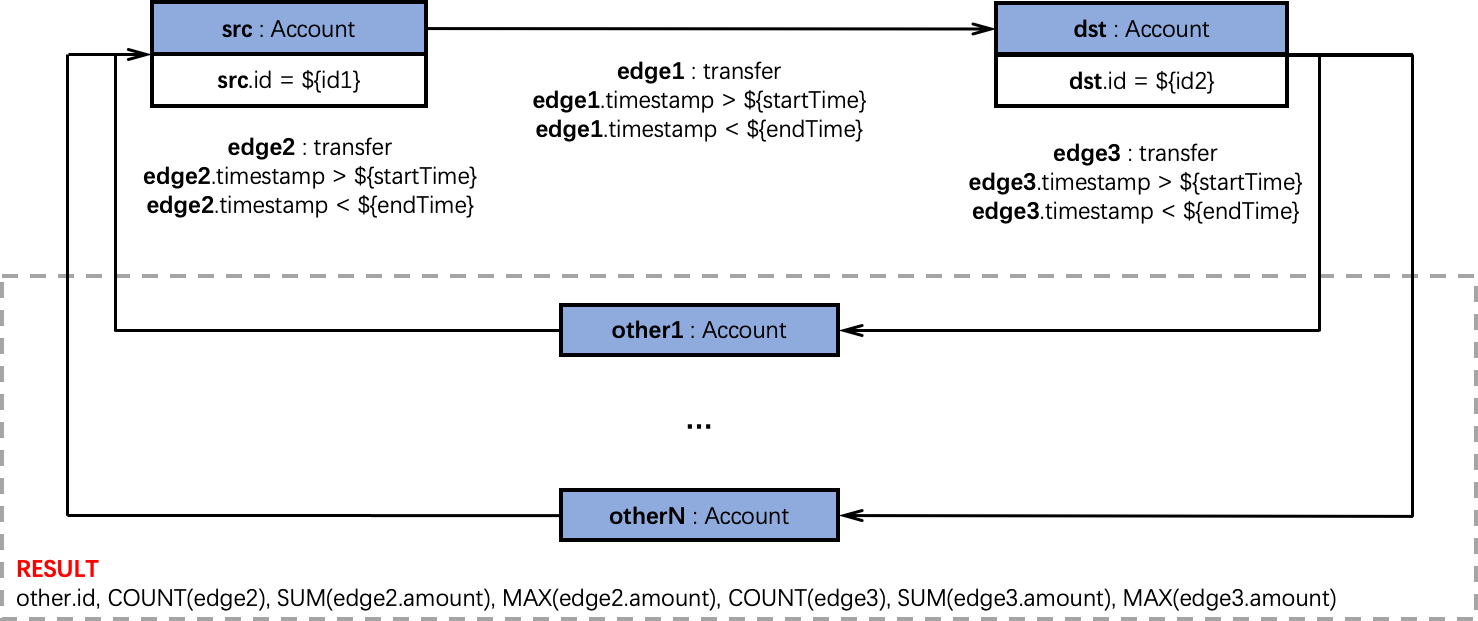} \tabularnewline \hline

	desc. & Given two accounts \emph{src} and \emph{dst}, and a specified time
window between \emph{startTime} and \emph{endTime},

\begin{enumerate}
\def\labelenumi{(\arabic{enumi})}
\item
  check whether \emph{src} transferred money to \emph{dst} in the given
  time window (\emph{edge1}). If \emph{edge1} does not exist, return
  with empty results (the result size is 0).
\item
  find all other accounts (other1, \ldots, otherN) which received money
  from dst (\emph{edge2}) and transferred money to src (\emph{edge3}) in
  a specific time.
\end{enumerate}

For each of these other accounts, return the id of the account, the sum
and max of the transfer amount (\emph{edge2} and \emph{edge3}). \\ \hline

		params &
		\innerCardVSpace{\begin{tabularx}{\attributeCardWidth}{|>{\paramNumberCell}C{\attributeNumberWidth}|>{\varNameCell}M|>{\typeCell}m{\typeWidth}|Y|} \hline
		$\mathsf{1}$ & id1 & ID & id of the src Account \\ \hline
		$\mathsf{2}$ & id2 & ID & id of the dst Account \\ \hline
		$\mathsf{3}$ & startTime & DateTime & begin of the time window \\ \hline
		$\mathsf{4}$ & endTime & DateTime & end of the time window \\ \hline
		\end{tabularx}}\innerCardVSpace \\ \hline

		result &
		\innerCardVSpace{\begin{tabularx}{\attributeCardWidth}{|>{\resultNumberCell}C{\attributeNumberWidth}|>{\varNameCell}M|>{\typeCell}m{\typeWidth}|>{\resultOriginCell}c|Y|} \hline
		$\mathsf{1}$ & otherId & ID & R &
				the id of the other account \\ \hline
		$\mathsf{2}$ & numEdge2 & 64-bit Integer & A &
				transfers' count from otherAccount to srcAccount \\ \hline
		$\mathsf{3}$ & sumEdge2Amount & 64-bit Float & A &
				sum of transfers from otherAccount to srcAccount (rounded to 3 decimal
places) \\ \hline
		$\mathsf{4}$ & maxEdge2Amount & 64-bit Float & A &
				max of transfers from otherAccount to srcAccount (rounded to 3 decimal
places) \\ \hline
		$\mathsf{5}$ & numEdge3 & 64-bit Integer & A &
				transfers' count from dstAccount to otherAccount \\ \hline
		$\mathsf{6}$ & sumEdge3Amount & 64-bit Float & A &
				sum of transfers from dstAccount to otherAccount (rounded to 3 decimal
places) \\ \hline
		$\mathsf{7}$ & maxEdge3Amount & 64-bit Float & A &
				max of transfers from dstAccount to otherAccount (rounded to 3 decimal
places) \\ \hline
		\end{tabularx}}\innerCardVSpace \\ \hline

		sort		&
		\innerCardVSpace{\begin{tabularx}{\attributeCardWidth}{|>{\sortNumberCell}C{\attributeNumberWidth}|>{\varNameCell}X|>{\directionCell}c|Y|} \hline
		$\mathsf{1}$ & sumEdge2Amount & $\desc$ &  \\ \hline
		$\mathsf{2}$ & sumEdge3Amount & $\desc$ &  \\ \hline
		$\mathsf{3}$ & otherId & $\asc$ &  \\ \hline
		\end{tabularx}}\innerCardVSpace \\ \hline
	CPs &
	\multicolumn{1}{>{\raggedright}l|}{
		\chokePoint{3.2}, 
		\chokePoint{3.4}, 
		\chokePoint{6.2}, 
		\chokePoint{8.7}
		} \\ \hline
	%
\end{tabularx}
\queryCardVSpace

\let\emph\oldemph
\let\textbf\oldtextbf

\renewcommand{\currentQueryCard}{0}
\renewcommand*{\arraystretch}{1.1}

\subsection*{Transaction / complex-read / 5}
\label{sec:transaction-complex-read-05}

\let\oldemph\emph
\renewcommand{\emph}[1]{{\footnotesize \sf #1}}
\let\oldtextbf\textbf
\renewcommand{\textbf}[1]{{\it #1}}\renewcommand{\currentQueryCard}{5}
\marginpar{
	\raggedleft
	\scriptsize

	\queryRefCard{transaction-complex-read-01}{TCR}{1}\\
	\queryRefCard{transaction-complex-read-02}{TCR}{2}\\
	\queryRefCard{transaction-complex-read-03}{TCR}{3}\\
	\queryRefCard{transaction-complex-read-04}{TCR}{4}\\
	\queryRefCard{transaction-complex-read-05}{TCR}{5}\\
	\queryRefCard{transaction-complex-read-06}{TCR}{6}\\
	\queryRefCard{transaction-complex-read-07}{TCR}{7}\\
	\queryRefCard{transaction-complex-read-08}{TCR}{8}\\
	\queryRefCard{transaction-complex-read-09}{TCR}{9}\\
	\queryRefCard{transaction-complex-read-10}{TCR}{10}\\
	\queryRefCard{transaction-complex-read-11}{TCR}{11}\\
	\queryRefCard{transaction-complex-read-12}{TCR}{12}\\
}

\noindent\begin{tabularx}{\queryCardWidth}{|>{\queryPropertyCell}p{\queryPropertyCellWidth}|X|}
	\hline
	query & Transaction / complex-read / 5 \\ \hline
	title & Exact Account Transfer Trace \\ \hline

		pattern & \centering \includegraphics[scale=\yedscale,margin=0cm .2cm]{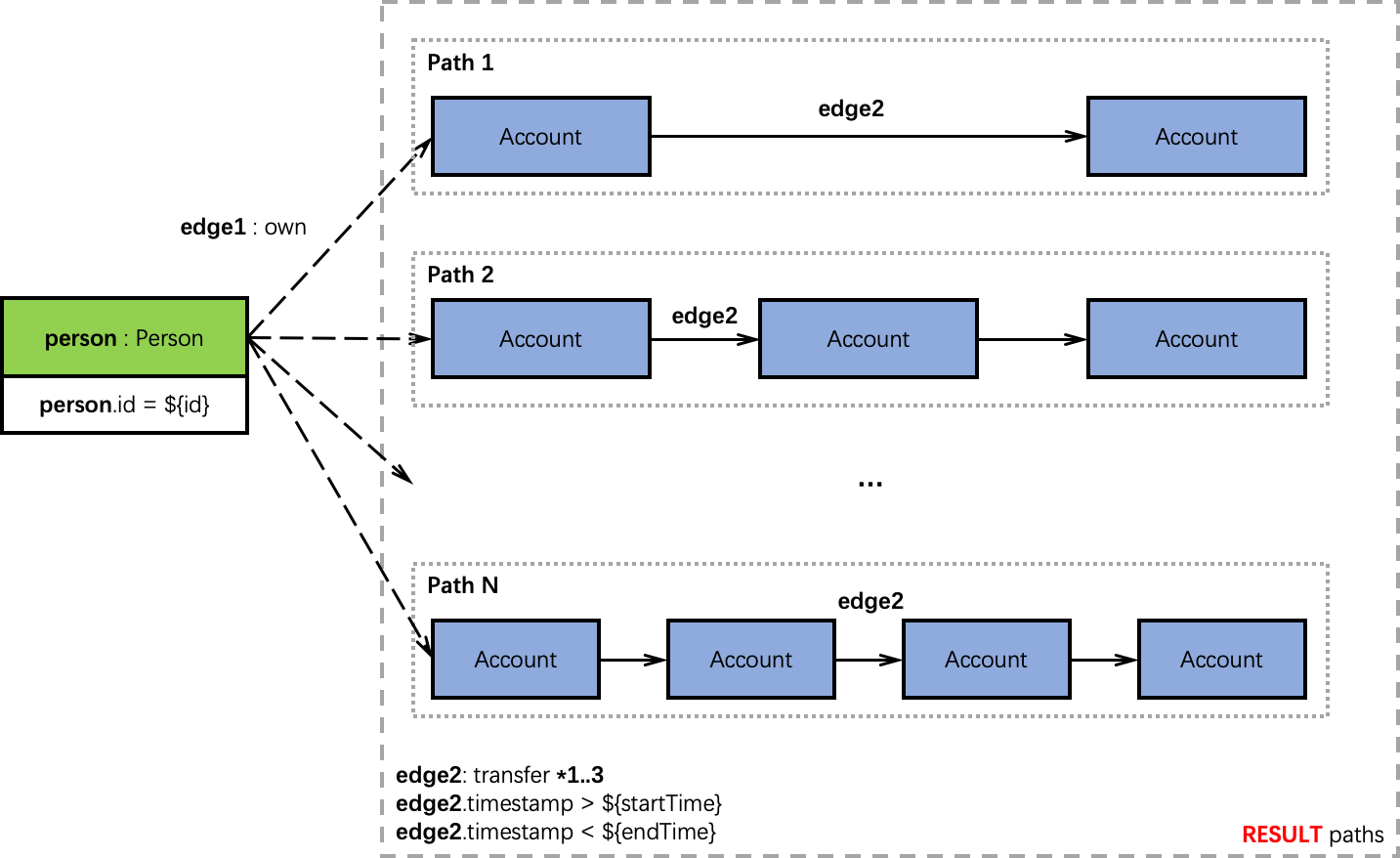} \tabularnewline \hline

	desc. & Given a \emph{Person} and a specified time window between
\emph{startTime} and \emph{endTime}, find the transfer trace from the
account (\emph{src}) owned by the \emph{Person} to another account
(\emph{dst}) by at most 3 steps. Note that the trace (\emph{edge2}) must
be ascending order(only greater than) of their timestamps. Return all
the transfer traces.

Note: Multiple edges of from the same src to the same dst should be seen
as identical path. And the resulting paths shall not include recurring
accounts (cycles in the trace are not allowed). The results may not be
in a deterministic order since they are only sorted by the length of the
path. Driver will validate the results after sorting. \\ \hline

		params &
		\innerCardVSpace{\begin{tabularx}{\attributeCardWidth}{|>{\paramNumberCell}C{\attributeNumberWidth}|>{\varNameCell}M|>{\typeCell}m{\typeWidth}|Y|} \hline
		$\mathsf{1}$ & id & ID & id of the start Person \\ \hline
		$\mathsf{2}$ & startTime & DateTime & begin of the time window \\ \hline
		$\mathsf{3}$ & endTime & DateTime & end of the time window \\ \hline
		$\mathsf{4}$ & truncationLimit & 32-bit Integer & maximum edges traversed at each step \\ \hline
		$\mathsf{5}$ & truncationOrder & Enum & the sort order before truncation at each step \\ \hline
		\end{tabularx}}\innerCardVSpace \\ \hline

		result &
		\innerCardVSpace{\begin{tabularx}{\attributeCardWidth}{|>{\resultNumberCell}C{\attributeNumberWidth}|>{\varNameCell}M|>{\typeCell}m{\typeWidth}|>{\resultOriginCell}c|Y|} \hline
		$\mathsf{1}$ & path & Path & S &
				a transfer trace. See the requirements in
\autoref{subsec:returned-values} \\ \hline
		\end{tabularx}}\innerCardVSpace \\ \hline

		sort		&
		\innerCardVSpace{\begin{tabularx}{\attributeCardWidth}{|>{\sortNumberCell}C{\attributeNumberWidth}|>{\varNameCell}X|>{\directionCell}c|Y|} \hline
		$\mathsf{1}$ & pathLength & $\desc$ &  \\ \hline
		\end{tabularx}}\innerCardVSpace \\ \hline
	CPs &
	\multicolumn{1}{>{\raggedright}l|}{
		\chokePoint{1.1}, 
		\chokePoint{3.2}, 
		\chokePoint{3.4}, 
		\chokePoint{6.2}, 
		\chokePoint{7.1}, 
		\chokePoint{7.4}, 
		\chokePoint{8.7}, 
		\chokePoint{8.8}
		} \\ \hline
	%
\end{tabularx}
\queryCardVSpace

\let\emph\oldemph
\let\textbf\oldtextbf

\renewcommand{\currentQueryCard}{0}
\renewcommand*{\arraystretch}{1.1}

\subsection*{Transaction / complex-read / 6}
\label{sec:transaction-complex-read-06}

\let\oldemph\emph
\renewcommand{\emph}[1]{{\footnotesize \sf #1}}
\let\oldtextbf\textbf
\renewcommand{\textbf}[1]{{\it #1}}\renewcommand{\currentQueryCard}{6}
\marginpar{
	\raggedleft
	\scriptsize

	\queryRefCard{transaction-complex-read-01}{TCR}{1}\\
	\queryRefCard{transaction-complex-read-02}{TCR}{2}\\
	\queryRefCard{transaction-complex-read-03}{TCR}{3}\\
	\queryRefCard{transaction-complex-read-04}{TCR}{4}\\
	\queryRefCard{transaction-complex-read-05}{TCR}{5}\\
	\queryRefCard{transaction-complex-read-06}{TCR}{6}\\
	\queryRefCard{transaction-complex-read-07}{TCR}{7}\\
	\queryRefCard{transaction-complex-read-08}{TCR}{8}\\
	\queryRefCard{transaction-complex-read-09}{TCR}{9}\\
	\queryRefCard{transaction-complex-read-10}{TCR}{10}\\
	\queryRefCard{transaction-complex-read-11}{TCR}{11}\\
	\queryRefCard{transaction-complex-read-12}{TCR}{12}\\
}

\noindent\begin{tabularx}{\queryCardWidth}{|>{\queryPropertyCell}p{\queryPropertyCellWidth}|X|}
	\hline
	query & Transaction / complex-read / 6 \\ \hline
	title & Withdrawal after Many-to-One transfer \\ \hline

		pattern & \centering \includegraphics[scale=\yedscale,margin=0cm .2cm]{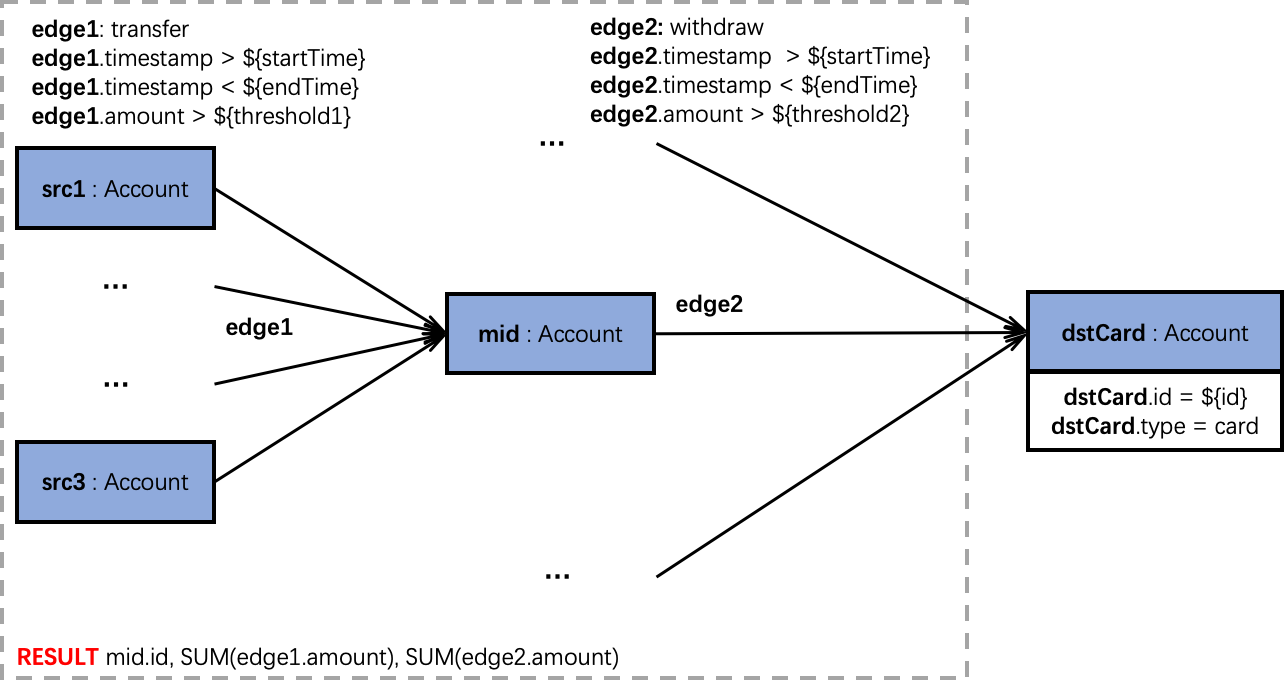} \tabularnewline \hline

	desc. & Given an \emph{account} of type card and a specified time window between
\emph{startTime} and \emph{endTime}, find all the connected
\emph{accounts} (\emph{mid}) via withdrawal (\emph{edge2}) satisfying,
(1) More than 3 \emph{transfer-ins} (\emph{edge1}) from other
\emph{accounts} (\emph{src}) whose amount exceeds \emph{threshold1}. (2)
The amount of \emph{withdrawal} (\emph{edge2}) from \emph{mid} to
\emph{dstCard} whose exceeds \emph{threshold2}. Return the sum of
transfer amount from \emph{src} to \emph{mid}, the amount from
\emph{mid} to \emph{dstCard} grouped by \emph{mid}. \\ \hline

		params &
		\innerCardVSpace{\begin{tabularx}{\attributeCardWidth}{|>{\paramNumberCell}C{\attributeNumberWidth}|>{\varNameCell}M|>{\typeCell}m{\typeWidth}|Y|} \hline
		$\mathsf{1}$ & id & ID & id of the card account \\ \hline
		$\mathsf{2}$ & threshold1 & 64-bit Float & threshold of transfer amount \\ \hline
		$\mathsf{3}$ & threshold2 & 64-bit Float & threshold of transfer amount \\ \hline
		$\mathsf{4}$ & startTime & DateTime & begin of the time window \\ \hline
		$\mathsf{5}$ & endTime & DateTime & end of the time window \\ \hline
		$\mathsf{6}$ & truncationLimit & 32-bit Integer & maximum edges traversed at each step \\ \hline
		$\mathsf{7}$ & truncationOrder & Enum & the sort order before truncation at each step \\ \hline
		\end{tabularx}}\innerCardVSpace \\ \hline

		result &
		\innerCardVSpace{\begin{tabularx}{\attributeCardWidth}{|>{\resultNumberCell}C{\attributeNumberWidth}|>{\varNameCell}M|>{\typeCell}m{\typeWidth}|>{\resultOriginCell}c|Y|} \hline
		$\mathsf{1}$ & midId & ID & R &
				the id of the middle account \\ \hline
		$\mathsf{2}$ & sumEdge1Amount & 64-bit Float & A &
				the amount of transfer from src accounts to mid (rounded to 3 decimal
places) \\ \hline
		$\mathsf{3}$ & sumEdge2Amount & 64-bit Float & A &
				the amount of withdrawal from mid to dstCard (rounded to 3 decimal
places) \\ \hline
		\end{tabularx}}\innerCardVSpace \\ \hline

		sort		&
		\innerCardVSpace{\begin{tabularx}{\attributeCardWidth}{|>{\sortNumberCell}C{\attributeNumberWidth}|>{\varNameCell}X|>{\directionCell}c|Y|} \hline
		$\mathsf{1}$ & sumEdge2Amount & $\desc$ &  \\ \hline
		$\mathsf{2}$ & midId & $\asc$ &  \\ \hline
		\end{tabularx}}\innerCardVSpace \\ \hline
	CPs &
	\multicolumn{1}{>{\raggedright}l|}{
		\chokePoint{3.2}, 
		\chokePoint{3.4}, 
		\chokePoint{6.2}, 
		\chokePoint{8.7}
		} \\ \hline
	%
\end{tabularx}
\queryCardVSpace

\let\emph\oldemph
\let\textbf\oldtextbf

\renewcommand{\currentQueryCard}{0}
\renewcommand*{\arraystretch}{1.1}

\subsection*{Transaction / complex-read / 7}
\label{sec:transaction-complex-read-07}

\let\oldemph\emph
\renewcommand{\emph}[1]{{\footnotesize \sf #1}}
\let\oldtextbf\textbf
\renewcommand{\textbf}[1]{{\it #1}}\renewcommand{\currentQueryCard}{7}
\marginpar{
	\raggedleft
	\scriptsize

	\queryRefCard{transaction-complex-read-01}{TCR}{1}\\
	\queryRefCard{transaction-complex-read-02}{TCR}{2}\\
	\queryRefCard{transaction-complex-read-03}{TCR}{3}\\
	\queryRefCard{transaction-complex-read-04}{TCR}{4}\\
	\queryRefCard{transaction-complex-read-05}{TCR}{5}\\
	\queryRefCard{transaction-complex-read-06}{TCR}{6}\\
	\queryRefCard{transaction-complex-read-07}{TCR}{7}\\
	\queryRefCard{transaction-complex-read-08}{TCR}{8}\\
	\queryRefCard{transaction-complex-read-09}{TCR}{9}\\
	\queryRefCard{transaction-complex-read-10}{TCR}{10}\\
	\queryRefCard{transaction-complex-read-11}{TCR}{11}\\
	\queryRefCard{transaction-complex-read-12}{TCR}{12}\\
}

\noindent\begin{tabularx}{\queryCardWidth}{|>{\queryPropertyCell}p{\queryPropertyCellWidth}|X|}
	\hline
	query & Transaction / complex-read / 7 \\ \hline
	title & Transfer in/out ratio \\ \hline

		pattern & \centering \includegraphics[scale=\yedscale,margin=0cm .2cm]{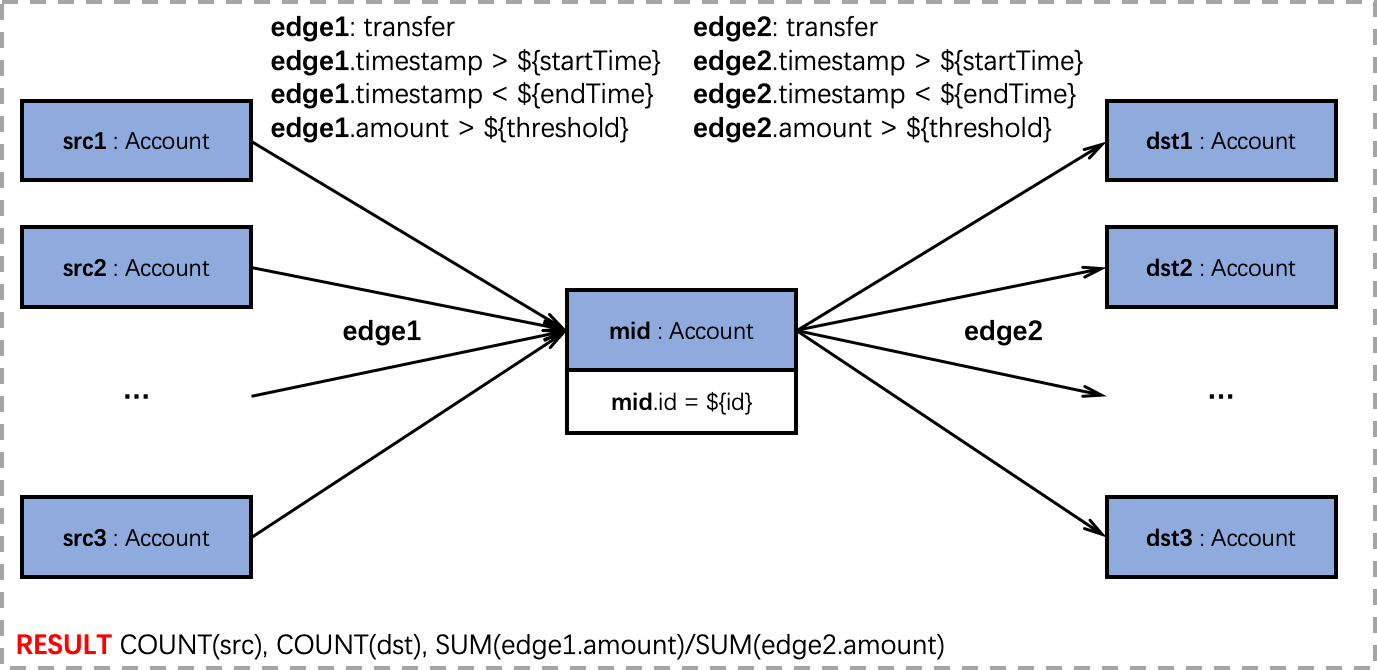} \tabularnewline \hline

	desc. & Given an \emph{Account} and a specified time window between
\emph{startTime} and \emph{endTime}, find all the \emph{transfer-in}
(\emph{edge1}) and \emph{transfer-out} (\emph{edge2}) whose amount
exceeds \emph{threshold}. Return the count of \emph{src} and \emph{dst}
accounts and the ratio of \emph{transfer-in} amount over
\emph{transfer-out} amount. The fast-in and fash-out means a tight
window between \emph{startTime} and \emph{endTime}. Return the ratio as
-1 if there is no \emph{edge2}. \\ \hline

		params &
		\innerCardVSpace{\begin{tabularx}{\attributeCardWidth}{|>{\paramNumberCell}C{\attributeNumberWidth}|>{\varNameCell}M|>{\typeCell}m{\typeWidth}|Y|} \hline
		$\mathsf{1}$ & id & ID & id of mid account \\ \hline
		$\mathsf{2}$ & threshold & 64-bit Float & transfer amount threshold \\ \hline
		$\mathsf{3}$ & startTime & DateTime & begin of the time window \\ \hline
		$\mathsf{4}$ & endTime & DateTime & end of the time window \\ \hline
		$\mathsf{5}$ & truncationLimit & 32-bit Integer & maximum edges traversed at each step \\ \hline
		$\mathsf{6}$ & truncationOrder & Enum & the sort order before truncation at each step \\ \hline
		\end{tabularx}}\innerCardVSpace \\ \hline

		result &
		\innerCardVSpace{\begin{tabularx}{\attributeCardWidth}{|>{\resultNumberCell}C{\attributeNumberWidth}|>{\varNameCell}M|>{\typeCell}m{\typeWidth}|>{\resultOriginCell}c|Y|} \hline
		$\mathsf{1}$ & numSrc & 32-bit Integer & A &
				num of the distinct src accounts \\ \hline
		$\mathsf{2}$ & numDst & 32-bit Integer & A &
				num of the distinct dst accounts \\ \hline
		$\mathsf{3}$ & inOutRatio & 32-bit Float & C &
				the amount ratio of transfers-in over transfers-out (rounded to 3
decimal places) \\ \hline
		\end{tabularx}}\innerCardVSpace \\ \hline

	CPs &
	\multicolumn{1}{>{\raggedright}l|}{
		\chokePoint{1.2}, 
		\chokePoint{3.2}, 
		\chokePoint{3.4}, 
		\chokePoint{6.2}, 
		\chokePoint{8.7}
		} \\ \hline
	%
\end{tabularx}
\queryCardVSpace

\let\emph\oldemph
\let\textbf\oldtextbf

\renewcommand{\currentQueryCard}{0}
\renewcommand*{\arraystretch}{1.1}

\subsection*{Transaction / complex-read / 8}
\label{sec:transaction-complex-read-08}

\let\oldemph\emph
\renewcommand{\emph}[1]{{\footnotesize \sf #1}}
\let\oldtextbf\textbf
\renewcommand{\textbf}[1]{{\it #1}}\renewcommand{\currentQueryCard}{8}
\marginpar{
	\raggedleft
	\scriptsize

	\queryRefCard{transaction-complex-read-01}{TCR}{1}\\
	\queryRefCard{transaction-complex-read-02}{TCR}{2}\\
	\queryRefCard{transaction-complex-read-03}{TCR}{3}\\
	\queryRefCard{transaction-complex-read-04}{TCR}{4}\\
	\queryRefCard{transaction-complex-read-05}{TCR}{5}\\
	\queryRefCard{transaction-complex-read-06}{TCR}{6}\\
	\queryRefCard{transaction-complex-read-07}{TCR}{7}\\
	\queryRefCard{transaction-complex-read-08}{TCR}{8}\\
	\queryRefCard{transaction-complex-read-09}{TCR}{9}\\
	\queryRefCard{transaction-complex-read-10}{TCR}{10}\\
	\queryRefCard{transaction-complex-read-11}{TCR}{11}\\
	\queryRefCard{transaction-complex-read-12}{TCR}{12}\\
}

\noindent\begin{tabularx}{\queryCardWidth}{|>{\queryPropertyCell}p{\queryPropertyCellWidth}|X|}
	\hline
	query & Transaction / complex-read / 8 \\ \hline
	title & Transfer trace after loan applied \\ \hline

		pattern & \centering \includegraphics[scale=\yedscale,margin=0cm .2cm]{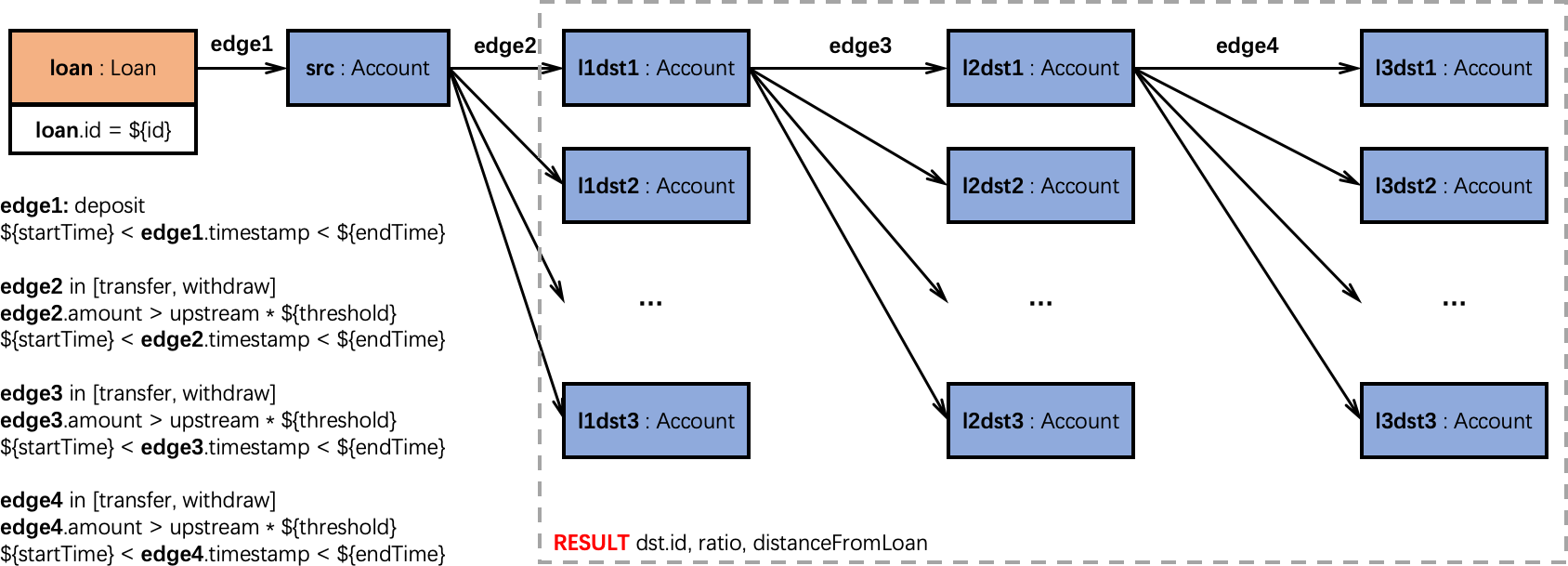} \tabularnewline \hline

	desc. & Given a \emph{Loan} and a specified time window between \emph{startTime}
and \emph{endTime}, trace the fund \emph{transfer} or \emph{withdraw} by
at most 3 steps from the \emph{account} the \emph{Loan} \emph{deposits}.
Note that the transfer paths of \emph{edge1}, \emph{edge2}, \emph{edge3}
and \emph{edge4} are in a specific time range between \emph{startTime}
and \emph{endTime}. Amount of each transfers or withdrawals between the
account and the upstream account should exceed a specified
\emph{threshold} of the upstream transfer. Return all the accounts' id
in the downstream of loan with the final ratio and distanceFromLoan.

Note: Upstream of an edge refers to the aggregated total amounts of all
\emph{transfer-in} edges of its source \emph{Account}. \\ \hline

		params &
		\innerCardVSpace{\begin{tabularx}{\attributeCardWidth}{|>{\paramNumberCell}C{\attributeNumberWidth}|>{\varNameCell}M|>{\typeCell}m{\typeWidth}|Y|} \hline
		$\mathsf{1}$ & id & ID & id of the Loan \\ \hline
		$\mathsf{2}$ & threshold & 32-bit Float & threshold of the amount over the upstream's \\ \hline
		$\mathsf{3}$ & startTime & DateTime & begin of the time window \\ \hline
		$\mathsf{4}$ & endTime & DateTime & end of the time window \\ \hline
		$\mathsf{5}$ & truncationLimit & 32-bit Integer & maximum edges traversed at each step \\ \hline
		$\mathsf{6}$ & truncationOrder & Enum & the sort order before truncation at each step \\ \hline
		\end{tabularx}}\innerCardVSpace \\ \hline

		result &
		\innerCardVSpace{\begin{tabularx}{\attributeCardWidth}{|>{\resultNumberCell}C{\attributeNumberWidth}|>{\varNameCell}M|>{\typeCell}m{\typeWidth}|>{\resultOriginCell}c|Y|} \hline
		$\mathsf{1}$ & dstId & ID & R &
				the id of the account in transfer traces \\ \hline
		$\mathsf{2}$ & ratio & 32-bit Float & C &
				the final ratio of the inflow's amount of each account over the loan
(rounded to 3 decimal places) \\ \hline
		$\mathsf{3}$ & minDistanceFromLoan & 32-bit Integer & C &
				the min distance from the account to the loan \\ \hline
		\end{tabularx}}\innerCardVSpace \\ \hline

		sort		&
		\innerCardVSpace{\begin{tabularx}{\attributeCardWidth}{|>{\sortNumberCell}C{\attributeNumberWidth}|>{\varNameCell}X|>{\directionCell}c|Y|} \hline
		$\mathsf{1}$ & distanceFromLoan & $\desc$ &  \\ \hline
		$\mathsf{2}$ & ratio & $\desc$ &  \\ \hline
		$\mathsf{3}$ & dstId & $\asc$ &  \\ \hline
		\end{tabularx}}\innerCardVSpace \\ \hline
	CPs &
	\multicolumn{1}{>{\raggedright}l|}{
		\chokePoint{3.2}, 
		\chokePoint{3.4}, 
		\chokePoint{6.2}, 
		\chokePoint{7.1}, 
		\chokePoint{8.7}
		} \\ \hline
	%
\end{tabularx}
\queryCardVSpace

\let\emph\oldemph
\let\textbf\oldtextbf

\renewcommand{\currentQueryCard}{0}
\renewcommand*{\arraystretch}{1.1}

\subsection*{Transaction / complex-read / 9}
\label{sec:transaction-complex-read-09}

\let\oldemph\emph
\renewcommand{\emph}[1]{{\footnotesize \sf #1}}
\let\oldtextbf\textbf
\renewcommand{\textbf}[1]{{\it #1}}\renewcommand{\currentQueryCard}{9}
\marginpar{
	\raggedleft
	\scriptsize

	\queryRefCard{transaction-complex-read-01}{TCR}{1}\\
	\queryRefCard{transaction-complex-read-02}{TCR}{2}\\
	\queryRefCard{transaction-complex-read-03}{TCR}{3}\\
	\queryRefCard{transaction-complex-read-04}{TCR}{4}\\
	\queryRefCard{transaction-complex-read-05}{TCR}{5}\\
	\queryRefCard{transaction-complex-read-06}{TCR}{6}\\
	\queryRefCard{transaction-complex-read-07}{TCR}{7}\\
	\queryRefCard{transaction-complex-read-08}{TCR}{8}\\
	\queryRefCard{transaction-complex-read-09}{TCR}{9}\\
	\queryRefCard{transaction-complex-read-10}{TCR}{10}\\
	\queryRefCard{transaction-complex-read-11}{TCR}{11}\\
	\queryRefCard{transaction-complex-read-12}{TCR}{12}\\
}

\noindent\begin{tabularx}{\queryCardWidth}{|>{\queryPropertyCell}p{\queryPropertyCellWidth}|X|}
	\hline
	query & Transaction / complex-read / 9 \\ \hline
	title & Money laundering with loan involved \\ \hline

		pattern & \centering \includegraphics[scale=\yedscale,margin=0cm .2cm]{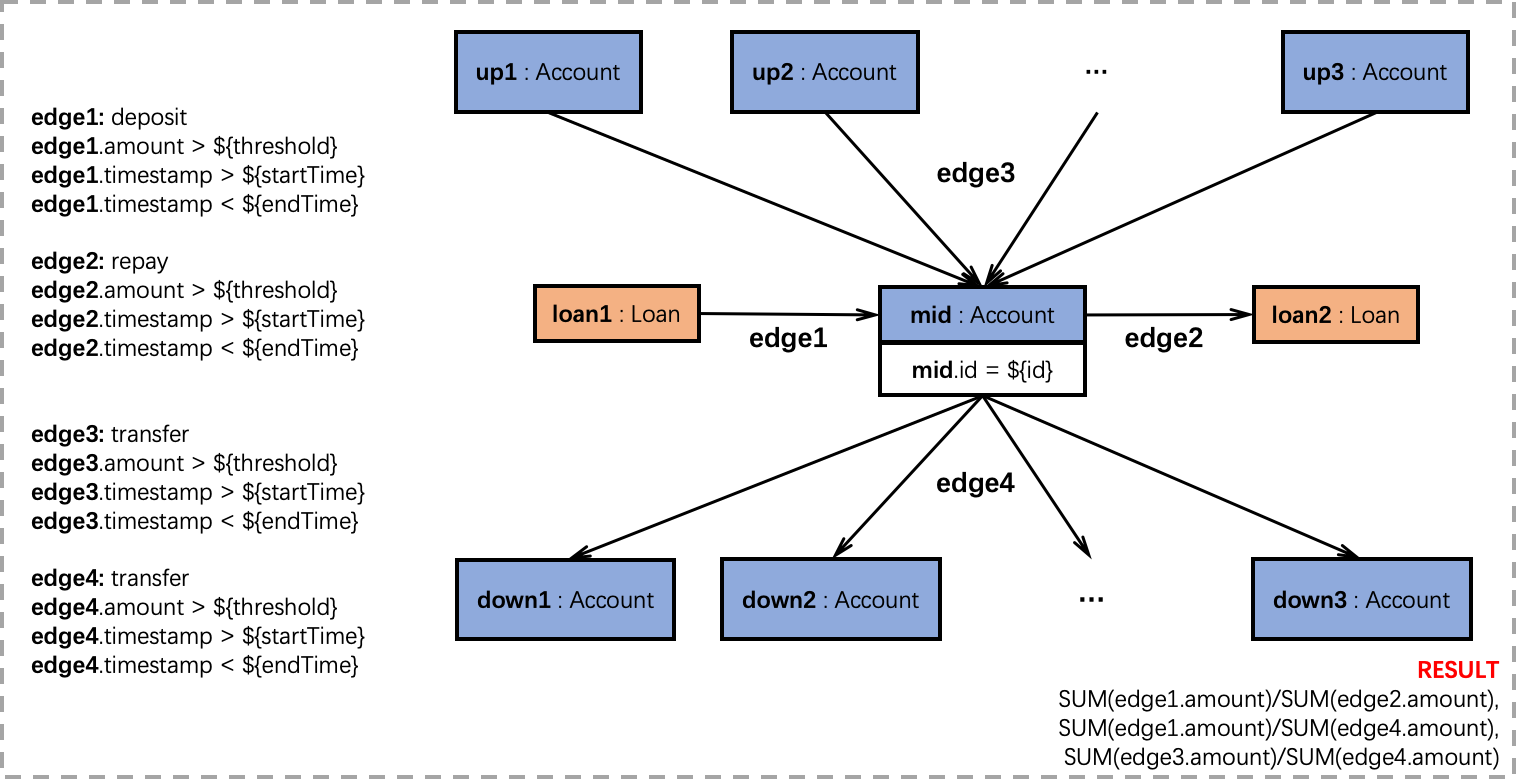} \tabularnewline \hline

	desc. & Given an \emph{account}, a bound of transfer amount and a specified time
window between \emph{startTime} and \emph{endTime}, find the
\emph{deposit} and \emph{repay} edge between the \emph{account} and a
\emph{loan}, the \emph{transfers-in} and \emph{transfers-out}. Return
\emph{ratioRepay} (sum of all the \emph{edge1} over sum of all the
\emph{edge2}), \emph{ratioDeposit} (sum of \emph{edge1} over sum of
\emph{edge4}), \emph{ratioTransfer} (sum of \emph{edge3} over sum of
\emph{edge4}). Return -1 for \emph{ratioRepay} if there is no
\emph{edge2} found. Return -1 for \emph{ratioDeposit} and
\emph{ratioTransfer} if there is no \emph{edge4} found.

Note: There may be multiple loans that the given account is related to. \\ \hline

		params &
		\innerCardVSpace{\begin{tabularx}{\attributeCardWidth}{|>{\paramNumberCell}C{\attributeNumberWidth}|>{\varNameCell}M|>{\typeCell}m{\typeWidth}|Y|} \hline
		$\mathsf{1}$ & id & ID & id of the Account \\ \hline
		$\mathsf{2}$ & threshold & 64-bit Float & threshold of amount \\ \hline
		$\mathsf{3}$ & startTime & DateTime & begin of the time window \\ \hline
		$\mathsf{4}$ & endTime & DateTime & end of the time window \\ \hline
		$\mathsf{5}$ & truncationLimit & 32-bit Integer & maximum edges traversed at each step \\ \hline
		$\mathsf{6}$ & truncationOrder & Enum & the sort order before truncation at each step \\ \hline
		\end{tabularx}}\innerCardVSpace \\ \hline

		result &
		\innerCardVSpace{\begin{tabularx}{\attributeCardWidth}{|>{\resultNumberCell}C{\attributeNumberWidth}|>{\varNameCell}M|>{\typeCell}m{\typeWidth}|>{\resultOriginCell}c|Y|} \hline
		$\mathsf{1}$ & ratioRepay & 32-bit Float & C &
				sumEdge1Amount/sumEdge2Amount (rounded to 3 decimal places) \\ \hline
		$\mathsf{2}$ & ratioDeposit & 32-bit Float & C &
				sumEdge1Amount/sumEdge4Amount (rounded to 3 decimal places) \\ \hline
		$\mathsf{3}$ & ratioTransfer & 32-bit Float & C &
				sumEdge3Amount/sumEdge4Amount (rounded to 3 decimal places) \\ \hline
		\end{tabularx}}\innerCardVSpace \\ \hline

	CPs &
	\multicolumn{1}{>{\raggedright}l|}{
		\chokePoint{3.2}, 
		\chokePoint{3.4}, 
		\chokePoint{6.2}, 
		\chokePoint{8.7}
		} \\ \hline
	%
\end{tabularx}
\queryCardVSpace

\let\emph\oldemph
\let\textbf\oldtextbf

\renewcommand{\currentQueryCard}{0}
\renewcommand*{\arraystretch}{1.1}

\subsection*{Transaction / complex-read / 10}
\label{sec:transaction-complex-read-10}

\let\oldemph\emph
\renewcommand{\emph}[1]{{\footnotesize \sf #1}}
\let\oldtextbf\textbf
\renewcommand{\textbf}[1]{{\it #1}}\renewcommand{\currentQueryCard}{10}
\marginpar{
	\raggedleft
	\scriptsize

	\queryRefCard{transaction-complex-read-01}{TCR}{1}\\
	\queryRefCard{transaction-complex-read-02}{TCR}{2}\\
	\queryRefCard{transaction-complex-read-03}{TCR}{3}\\
	\queryRefCard{transaction-complex-read-04}{TCR}{4}\\
	\queryRefCard{transaction-complex-read-05}{TCR}{5}\\
	\queryRefCard{transaction-complex-read-06}{TCR}{6}\\
	\queryRefCard{transaction-complex-read-07}{TCR}{7}\\
	\queryRefCard{transaction-complex-read-08}{TCR}{8}\\
	\queryRefCard{transaction-complex-read-09}{TCR}{9}\\
	\queryRefCard{transaction-complex-read-10}{TCR}{10}\\
	\queryRefCard{transaction-complex-read-11}{TCR}{11}\\
	\queryRefCard{transaction-complex-read-12}{TCR}{12}\\
}

\noindent\begin{tabularx}{\queryCardWidth}{|>{\queryPropertyCell}p{\queryPropertyCellWidth}|X|}
	\hline
	query & Transaction / complex-read / 10 \\ \hline
	title & Similarity of investor relationship \\ \hline

		pattern & \centering \includegraphics[scale=\yedscale,margin=0cm .2cm]{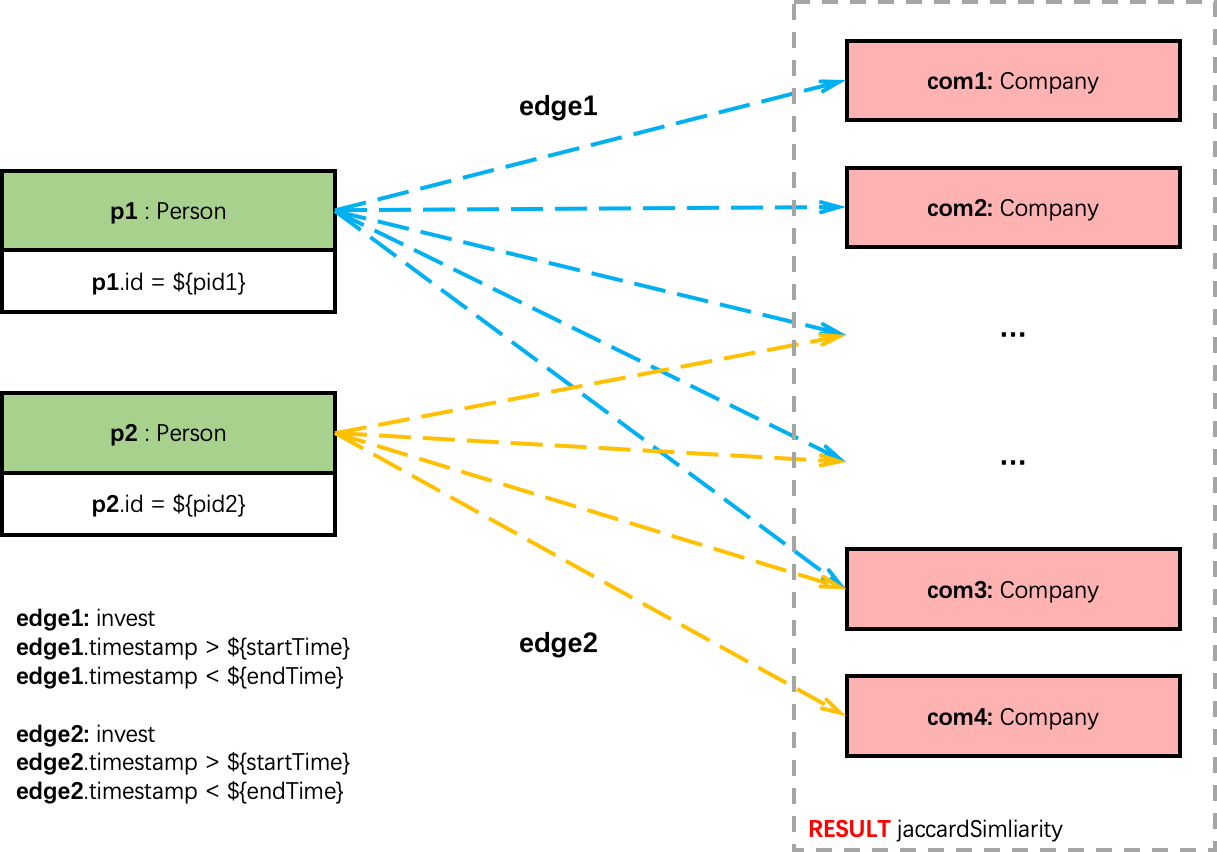} \tabularnewline \hline

	desc. & Given two \emph{Persons} and a specified time window between
\emph{startTime} and \emph{endTime}, find all the \emph{Companies} the
two \emph{Persons} invest in. Return the Jaccard similarity between the
two companies set. Return 0 if there is no edges found connecting to any
of these two persons. \\ \hline

		params &
		\innerCardVSpace{\begin{tabularx}{\attributeCardWidth}{|>{\paramNumberCell}C{\attributeNumberWidth}|>{\varNameCell}M|>{\typeCell}m{\typeWidth}|Y|} \hline
		$\mathsf{1}$ & pid1 & ID & id of Person1 \\ \hline
		$\mathsf{2}$ & pid2 & ID & id of Person2 \\ \hline
		$\mathsf{3}$ & startTime & DateTime & begin of the time window \\ \hline
		$\mathsf{4}$ & endTime & DateTime & end of the time window \\ \hline
		\end{tabularx}}\innerCardVSpace \\ \hline

		result &
		\innerCardVSpace{\begin{tabularx}{\attributeCardWidth}{|>{\resultNumberCell}C{\attributeNumberWidth}|>{\varNameCell}M|>{\typeCell}m{\typeWidth}|>{\resultOriginCell}c|Y|} \hline
		$\mathsf{1}$ & jaccardSimilarity & 32-bit Float & C &
				Jaccard similarity between two sets (rounded to 3 decimal places) \\ \hline
		\end{tabularx}}\innerCardVSpace \\ \hline

	CPs &
	\multicolumn{1}{>{\raggedright}l|}{
		\chokePoint{3.2}, 
		\chokePoint{3.4}, 
		\chokePoint{6.2}, 
		\chokePoint{8.7}
		} \\ \hline
	%
\end{tabularx}
\queryCardVSpace

\let\emph\oldemph
\let\textbf\oldtextbf

\renewcommand{\currentQueryCard}{0}
\renewcommand*{\arraystretch}{1.1}

\subsection*{Transaction / complex-read / 11}
\label{sec:transaction-complex-read-11}

\let\oldemph\emph
\renewcommand{\emph}[1]{{\footnotesize \sf #1}}
\let\oldtextbf\textbf
\renewcommand{\textbf}[1]{{\it #1}}\renewcommand{\currentQueryCard}{11}
\marginpar{
	\raggedleft
	\scriptsize

	\queryRefCard{transaction-complex-read-01}{TCR}{1}\\
	\queryRefCard{transaction-complex-read-02}{TCR}{2}\\
	\queryRefCard{transaction-complex-read-03}{TCR}{3}\\
	\queryRefCard{transaction-complex-read-04}{TCR}{4}\\
	\queryRefCard{transaction-complex-read-05}{TCR}{5}\\
	\queryRefCard{transaction-complex-read-06}{TCR}{6}\\
	\queryRefCard{transaction-complex-read-07}{TCR}{7}\\
	\queryRefCard{transaction-complex-read-08}{TCR}{8}\\
	\queryRefCard{transaction-complex-read-09}{TCR}{9}\\
	\queryRefCard{transaction-complex-read-10}{TCR}{10}\\
	\queryRefCard{transaction-complex-read-11}{TCR}{11}\\
	\queryRefCard{transaction-complex-read-12}{TCR}{12}\\
}

\noindent\begin{tabularx}{\queryCardWidth}{|>{\queryPropertyCell}p{\queryPropertyCellWidth}|X|}
	\hline
	query & Transaction / complex-read / 11 \\ \hline
	title & Guarantee Chain Detection \\ \hline

		pattern & \centering \includegraphics[scale=\yedscale,margin=0cm .2cm]{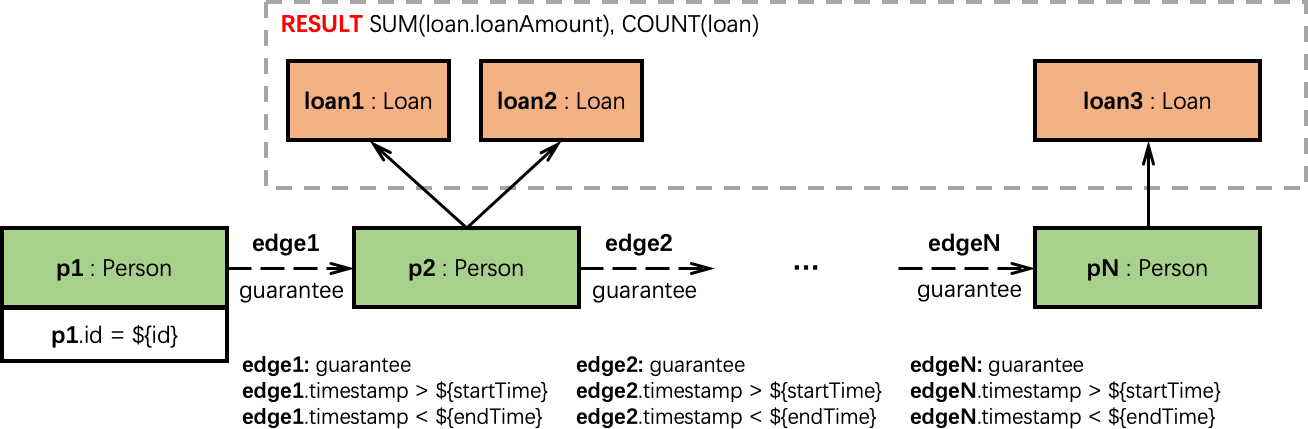} \tabularnewline \hline

	desc. & Given a \emph{Person} and a specified time window between
\emph{startTime} and \emph{endTime}, find all the persons in the
guarantee chain until end and their loans applied. Return the sum of
\emph{loan} amount and the count of distinct \emph{loan}s. \\ \hline

		params &
		\innerCardVSpace{\begin{tabularx}{\attributeCardWidth}{|>{\paramNumberCell}C{\attributeNumberWidth}|>{\varNameCell}M|>{\typeCell}m{\typeWidth}|Y|} \hline
		$\mathsf{1}$ & id & ID & id of the Person \\ \hline
		$\mathsf{2}$ & startTime & DateTime & begin of the time window \\ \hline
		$\mathsf{3}$ & endTime & DateTime & end of the time window \\ \hline
		$\mathsf{4}$ & truncationLimit & 32-bit Integer & maximum edges traversed at each step \\ \hline
		$\mathsf{5}$ & truncationOrder & Enum & the sort order before truncation at each step \\ \hline
		\end{tabularx}}\innerCardVSpace \\ \hline

		result &
		\innerCardVSpace{\begin{tabularx}{\attributeCardWidth}{|>{\resultNumberCell}C{\attributeNumberWidth}|>{\varNameCell}M|>{\typeCell}m{\typeWidth}|>{\resultOriginCell}c|Y|} \hline
		$\mathsf{1}$ & sumLoanAmount & 64-bit Float & A &
				sum of the loans' amount (rounded to 3 decimal places) \\ \hline
		$\mathsf{2}$ & numLoans & 32-bit Integer & A &
				num of the loans \\ \hline
		\end{tabularx}}\innerCardVSpace \\ \hline

	CPs &
	\multicolumn{1}{>{\raggedright}l|}{
		\chokePoint{3.2}, 
		\chokePoint{3.4}, 
		\chokePoint{6.2}, 
		\chokePoint{7.4}, 
		\chokePoint{8.7}
		} \\ \hline
	%
\end{tabularx}
\queryCardVSpace

\let\emph\oldemph
\let\textbf\oldtextbf

\renewcommand{\currentQueryCard}{0}
\renewcommand*{\arraystretch}{1.1}

\subsection*{Transaction / complex-read / 12}
\label{sec:transaction-complex-read-12}

\let\oldemph\emph
\renewcommand{\emph}[1]{{\footnotesize \sf #1}}
\let\oldtextbf\textbf
\renewcommand{\textbf}[1]{{\it #1}}\renewcommand{\currentQueryCard}{12}
\marginpar{
	\raggedleft
	\scriptsize

	\queryRefCard{transaction-complex-read-01}{TCR}{1}\\
	\queryRefCard{transaction-complex-read-02}{TCR}{2}\\
	\queryRefCard{transaction-complex-read-03}{TCR}{3}\\
	\queryRefCard{transaction-complex-read-04}{TCR}{4}\\
	\queryRefCard{transaction-complex-read-05}{TCR}{5}\\
	\queryRefCard{transaction-complex-read-06}{TCR}{6}\\
	\queryRefCard{transaction-complex-read-07}{TCR}{7}\\
	\queryRefCard{transaction-complex-read-08}{TCR}{8}\\
	\queryRefCard{transaction-complex-read-09}{TCR}{9}\\
	\queryRefCard{transaction-complex-read-10}{TCR}{10}\\
	\queryRefCard{transaction-complex-read-11}{TCR}{11}\\
	\queryRefCard{transaction-complex-read-12}{TCR}{12}\\
}

\noindent\begin{tabularx}{\queryCardWidth}{|>{\queryPropertyCell}p{\queryPropertyCellWidth}|X|}
	\hline
	query & Transaction / complex-read / 12 \\ \hline
	title & Transfer to company amount statistics \\ \hline

		pattern & \centering \includegraphics[scale=\yedscale,margin=0cm .2cm]{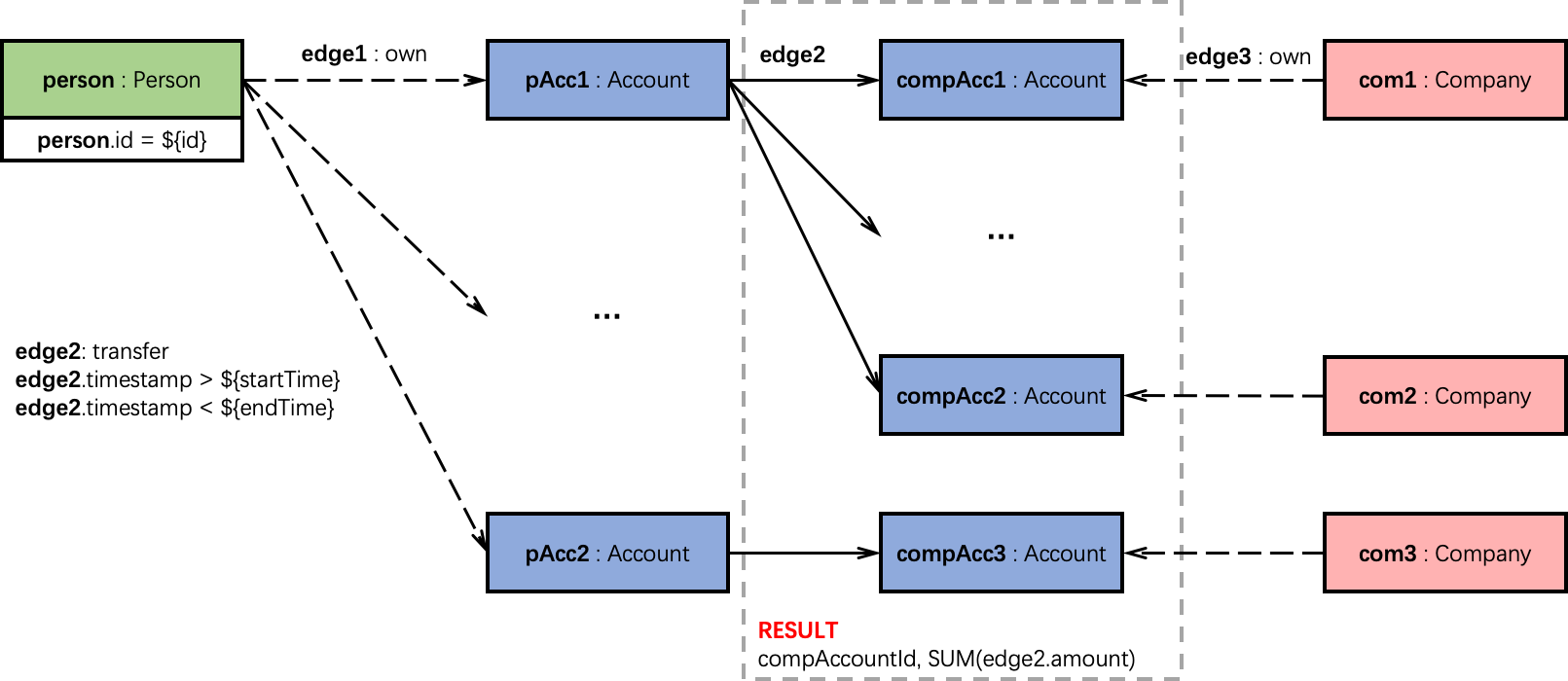} \tabularnewline \hline

	desc. & Given a \emph{Person} and a specified time window between
\emph{startTime} and \emph{endTime}, find all the company accounts that
s/he has transferred to. Return the ids of the companies' accounts and
the sum of their transfer amount. \\ \hline

		params &
		\innerCardVSpace{\begin{tabularx}{\attributeCardWidth}{|>{\paramNumberCell}C{\attributeNumberWidth}|>{\varNameCell}M|>{\typeCell}m{\typeWidth}|Y|} \hline
		$\mathsf{1}$ & id & ID & id of the person \\ \hline
		$\mathsf{2}$ & startTime & DateTime & begin of the time window \\ \hline
		$\mathsf{3}$ & endTime & DateTime & end of the time window \\ \hline
		$\mathsf{4}$ & truncationLimit & 32-bit Integer & maximum edges traversed at each step \\ \hline
		$\mathsf{5}$ & truncationOrder & Enum & the sort order before truncation at each step \\ \hline
		\end{tabularx}}\innerCardVSpace \\ \hline

		result &
		\innerCardVSpace{\begin{tabularx}{\attributeCardWidth}{|>{\resultNumberCell}C{\attributeNumberWidth}|>{\varNameCell}M|>{\typeCell}m{\typeWidth}|>{\resultOriginCell}c|Y|} \hline
		$\mathsf{1}$ & compAccountId & ID & R &
				the id of the company account \\ \hline
		$\mathsf{2}$ & sumEdge2Amount & 64-bit Float & A &
				the amount sum transferred to company's account (rounded to 3 decimal
places) \\ \hline
		\end{tabularx}}\innerCardVSpace \\ \hline

		sort		&
		\innerCardVSpace{\begin{tabularx}{\attributeCardWidth}{|>{\sortNumberCell}C{\attributeNumberWidth}|>{\varNameCell}X|>{\directionCell}c|Y|} \hline
		$\mathsf{1}$ & sumEdge2Amount & $\desc$ &  \\ \hline
		$\mathsf{2}$ & compAccountId & $\asc$ &  \\ \hline
		\end{tabularx}}\innerCardVSpace \\ \hline
	CPs &
	\multicolumn{1}{>{\raggedright}l|}{
		\chokePoint{3.2}, 
		\chokePoint{3.4}, 
		\chokePoint{6.2}, 
		\chokePoint{7.1}, 
		\chokePoint{8.7}
		} \\ \hline
	%
\end{tabularx}
\queryCardVSpace

\let\emph\oldemph
\let\textbf\oldtextbf

\renewcommand{\currentQueryCard}{0}


\section{Simple Read Queries}
\label{sec:simple-read-queries}

\renewcommand*{\arraystretch}{1.1}

\subsection*{Transaction / simple-read / 1}
\label{sec:transaction-simple-read-01}

\let\oldemph\emph
\renewcommand{\emph}[1]{{\footnotesize \sf #1}}
\let\oldtextbf\textbf
\renewcommand{\textbf}[1]{{\it #1}}\renewcommand{\currentQueryCard}{1}
\marginpar{
	\raggedleft
	\scriptsize

	\queryRefCard{transaction-simple-read-01}{TSR}{1}\\
	\queryRefCard{transaction-simple-read-02}{TSR}{2}\\
	\queryRefCard{transaction-simple-read-03}{TSR}{3}\\
	\queryRefCard{transaction-simple-read-04}{TSR}{4}\\
	\queryRefCard{transaction-simple-read-05}{TSR}{5}\\
	\queryRefCard{transaction-simple-read-06}{TSR}{6}\\
}

\noindent\begin{tabularx}{\queryCardWidth}{|>{\queryPropertyCell}p{\queryPropertyCellWidth}|X|}
	\hline
	query & Transaction / simple-read / 1 \\ \hline
	title & Exact account query \\ \hline

		pattern & \centering \includegraphics[scale=\yedscale,margin=0cm .2cm]{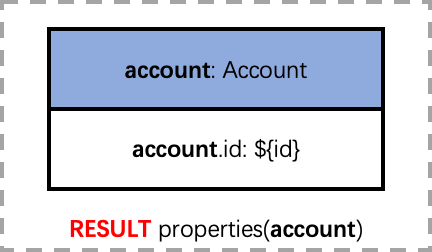} \tabularnewline \hline

	desc. & Given an id of an \emph{Account}, find the properties of the specific
\emph{Account}. \\ \hline

		params &
		\innerCardVSpace{\begin{tabularx}{\attributeCardWidth}{|>{\paramNumberCell}C{\attributeNumberWidth}|>{\varNameCell}M|>{\typeCell}m{\typeWidth}|Y|} \hline
		$\mathsf{1}$ & id & ID & id of the Account \\ \hline
		\end{tabularx}}\innerCardVSpace \\ \hline

		result &
		\innerCardVSpace{\begin{tabularx}{\attributeCardWidth}{|>{\resultNumberCell}C{\attributeNumberWidth}|>{\varNameCell}M|>{\typeCell}m{\typeWidth}|>{\resultOriginCell}c|Y|} \hline
		$\mathsf{1}$ & createTime & DateTime & R &
				the time when the account created \\ \hline
		$\mathsf{2}$ & isBlocked & Boolean & R &
				if the account is blocked \\ \hline
		$\mathsf{3}$ & type & String & R &
				the account type \\ \hline
		\end{tabularx}}\innerCardVSpace \\ \hline
	
%
	%
	%
	%
	%
\end{tabularx}
\queryCardVSpace

\let\emph\oldemph
\let\textbf\oldtextbf

\renewcommand{\currentQueryCard}{0}
\renewcommand*{\arraystretch}{1.1}

\subsection*{Transaction / simple-read / 2}
\label{sec:transaction-simple-read-02}

\let\oldemph\emph
\renewcommand{\emph}[1]{{\footnotesize \sf #1}}
\let\oldtextbf\textbf
\renewcommand{\textbf}[1]{{\it #1}}\renewcommand{\currentQueryCard}{2}
\marginpar{
	\raggedleft
	\scriptsize

	\queryRefCard{transaction-simple-read-01}{TSR}{1}\\
	\queryRefCard{transaction-simple-read-02}{TSR}{2}\\
	\queryRefCard{transaction-simple-read-03}{TSR}{3}\\
	\queryRefCard{transaction-simple-read-04}{TSR}{4}\\
	\queryRefCard{transaction-simple-read-05}{TSR}{5}\\
	\queryRefCard{transaction-simple-read-06}{TSR}{6}\\
}

\noindent\begin{tabularx}{\queryCardWidth}{|>{\queryPropertyCell}p{\queryPropertyCellWidth}|X|}
	\hline
	query & Transaction / simple-read / 2 \\ \hline
	title & Transfer-ins and transfer-outs \\ \hline

		pattern & \centering \includegraphics[scale=\yedscale,margin=0cm .2cm]{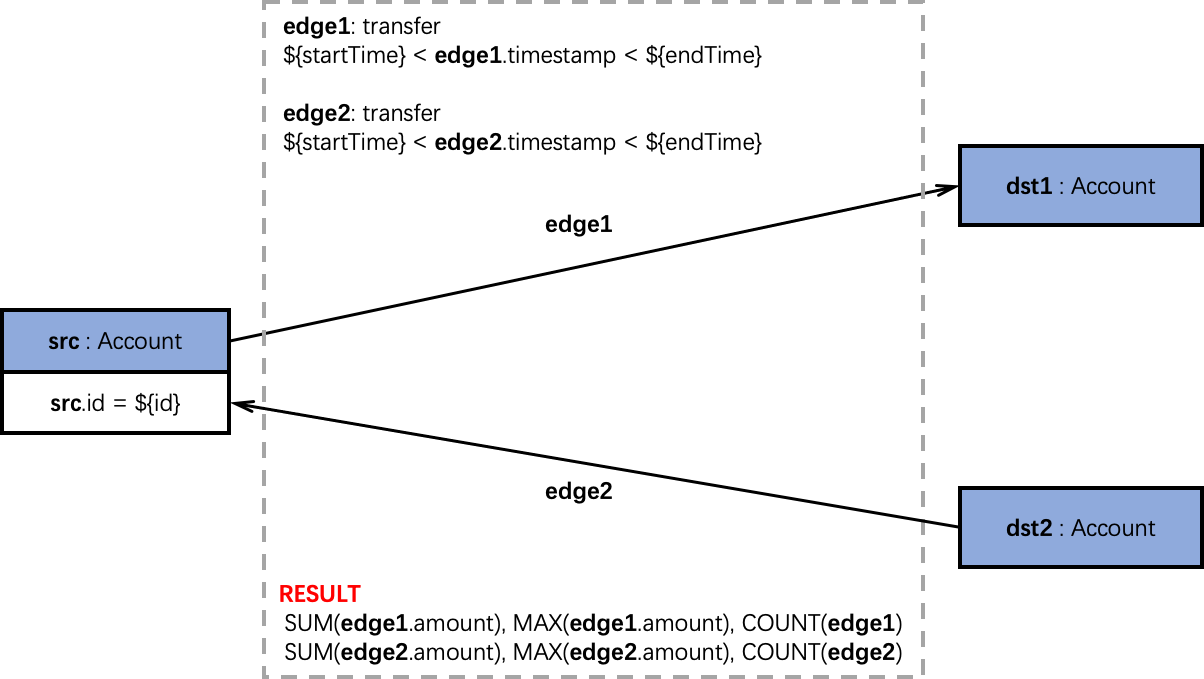} \tabularnewline \hline

	desc. & Given an \emph{account}, find the sum and max of fund amount in
\emph{transfer-ins} and \emph{transfer-outs} between them in a specific
time range between \emph{startTime} and \emph{endTime}. Return the sum
and max of amount. For \emph{edge1} and \emph{edge2}, return -1 for the
max (\emph{maxEdge1Amount} and \emph{maxEdge2Amount}) if there is no
transfer. \\ \hline

		params &
		\innerCardVSpace{\begin{tabularx}{\attributeCardWidth}{|>{\paramNumberCell}C{\attributeNumberWidth}|>{\varNameCell}M|>{\typeCell}m{\typeWidth}|Y|} \hline
		$\mathsf{1}$ & id & ID & id of the account \\ \hline
		$\mathsf{2}$ & startTime & DateTime & begin of the time window \\ \hline
		$\mathsf{3}$ & endTime & DateTime & end of the time window \\ \hline
		\end{tabularx}}\innerCardVSpace \\ \hline

		result &
		\innerCardVSpace{\begin{tabularx}{\attributeCardWidth}{|>{\resultNumberCell}C{\attributeNumberWidth}|>{\varNameCell}M|>{\typeCell}m{\typeWidth}|>{\resultOriginCell}c|Y|} \hline
		$\mathsf{1}$ & sumEdge1Amount & 64-bit Float & A &
				sum of transfer-outs (rounded to 3 decimal places) \\ \hline
		$\mathsf{2}$ & maxEdge1Amount & 64-bit Float & A &
				max of transfer-outs (rounded to 3 decimal places) \\ \hline
		$\mathsf{3}$ & numEdge1 & 64-bit Integer & A &
				count of transfer-outs \\ \hline
		$\mathsf{4}$ & sumEdge2Amount & 64-bit Float & A &
				sum of transfer-ins (rounded to 3 decimal places) \\ \hline
		$\mathsf{5}$ & maxEdge2Amount & 64-bit Float & A &
				max of transfer-ins (rounded to 3 decimal places) \\ \hline
		$\mathsf{6}$ & numEdge2 & 64-bit Integer & A &
				count of transfer-outs \\ \hline
		\end{tabularx}}\innerCardVSpace \\ \hline
	
%
	%
	%
	%
	%
\end{tabularx}
\queryCardVSpace

\let\emph\oldemph
\let\textbf\oldtextbf

\renewcommand{\currentQueryCard}{0}
\renewcommand*{\arraystretch}{1.1}

\subsection*{Transaction / simple-read / 3}
\label{sec:transaction-simple-read-03}

\let\oldemph\emph
\renewcommand{\emph}[1]{{\footnotesize \sf #1}}
\let\oldtextbf\textbf
\renewcommand{\textbf}[1]{{\it #1}}\renewcommand{\currentQueryCard}{3}
\marginpar{
	\raggedleft
	\scriptsize

	\queryRefCard{transaction-simple-read-01}{TSR}{1}\\
	\queryRefCard{transaction-simple-read-02}{TSR}{2}\\
	\queryRefCard{transaction-simple-read-03}{TSR}{3}\\
	\queryRefCard{transaction-simple-read-04}{TSR}{4}\\
	\queryRefCard{transaction-simple-read-05}{TSR}{5}\\
	\queryRefCard{transaction-simple-read-06}{TSR}{6}\\
}

\noindent\begin{tabularx}{\queryCardWidth}{|>{\queryPropertyCell}p{\queryPropertyCellWidth}|X|}
	\hline
	query & Transaction / simple-read / 3 \\ \hline
	title & Many-to-one blocked account monitoring \\ \hline

		pattern & \centering \includegraphics[scale=\yedscale,margin=0cm .2cm]{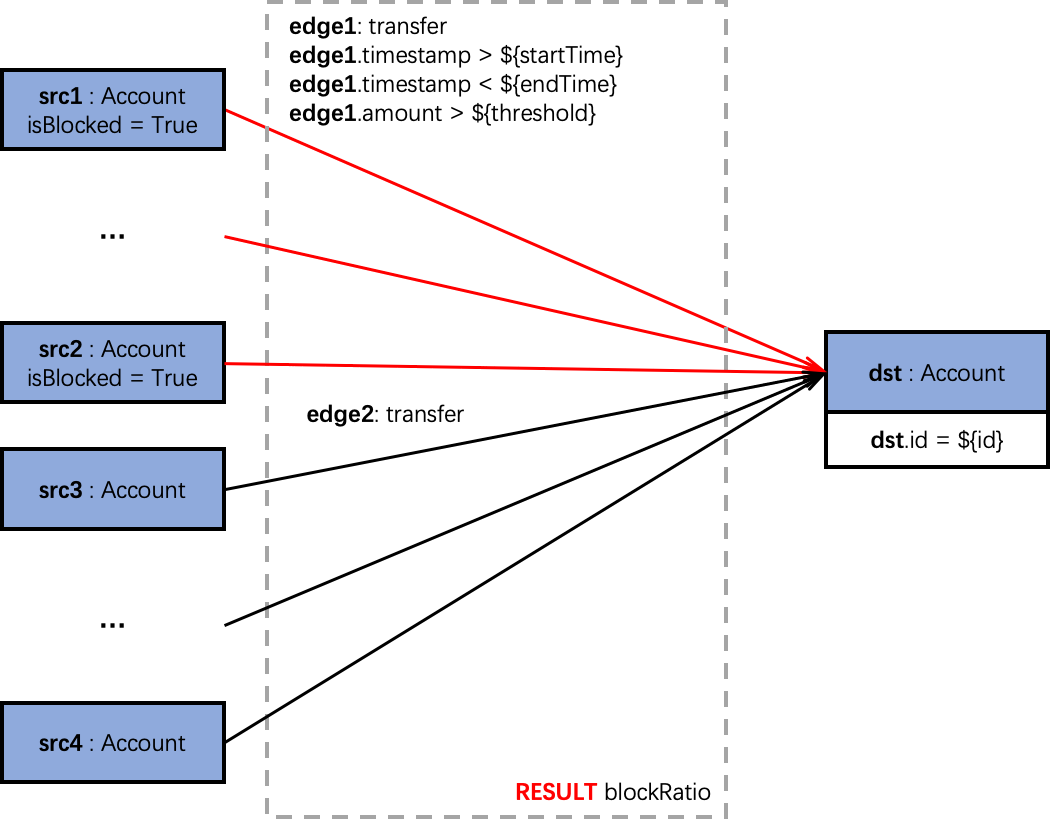} \tabularnewline \hline

	desc. & Given an \emph{Account}, find the ratio of \emph{transfer-ins} from
blocked \emph{Accounts} in all its \emph{transfer-ins} in a specific
time range between \emph{startTime} and \emph{endTime}. Return the
ratio. Return -1 if there is no \emph{transfer-ins} to the given
account. \\ \hline

		params &
		\innerCardVSpace{\begin{tabularx}{\attributeCardWidth}{|>{\paramNumberCell}C{\attributeNumberWidth}|>{\varNameCell}M|>{\typeCell}m{\typeWidth}|Y|} \hline
		$\mathsf{1}$ & id & ID & id of the dstAccount \\ \hline
		$\mathsf{2}$ & threshold & 64-bit Float & threshold of transfer amount \\ \hline
		$\mathsf{3}$ & startTime & DateTime & begin of the time window \\ \hline
		$\mathsf{4}$ & endTime & DateTime & end of the time window \\ \hline
		\end{tabularx}}\innerCardVSpace \\ \hline

		result &
		\innerCardVSpace{\begin{tabularx}{\attributeCardWidth}{|>{\resultNumberCell}C{\attributeNumberWidth}|>{\varNameCell}M|>{\typeCell}m{\typeWidth}|>{\resultOriginCell}c|Y|} \hline
		$\mathsf{1}$ & blockRatio & 32-bit Float & A &
				count(edge1) over count(edge2) (rounded to 3 decimal places) \\ \hline
		\end{tabularx}}\innerCardVSpace \\ \hline
	
%
	%
	%
	%
	%
\end{tabularx}
\queryCardVSpace

\let\emph\oldemph
\let\textbf\oldtextbf

\renewcommand{\currentQueryCard}{0}
\renewcommand*{\arraystretch}{1.1}

\subsection*{Transaction / simple-read / 4}
\label{sec:transaction-simple-read-04}

\let\oldemph\emph
\renewcommand{\emph}[1]{{\footnotesize \sf #1}}
\let\oldtextbf\textbf
\renewcommand{\textbf}[1]{{\it #1}}\renewcommand{\currentQueryCard}{4}
\marginpar{
	\raggedleft
	\scriptsize

	\queryRefCard{transaction-simple-read-01}{TSR}{1}\\
	\queryRefCard{transaction-simple-read-02}{TSR}{2}\\
	\queryRefCard{transaction-simple-read-03}{TSR}{3}\\
	\queryRefCard{transaction-simple-read-04}{TSR}{4}\\
	\queryRefCard{transaction-simple-read-05}{TSR}{5}\\
	\queryRefCard{transaction-simple-read-06}{TSR}{6}\\
}

\noindent\begin{tabularx}{\queryCardWidth}{|>{\queryPropertyCell}p{\queryPropertyCellWidth}|X|}
	\hline
	query & Transaction / simple-read / 4 \\ \hline
	title & Account transfer-outs over threshold \\ \hline

		pattern & \centering \includegraphics[scale=\yedscale,margin=0cm .2cm]{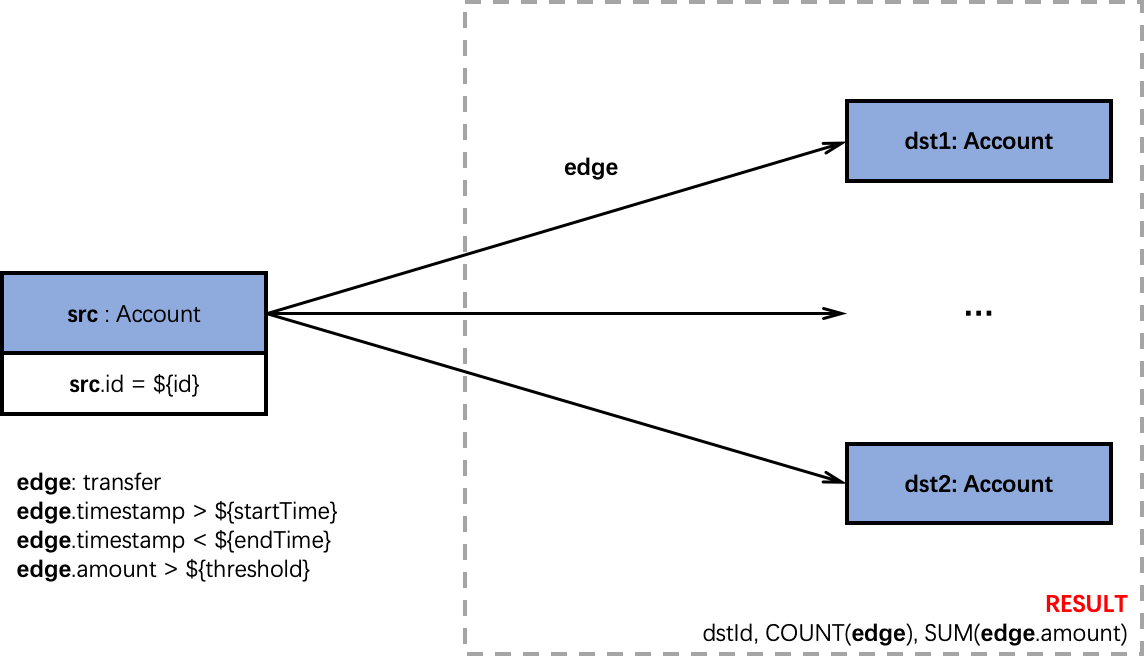} \tabularnewline \hline

	desc. & Given an account (\emph{src}), find all the transfer-outs (\emph{edge})
from the \emph{src} to a \emph{dst} where the amount exceeds
\emph{threshold} in a specific time range between \emph{startTime} and
\emph{endTime}. Return the count of \emph{transfer-outs} and the amount
sum. \\ \hline

		params &
		\innerCardVSpace{\begin{tabularx}{\attributeCardWidth}{|>{\paramNumberCell}C{\attributeNumberWidth}|>{\varNameCell}M|>{\typeCell}m{\typeWidth}|Y|} \hline
		$\mathsf{1}$ & id & ID & id of the dstAccount \\ \hline
		$\mathsf{2}$ & threshold & 64-bit Float & threshold of transfer amount \\ \hline
		$\mathsf{3}$ & startTime & DateTime & begin of the time window \\ \hline
		$\mathsf{4}$ & endTime & DateTime & end of the time window \\ \hline
		\end{tabularx}}\innerCardVSpace \\ \hline

		result &
		\innerCardVSpace{\begin{tabularx}{\attributeCardWidth}{|>{\resultNumberCell}C{\attributeNumberWidth}|>{\varNameCell}M|>{\typeCell}m{\typeWidth}|>{\resultOriginCell}c|Y|} \hline
		$\mathsf{1}$ & dstId & ID & R &
				the id of the dst account \\ \hline
		$\mathsf{2}$ & numEdges & 32-bit Integer & A &
				num of the transfers from src to dst \\ \hline
		$\mathsf{3}$ & sumAmount & 64-bit Float & A &
				sum of the transfers from src to dst (rounded to 3 decimal places) \\ \hline
		\end{tabularx}}\innerCardVSpace \\ \hline

		sort		&
		\innerCardVSpace{\begin{tabularx}{\attributeCardWidth}{|>{\sortNumberCell}C{\attributeNumberWidth}|>{\varNameCell}X|>{\directionCell}c|Y|} \hline
		$\mathsf{1}$ & sumAmount & $\desc$ &  \\ \hline
		$\mathsf{2}$ & dstId & $\asc$ &  \\ \hline
		\end{tabularx}}\innerCardVSpace \\ \hline
	%
	%
	%
\end{tabularx}
\queryCardVSpace

\let\emph\oldemph
\let\textbf\oldtextbf

\renewcommand{\currentQueryCard}{0}
\renewcommand*{\arraystretch}{1.1}

\subsection*{Transaction / simple-read / 5}
\label{sec:transaction-simple-read-05}

\let\oldemph\emph
\renewcommand{\emph}[1]{{\footnotesize \sf #1}}
\let\oldtextbf\textbf
\renewcommand{\textbf}[1]{{\it #1}}\renewcommand{\currentQueryCard}{5}
\marginpar{
	\raggedleft
	\scriptsize

	\queryRefCard{transaction-simple-read-01}{TSR}{1}\\
	\queryRefCard{transaction-simple-read-02}{TSR}{2}\\
	\queryRefCard{transaction-simple-read-03}{TSR}{3}\\
	\queryRefCard{transaction-simple-read-04}{TSR}{4}\\
	\queryRefCard{transaction-simple-read-05}{TSR}{5}\\
	\queryRefCard{transaction-simple-read-06}{TSR}{6}\\
}

\noindent\begin{tabularx}{\queryCardWidth}{|>{\queryPropertyCell}p{\queryPropertyCellWidth}|X|}
	\hline
	query & Transaction / simple-read / 5 \\ \hline
	title & Account transfer-ins over threshold \\ \hline

		pattern & \centering \includegraphics[scale=\yedscale,margin=0cm .2cm]{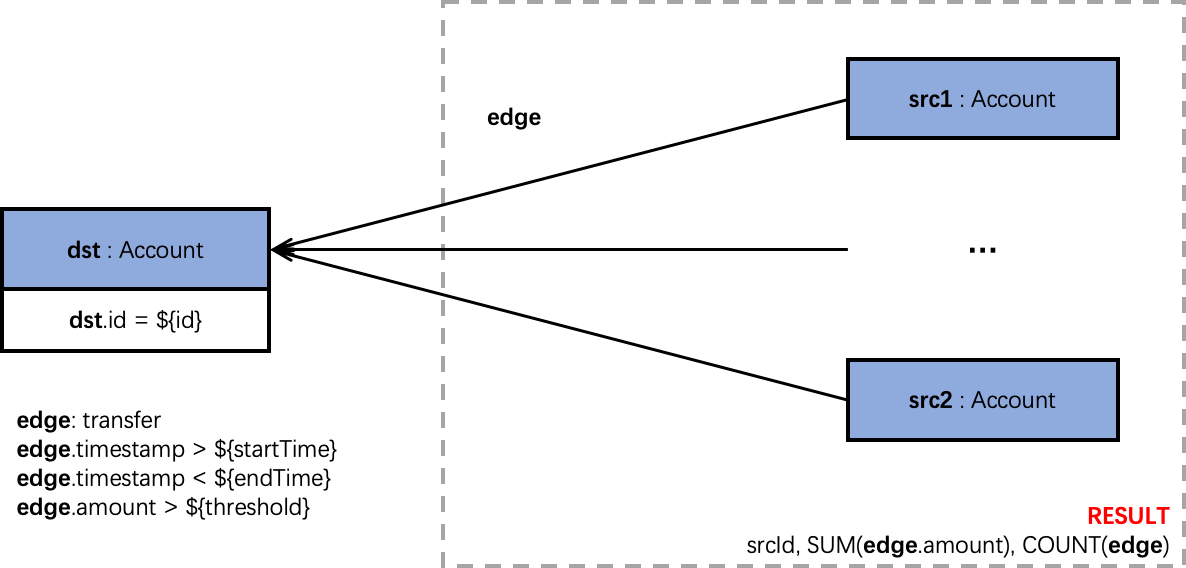} \tabularnewline \hline

	desc. & Given an account (\emph{dst}), find all the transfer-ins (\emph{edge})
from the \emph{src} to a \emph{dst} where the amount exceeds
\emph{threshold} in a specific time range between \emph{startTime} and
\emph{endTime}. Return the count of \emph{transfer-ins} and the amount
sum. \\ \hline

		params &
		\innerCardVSpace{\begin{tabularx}{\attributeCardWidth}{|>{\paramNumberCell}C{\attributeNumberWidth}|>{\varNameCell}M|>{\typeCell}m{\typeWidth}|Y|} \hline
		$\mathsf{1}$ & id & ID & id of the Account \\ \hline
		$\mathsf{2}$ & threshold & 64-bit Float & threshold of transfer amount \\ \hline
		$\mathsf{3}$ & startTime & DateTime & begin of the time window \\ \hline
		$\mathsf{4}$ & endTime & DateTime & end of the time window \\ \hline
		\end{tabularx}}\innerCardVSpace \\ \hline

		result &
		\innerCardVSpace{\begin{tabularx}{\attributeCardWidth}{|>{\resultNumberCell}C{\attributeNumberWidth}|>{\varNameCell}M|>{\typeCell}m{\typeWidth}|>{\resultOriginCell}c|Y|} \hline
		$\mathsf{1}$ & srcId & ID & R &
				the id of the src account \\ \hline
		$\mathsf{2}$ & numEdges & 32-bit Integer & A &
				num of the transfers from src to dst \\ \hline
		$\mathsf{3}$ & sumAmount & 64-bit Float & A &
				sum of the transfers from src to dst (rounded to 3 decimal places) \\ \hline
		\end{tabularx}}\innerCardVSpace \\ \hline

		sort		&
		\innerCardVSpace{\begin{tabularx}{\attributeCardWidth}{|>{\sortNumberCell}C{\attributeNumberWidth}|>{\varNameCell}X|>{\directionCell}c|Y|} \hline
		$\mathsf{1}$ & sumAmount & $\desc$ &  \\ \hline
		$\mathsf{2}$ & srcId & $\asc$ &  \\ \hline
		\end{tabularx}}\innerCardVSpace \\ \hline
	%
	%
	%
\end{tabularx}
\queryCardVSpace

\let\emph\oldemph
\let\textbf\oldtextbf

\renewcommand{\currentQueryCard}{0}
\renewcommand*{\arraystretch}{1.1}

\subsection*{Transaction / simple-read / 6}
\label{sec:transaction-simple-read-06}

\let\oldemph\emph
\renewcommand{\emph}[1]{{\footnotesize \sf #1}}
\let\oldtextbf\textbf
\renewcommand{\textbf}[1]{{\it #1}}\renewcommand{\currentQueryCard}{6}
\marginpar{
	\raggedleft
	\scriptsize

	\queryRefCard{transaction-simple-read-01}{TSR}{1}\\
	\queryRefCard{transaction-simple-read-02}{TSR}{2}\\
	\queryRefCard{transaction-simple-read-03}{TSR}{3}\\
	\queryRefCard{transaction-simple-read-04}{TSR}{4}\\
	\queryRefCard{transaction-simple-read-05}{TSR}{5}\\
	\queryRefCard{transaction-simple-read-06}{TSR}{6}\\
}

\noindent\begin{tabularx}{\queryCardWidth}{|>{\queryPropertyCell}p{\queryPropertyCellWidth}|X|}
	\hline
	query & Transaction / simple-read / 6 \\ \hline
	title & Accounts with the same transfer sources of exact account \\ \hline

		pattern & \centering \includegraphics[scale=\yedscale,margin=0cm .2cm]{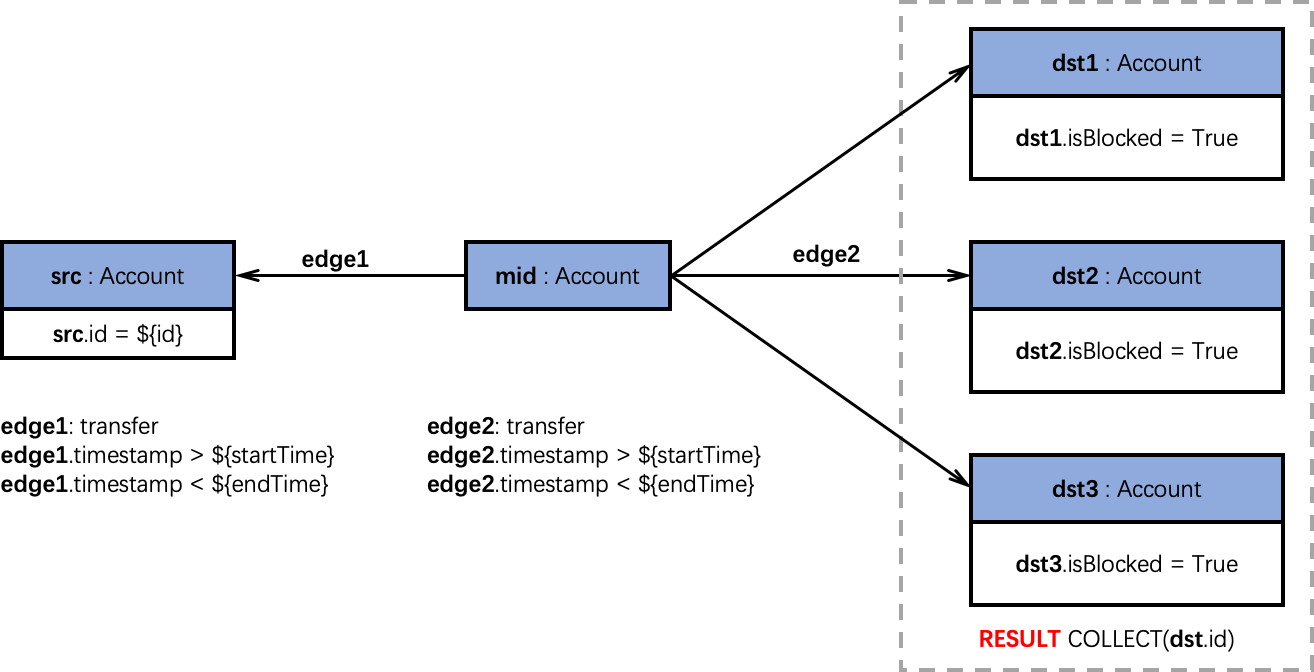} \tabularnewline \hline

	desc. & Given an \emph{Account} (\emph{account}), find all the blocked
\emph{Accounts} (\emph{dstAccounts}) that connect to a common account
(\emph{midAccoun}t) with the given \emph{Account} (\emph{account}).
Return all the \emph{accounts'} id. \\ \hline

		params &
		\innerCardVSpace{\begin{tabularx}{\attributeCardWidth}{|>{\paramNumberCell}C{\attributeNumberWidth}|>{\varNameCell}M|>{\typeCell}m{\typeWidth}|Y|} \hline
		$\mathsf{1}$ & id & ID & id of the Account \\ \hline
		$\mathsf{2}$ & startTime & DateTime & begin of the time window \\ \hline
		$\mathsf{3}$ & endTime & DateTime & end of the time window \\ \hline
		\end{tabularx}}\innerCardVSpace \\ \hline

		result &
		\innerCardVSpace{\begin{tabularx}{\attributeCardWidth}{|>{\resultNumberCell}C{\attributeNumberWidth}|>{\varNameCell}M|>{\typeCell}m{\typeWidth}|>{\resultOriginCell}c|Y|} \hline
		$\mathsf{1}$ & dstId & ID & R &
				ids of the accounts having same upstream account as the given account \\ \hline
		\end{tabularx}}\innerCardVSpace \\ \hline

		sort		&
		\innerCardVSpace{\begin{tabularx}{\attributeCardWidth}{|>{\sortNumberCell}C{\attributeNumberWidth}|>{\varNameCell}X|>{\directionCell}c|Y|} \hline
		$\mathsf{1}$ & dstId & $\asc$ &  \\ \hline
		\end{tabularx}}\innerCardVSpace \\ \hline
	%
	%
	%
\end{tabularx}
\queryCardVSpace

\let\emph\oldemph
\let\textbf\oldtextbf

\renewcommand{\currentQueryCard}{0}


\section{Write Queries}
\label{sec:write-queries}

In write queries, there are mainly two types of queries, inserts and deletes. In
real systems, there are deletion operations besides delete operations. Deletion
operations limit the architecture that can be used by a system. On the other
hand, systems are supposed to provide API for users to express delete operations
no matter with high-level structured languages like GQL and openCypher or
low-level storage layer API.

\renewcommand*{\arraystretch}{1.1}

\subsection*{Transaction / write / 1}
\label{sec:transaction-write-01}

\let\oldemph\emph
\renewcommand{\emph}[1]{{\footnotesize \sf #1}}
\let\oldtextbf\textbf
\renewcommand{\textbf}[1]{{\it #1}}\renewcommand{\currentQueryCard}{1}
\marginpar{
	\raggedleft
	\scriptsize
	
	\queryRefCard{transaction-write-01}{TW}{1}\\
	\queryRefCard{transaction-write-02}{TW}{2}\\
	\queryRefCard{transaction-write-03}{TW}{3}\\
	\queryRefCard{transaction-write-04}{TW}{4}\\
	\queryRefCard{transaction-write-05}{TW}{5}\\
	\queryRefCard{transaction-write-06}{TW}{6}\\
	\queryRefCard{transaction-write-07}{TW}{7}\\
	\queryRefCard{transaction-write-08}{TW}{8}\\
	\queryRefCard{transaction-write-09}{TW}{9}\\
	\queryRefCard{transaction-write-10}{TW}{10}\\
	\queryRefCard{transaction-write-11}{TW}{11}\\
	\queryRefCard{transaction-write-12}{TW}{12}\\
	\queryRefCard{transaction-write-13}{TW}{13}\\
	\queryRefCard{transaction-write-14}{TW}{14}\\
	\queryRefCard{transaction-write-15}{TW}{15}\\
	\queryRefCard{transaction-write-16}{TW}{16}\\
	\queryRefCard{transaction-write-17}{TW}{17}\\
	\queryRefCard{transaction-write-18}{TW}{18}\\
	\queryRefCard{transaction-write-19}{TW}{19}\\
}

\noindent\begin{tabularx}{\queryCardWidth}{|>{\queryPropertyCell}p{\queryPropertyCellWidth}|X|}
	\hline
	query & Transaction / write / 1 \\ \hline
	title & Add a Person \\ \hline

		pattern & \centering \includegraphics[scale=\yedscale,margin=0cm .2cm]{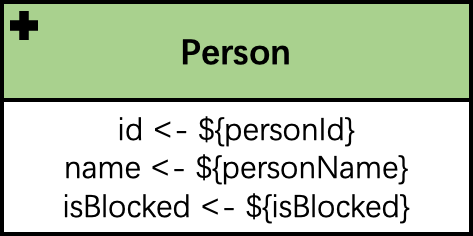} \tabularnewline \hline

	desc. & Add a \emph{Person}. \\ \hline

		params &
		\innerCardVSpace{\begin{tabularx}{\attributeCardWidth}{|>{\paramNumberCell}C{\attributeNumberWidth}|>{\varNameCell}M|>{\typeCell}m{\typeWidth}|Y|} \hline
		$\mathsf{1}$ & \$personId & ID &  \\ \hline
		$\mathsf{2}$ & \$personName & String &  \\ \hline
		$\mathsf{3}$ & \$isBlocked & Boolean &  \\ \hline
		\end{tabularx}}\innerCardVSpace \\ \hline
	
%
	
%
	%
	%
	%
	%
\end{tabularx}
\queryCardVSpace

\let\emph\oldemph
\let\textbf\oldtextbf

\renewcommand{\currentQueryCard}{0}
\renewcommand*{\arraystretch}{1.1}

\subsection*{Transaction / write / 2}
\label{sec:transaction-write-02}

\let\oldemph\emph
\renewcommand{\emph}[1]{{\footnotesize \sf #1}}
\let\oldtextbf\textbf
\renewcommand{\textbf}[1]{{\it #1}}\renewcommand{\currentQueryCard}{2}
\marginpar{
	\raggedleft
	\scriptsize
	
	\queryRefCard{transaction-write-01}{TW}{1}\\
	\queryRefCard{transaction-write-02}{TW}{2}\\
	\queryRefCard{transaction-write-03}{TW}{3}\\
	\queryRefCard{transaction-write-04}{TW}{4}\\
	\queryRefCard{transaction-write-05}{TW}{5}\\
	\queryRefCard{transaction-write-06}{TW}{6}\\
	\queryRefCard{transaction-write-07}{TW}{7}\\
	\queryRefCard{transaction-write-08}{TW}{8}\\
	\queryRefCard{transaction-write-09}{TW}{9}\\
	\queryRefCard{transaction-write-10}{TW}{10}\\
	\queryRefCard{transaction-write-11}{TW}{11}\\
	\queryRefCard{transaction-write-12}{TW}{12}\\
	\queryRefCard{transaction-write-13}{TW}{13}\\
	\queryRefCard{transaction-write-14}{TW}{14}\\
	\queryRefCard{transaction-write-15}{TW}{15}\\
	\queryRefCard{transaction-write-16}{TW}{16}\\
	\queryRefCard{transaction-write-17}{TW}{17}\\
	\queryRefCard{transaction-write-18}{TW}{18}\\
	\queryRefCard{transaction-write-19}{TW}{19}\\
}

\noindent\begin{tabularx}{\queryCardWidth}{|>{\queryPropertyCell}p{\queryPropertyCellWidth}|X|}
	\hline
	query & Transaction / write / 2 \\ \hline
	title & Add a Company \\ \hline

		pattern & \centering \includegraphics[scale=\yedscale,margin=0cm .2cm]{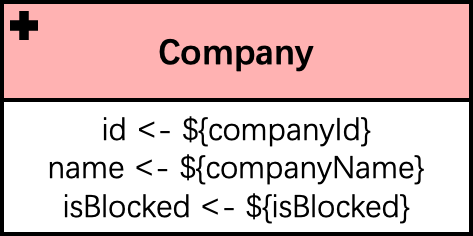} \tabularnewline \hline

	desc. & Add a \emph{Company}. \\ \hline

		params &
		\innerCardVSpace{\begin{tabularx}{\attributeCardWidth}{|>{\paramNumberCell}C{\attributeNumberWidth}|>{\varNameCell}M|>{\typeCell}m{\typeWidth}|Y|} \hline
		$\mathsf{1}$ & \$companyId & ID &  \\ \hline
		$\mathsf{2}$ & \$companyName & String &  \\ \hline
		$\mathsf{3}$ & \$isBlocked & Boolean &  \\ \hline
		\end{tabularx}}\innerCardVSpace \\ \hline
	
%
	
%
	%
	%
	%
	%
\end{tabularx}
\queryCardVSpace

\let\emph\oldemph
\let\textbf\oldtextbf

\renewcommand{\currentQueryCard}{0}
\renewcommand*{\arraystretch}{1.1}

\subsection*{Transaction / write / 3}
\label{sec:transaction-write-03}

\let\oldemph\emph
\renewcommand{\emph}[1]{{\footnotesize \sf #1}}
\let\oldtextbf\textbf
\renewcommand{\textbf}[1]{{\it #1}}\renewcommand{\currentQueryCard}{3}
\marginpar{
	\raggedleft
	\scriptsize
	
	\queryRefCard{transaction-write-01}{TW}{1}\\
	\queryRefCard{transaction-write-02}{TW}{2}\\
	\queryRefCard{transaction-write-03}{TW}{3}\\
	\queryRefCard{transaction-write-04}{TW}{4}\\
	\queryRefCard{transaction-write-05}{TW}{5}\\
	\queryRefCard{transaction-write-06}{TW}{6}\\
	\queryRefCard{transaction-write-07}{TW}{7}\\
	\queryRefCard{transaction-write-08}{TW}{8}\\
	\queryRefCard{transaction-write-09}{TW}{9}\\
	\queryRefCard{transaction-write-10}{TW}{10}\\
	\queryRefCard{transaction-write-11}{TW}{11}\\
	\queryRefCard{transaction-write-12}{TW}{12}\\
	\queryRefCard{transaction-write-13}{TW}{13}\\
	\queryRefCard{transaction-write-14}{TW}{14}\\
	\queryRefCard{transaction-write-15}{TW}{15}\\
	\queryRefCard{transaction-write-16}{TW}{16}\\
	\queryRefCard{transaction-write-17}{TW}{17}\\
	\queryRefCard{transaction-write-18}{TW}{18}\\
	\queryRefCard{transaction-write-19}{TW}{19}\\
}

\noindent\begin{tabularx}{\queryCardWidth}{|>{\queryPropertyCell}p{\queryPropertyCellWidth}|X|}
	\hline
	query & Transaction / write / 3 \\ \hline
	title & Add a Medium \\ \hline

		pattern & \centering \includegraphics[scale=\yedscale,margin=0cm .2cm]{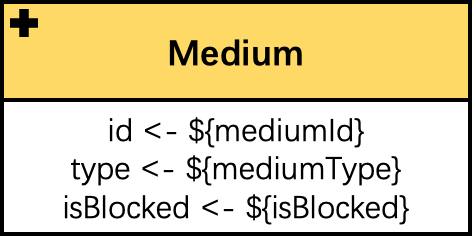} \tabularnewline \hline

	desc. & Add a \emph{Medium}. \\ \hline

		params &
		\innerCardVSpace{\begin{tabularx}{\attributeCardWidth}{|>{\paramNumberCell}C{\attributeNumberWidth}|>{\varNameCell}M|>{\typeCell}m{\typeWidth}|Y|} \hline
		$\mathsf{1}$ & \$mediumId & ID &  \\ \hline
		$\mathsf{2}$ & \$mediumType & String &  \\ \hline
		$\mathsf{3}$ & \$isBlocked & Boolean &  \\ \hline
		\end{tabularx}}\innerCardVSpace \\ \hline
	
%
	
%
	%
	%
	%
	%
\end{tabularx}
\queryCardVSpace

\let\emph\oldemph
\let\textbf\oldtextbf

\renewcommand{\currentQueryCard}{0}
\renewcommand*{\arraystretch}{1.1}

\subsection*{Transaction / write / 4}
\label{sec:transaction-write-04}

\let\oldemph\emph
\renewcommand{\emph}[1]{{\footnotesize \sf #1}}
\let\oldtextbf\textbf
\renewcommand{\textbf}[1]{{\it #1}}\renewcommand{\currentQueryCard}{4}
\marginpar{
	\raggedleft
	\scriptsize
	
	\queryRefCard{transaction-write-01}{TW}{1}\\
	\queryRefCard{transaction-write-02}{TW}{2}\\
	\queryRefCard{transaction-write-03}{TW}{3}\\
	\queryRefCard{transaction-write-04}{TW}{4}\\
	\queryRefCard{transaction-write-05}{TW}{5}\\
	\queryRefCard{transaction-write-06}{TW}{6}\\
	\queryRefCard{transaction-write-07}{TW}{7}\\
	\queryRefCard{transaction-write-08}{TW}{8}\\
	\queryRefCard{transaction-write-09}{TW}{9}\\
	\queryRefCard{transaction-write-10}{TW}{10}\\
	\queryRefCard{transaction-write-11}{TW}{11}\\
	\queryRefCard{transaction-write-12}{TW}{12}\\
	\queryRefCard{transaction-write-13}{TW}{13}\\
	\queryRefCard{transaction-write-14}{TW}{14}\\
	\queryRefCard{transaction-write-15}{TW}{15}\\
	\queryRefCard{transaction-write-16}{TW}{16}\\
	\queryRefCard{transaction-write-17}{TW}{17}\\
	\queryRefCard{transaction-write-18}{TW}{18}\\
	\queryRefCard{transaction-write-19}{TW}{19}\\
}

\noindent\begin{tabularx}{\queryCardWidth}{|>{\queryPropertyCell}p{\queryPropertyCellWidth}|X|}
	\hline
	query & Transaction / write / 4 \\ \hline
	title & Add an Account owned by Person \\ \hline

		pattern & \centering \includegraphics[scale=\yedscale,margin=0cm .2cm]{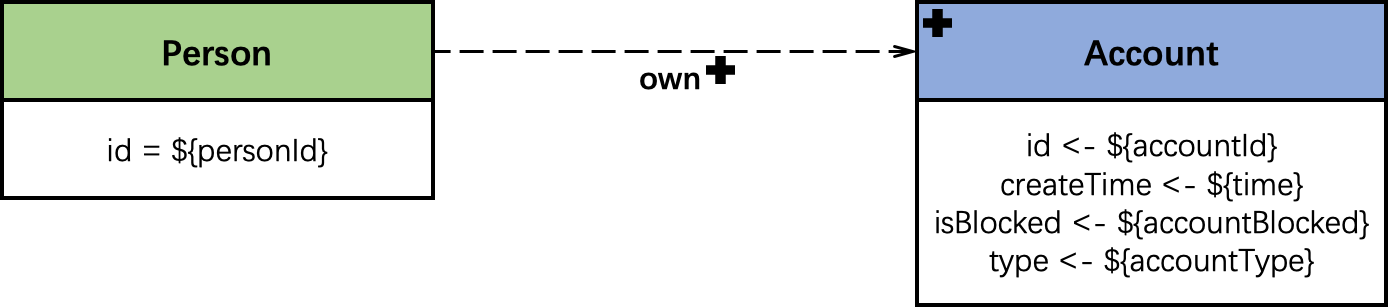} \tabularnewline \hline

	desc. & Add an \emph{Account} and an \emph{own} edge from \emph{Person} to the
\emph{Account}. \\ \hline

		params &
		\innerCardVSpace{\begin{tabularx}{\attributeCardWidth}{|>{\paramNumberCell}C{\attributeNumberWidth}|>{\varNameCell}M|>{\typeCell}m{\typeWidth}|Y|} \hline
		$\mathsf{1}$ & \$personId & ID &  \\ \hline
		$\mathsf{2}$ & \$accountId & ID &  \\ \hline
		$\mathsf{3}$ & \$time & DateTime &  \\ \hline
		$\mathsf{4}$ & \$accountBlocked & Boolean &  \\ \hline
		$\mathsf{5}$ & \$accountType & String &  \\ \hline
		\end{tabularx}}\innerCardVSpace \\ \hline
	
%
	
%
	%
	%
	%
	%
\end{tabularx}
\queryCardVSpace

\let\emph\oldemph
\let\textbf\oldtextbf

\renewcommand{\currentQueryCard}{0}
\renewcommand*{\arraystretch}{1.1}

\subsection*{Transaction / write / 5}
\label{sec:transaction-write-05}

\let\oldemph\emph
\renewcommand{\emph}[1]{{\footnotesize \sf #1}}
\let\oldtextbf\textbf
\renewcommand{\textbf}[1]{{\it #1}}\renewcommand{\currentQueryCard}{5}
\marginpar{
	\raggedleft
	\scriptsize
	
	\queryRefCard{transaction-write-01}{TW}{1}\\
	\queryRefCard{transaction-write-02}{TW}{2}\\
	\queryRefCard{transaction-write-03}{TW}{3}\\
	\queryRefCard{transaction-write-04}{TW}{4}\\
	\queryRefCard{transaction-write-05}{TW}{5}\\
	\queryRefCard{transaction-write-06}{TW}{6}\\
	\queryRefCard{transaction-write-07}{TW}{7}\\
	\queryRefCard{transaction-write-08}{TW}{8}\\
	\queryRefCard{transaction-write-09}{TW}{9}\\
	\queryRefCard{transaction-write-10}{TW}{10}\\
	\queryRefCard{transaction-write-11}{TW}{11}\\
	\queryRefCard{transaction-write-12}{TW}{12}\\
	\queryRefCard{transaction-write-13}{TW}{13}\\
	\queryRefCard{transaction-write-14}{TW}{14}\\
	\queryRefCard{transaction-write-15}{TW}{15}\\
	\queryRefCard{transaction-write-16}{TW}{16}\\
	\queryRefCard{transaction-write-17}{TW}{17}\\
	\queryRefCard{transaction-write-18}{TW}{18}\\
	\queryRefCard{transaction-write-19}{TW}{19}\\
}

\noindent\begin{tabularx}{\queryCardWidth}{|>{\queryPropertyCell}p{\queryPropertyCellWidth}|X|}
	\hline
	query & Transaction / write / 5 \\ \hline
	title & Add an Account owned by Company \\ \hline

		pattern & \centering \includegraphics[scale=\yedscale,margin=0cm .2cm]{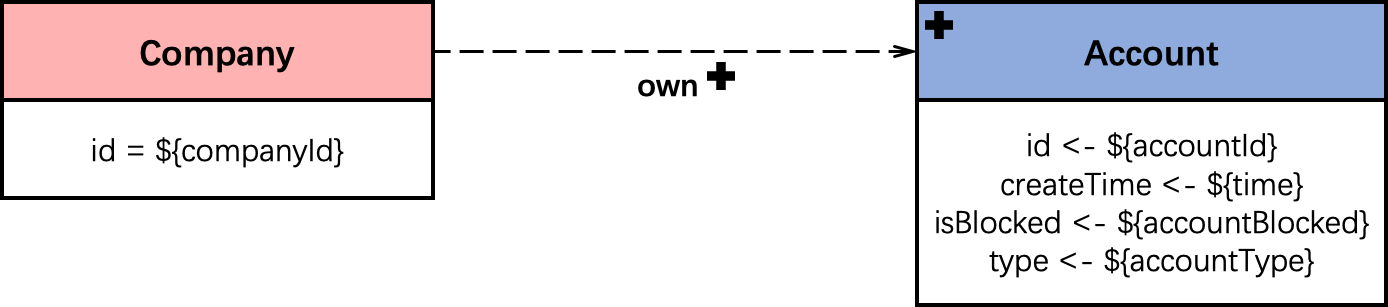} \tabularnewline \hline

	desc. & Add an \emph{Account} and an \emph{own} edge from \emph{Company} to the
\emph{Account}. \\ \hline

		params &
		\innerCardVSpace{\begin{tabularx}{\attributeCardWidth}{|>{\paramNumberCell}C{\attributeNumberWidth}|>{\varNameCell}M|>{\typeCell}m{\typeWidth}|Y|} \hline
		$\mathsf{1}$ & \$companyId & ID &  \\ \hline
		$\mathsf{2}$ & \$accountId & ID &  \\ \hline
		$\mathsf{3}$ & \$time & DateTime &  \\ \hline
		$\mathsf{4}$ & \$accountBlocked & Boolean &  \\ \hline
		$\mathsf{5}$ & \$accountType & String &  \\ \hline
		\end{tabularx}}\innerCardVSpace \\ \hline
	
%
	
%
	%
	%
	%
	%
\end{tabularx}
\queryCardVSpace

\let\emph\oldemph
\let\textbf\oldtextbf

\renewcommand{\currentQueryCard}{0}
\renewcommand*{\arraystretch}{1.1}

\subsection*{Transaction / write / 6}
\label{sec:transaction-write-06}

\let\oldemph\emph
\renewcommand{\emph}[1]{{\footnotesize \sf #1}}
\let\oldtextbf\textbf
\renewcommand{\textbf}[1]{{\it #1}}\renewcommand{\currentQueryCard}{6}
\marginpar{
	\raggedleft
	\scriptsize
	
	\queryRefCard{transaction-write-01}{TW}{1}\\
	\queryRefCard{transaction-write-02}{TW}{2}\\
	\queryRefCard{transaction-write-03}{TW}{3}\\
	\queryRefCard{transaction-write-04}{TW}{4}\\
	\queryRefCard{transaction-write-05}{TW}{5}\\
	\queryRefCard{transaction-write-06}{TW}{6}\\
	\queryRefCard{transaction-write-07}{TW}{7}\\
	\queryRefCard{transaction-write-08}{TW}{8}\\
	\queryRefCard{transaction-write-09}{TW}{9}\\
	\queryRefCard{transaction-write-10}{TW}{10}\\
	\queryRefCard{transaction-write-11}{TW}{11}\\
	\queryRefCard{transaction-write-12}{TW}{12}\\
	\queryRefCard{transaction-write-13}{TW}{13}\\
	\queryRefCard{transaction-write-14}{TW}{14}\\
	\queryRefCard{transaction-write-15}{TW}{15}\\
	\queryRefCard{transaction-write-16}{TW}{16}\\
	\queryRefCard{transaction-write-17}{TW}{17}\\
	\queryRefCard{transaction-write-18}{TW}{18}\\
	\queryRefCard{transaction-write-19}{TW}{19}\\
}

\noindent\begin{tabularx}{\queryCardWidth}{|>{\queryPropertyCell}p{\queryPropertyCellWidth}|X|}
	\hline
	query & Transaction / write / 6 \\ \hline
	title & Add Loan applied by Person \\ \hline

		pattern & \centering \includegraphics[scale=\yedscale,margin=0cm .2cm]{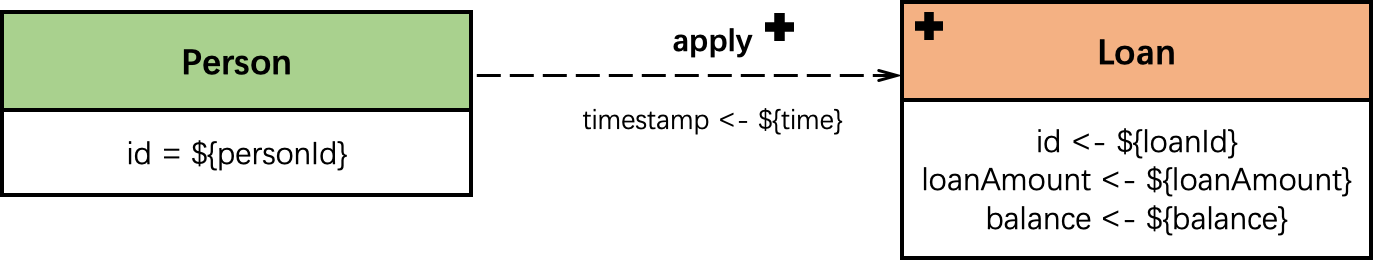} \tabularnewline \hline

	desc. & Add a \emph{Loan} and add an \emph{apply} edge from \emph{Person} to
\emph{Loan}. \\ \hline

		params &
		\innerCardVSpace{\begin{tabularx}{\attributeCardWidth}{|>{\paramNumberCell}C{\attributeNumberWidth}|>{\varNameCell}M|>{\typeCell}m{\typeWidth}|Y|} \hline
		$\mathsf{1}$ & \$personId & ID &  \\ \hline
		$\mathsf{2}$ & \$loanId & ID &  \\ \hline
		$\mathsf{3}$ & \$loanAmount & 64-bit Float &  \\ \hline
		$\mathsf{4}$ & \$balance & 64-bit Float &  \\ \hline
		$\mathsf{5}$ & \$time & DateTime &  \\ \hline
		\end{tabularx}}\innerCardVSpace \\ \hline
	
%
	
%
	%
	%
	%
	%
\end{tabularx}
\queryCardVSpace

\let\emph\oldemph
\let\textbf\oldtextbf

\renewcommand{\currentQueryCard}{0}
\renewcommand*{\arraystretch}{1.1}

\subsection*{Transaction / write / 7}
\label{sec:transaction-write-07}

\let\oldemph\emph
\renewcommand{\emph}[1]{{\footnotesize \sf #1}}
\let\oldtextbf\textbf
\renewcommand{\textbf}[1]{{\it #1}}\renewcommand{\currentQueryCard}{7}
\marginpar{
	\raggedleft
	\scriptsize
	
	\queryRefCard{transaction-write-01}{TW}{1}\\
	\queryRefCard{transaction-write-02}{TW}{2}\\
	\queryRefCard{transaction-write-03}{TW}{3}\\
	\queryRefCard{transaction-write-04}{TW}{4}\\
	\queryRefCard{transaction-write-05}{TW}{5}\\
	\queryRefCard{transaction-write-06}{TW}{6}\\
	\queryRefCard{transaction-write-07}{TW}{7}\\
	\queryRefCard{transaction-write-08}{TW}{8}\\
	\queryRefCard{transaction-write-09}{TW}{9}\\
	\queryRefCard{transaction-write-10}{TW}{10}\\
	\queryRefCard{transaction-write-11}{TW}{11}\\
	\queryRefCard{transaction-write-12}{TW}{12}\\
	\queryRefCard{transaction-write-13}{TW}{13}\\
	\queryRefCard{transaction-write-14}{TW}{14}\\
	\queryRefCard{transaction-write-15}{TW}{15}\\
	\queryRefCard{transaction-write-16}{TW}{16}\\
	\queryRefCard{transaction-write-17}{TW}{17}\\
	\queryRefCard{transaction-write-18}{TW}{18}\\
	\queryRefCard{transaction-write-19}{TW}{19}\\
}

\noindent\begin{tabularx}{\queryCardWidth}{|>{\queryPropertyCell}p{\queryPropertyCellWidth}|X|}
	\hline
	query & Transaction / write / 7 \\ \hline
	title & Add Loan applied by Company \\ \hline

		pattern & \centering \includegraphics[scale=\yedscale,margin=0cm .2cm]{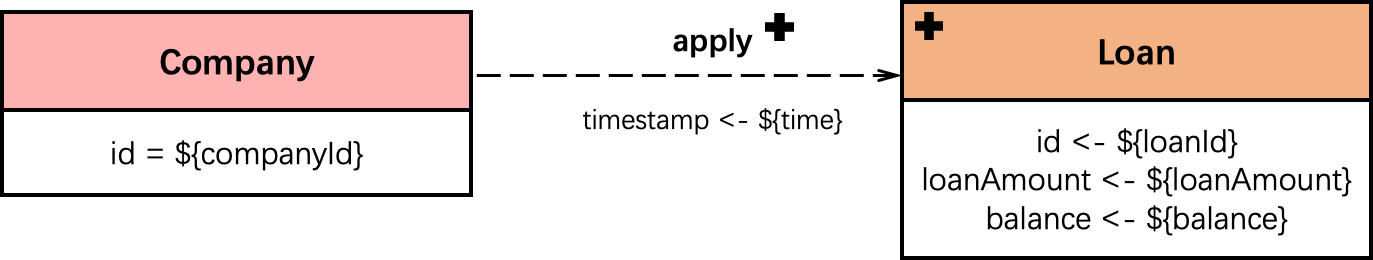} \tabularnewline \hline

	desc. & Add a \emph{Loan} and add an \emph{apply} edge from \emph{Company} to
\emph{Loan}. \\ \hline

		params &
		\innerCardVSpace{\begin{tabularx}{\attributeCardWidth}{|>{\paramNumberCell}C{\attributeNumberWidth}|>{\varNameCell}M|>{\typeCell}m{\typeWidth}|Y|} \hline
		$\mathsf{1}$ & \$companyId & ID &  \\ \hline
		$\mathsf{2}$ & \$loanId & ID &  \\ \hline
		$\mathsf{3}$ & \$loanAmount & 64-bit Float &  \\ \hline
		$\mathsf{4}$ & \$balance & 64-bit Float &  \\ \hline
		$\mathsf{5}$ & \$time & DateTime &  \\ \hline
		\end{tabularx}}\innerCardVSpace \\ \hline
	
%
	
%
	%
	%
	%
	%
\end{tabularx}
\queryCardVSpace

\let\emph\oldemph
\let\textbf\oldtextbf

\renewcommand{\currentQueryCard}{0}
\renewcommand*{\arraystretch}{1.1}

\subsection*{Transaction / write / 8}
\label{sec:transaction-write-08}

\let\oldemph\emph
\renewcommand{\emph}[1]{{\footnotesize \sf #1}}
\let\oldtextbf\textbf
\renewcommand{\textbf}[1]{{\it #1}}\renewcommand{\currentQueryCard}{8}
\marginpar{
	\raggedleft
	\scriptsize
	
	\queryRefCard{transaction-write-01}{TW}{1}\\
	\queryRefCard{transaction-write-02}{TW}{2}\\
	\queryRefCard{transaction-write-03}{TW}{3}\\
	\queryRefCard{transaction-write-04}{TW}{4}\\
	\queryRefCard{transaction-write-05}{TW}{5}\\
	\queryRefCard{transaction-write-06}{TW}{6}\\
	\queryRefCard{transaction-write-07}{TW}{7}\\
	\queryRefCard{transaction-write-08}{TW}{8}\\
	\queryRefCard{transaction-write-09}{TW}{9}\\
	\queryRefCard{transaction-write-10}{TW}{10}\\
	\queryRefCard{transaction-write-11}{TW}{11}\\
	\queryRefCard{transaction-write-12}{TW}{12}\\
	\queryRefCard{transaction-write-13}{TW}{13}\\
	\queryRefCard{transaction-write-14}{TW}{14}\\
	\queryRefCard{transaction-write-15}{TW}{15}\\
	\queryRefCard{transaction-write-16}{TW}{16}\\
	\queryRefCard{transaction-write-17}{TW}{17}\\
	\queryRefCard{transaction-write-18}{TW}{18}\\
	\queryRefCard{transaction-write-19}{TW}{19}\\
}

\noindent\begin{tabularx}{\queryCardWidth}{|>{\queryPropertyCell}p{\queryPropertyCellWidth}|X|}
	\hline
	query & Transaction / write / 8 \\ \hline
	title & Add invest between Person and Company \\ \hline

		pattern & \centering \includegraphics[scale=\yedscale,margin=0cm .2cm]{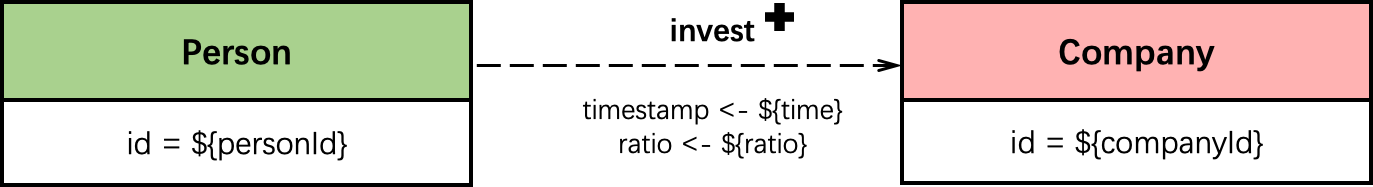} \tabularnewline \hline

	desc. & Add an \emph{invest} edge from a \emph{Person} to a \emph{Company}. \\ \hline

		params &
		\innerCardVSpace{\begin{tabularx}{\attributeCardWidth}{|>{\paramNumberCell}C{\attributeNumberWidth}|>{\varNameCell}M|>{\typeCell}m{\typeWidth}|Y|} \hline
		$\mathsf{1}$ & \$personId & ID &  \\ \hline
		$\mathsf{2}$ & \$companyId & ID &  \\ \hline
		$\mathsf{3}$ & \$time & DateTime &  \\ \hline
		$\mathsf{4}$ & \$ratio & 64-bit Float &  \\ \hline
		\end{tabularx}}\innerCardVSpace \\ \hline
	
%
	
%
	%
	%
	%
	%
\end{tabularx}
\queryCardVSpace

\let\emph\oldemph
\let\textbf\oldtextbf

\renewcommand{\currentQueryCard}{0}
\renewcommand*{\arraystretch}{1.1}

\subsection*{Transaction / write / 9}
\label{sec:transaction-write-09}

\let\oldemph\emph
\renewcommand{\emph}[1]{{\footnotesize \sf #1}}
\let\oldtextbf\textbf
\renewcommand{\textbf}[1]{{\it #1}}\renewcommand{\currentQueryCard}{9}
\marginpar{
	\raggedleft
	\scriptsize
	
	\queryRefCard{transaction-write-01}{TW}{1}\\
	\queryRefCard{transaction-write-02}{TW}{2}\\
	\queryRefCard{transaction-write-03}{TW}{3}\\
	\queryRefCard{transaction-write-04}{TW}{4}\\
	\queryRefCard{transaction-write-05}{TW}{5}\\
	\queryRefCard{transaction-write-06}{TW}{6}\\
	\queryRefCard{transaction-write-07}{TW}{7}\\
	\queryRefCard{transaction-write-08}{TW}{8}\\
	\queryRefCard{transaction-write-09}{TW}{9}\\
	\queryRefCard{transaction-write-10}{TW}{10}\\
	\queryRefCard{transaction-write-11}{TW}{11}\\
	\queryRefCard{transaction-write-12}{TW}{12}\\
	\queryRefCard{transaction-write-13}{TW}{13}\\
	\queryRefCard{transaction-write-14}{TW}{14}\\
	\queryRefCard{transaction-write-15}{TW}{15}\\
	\queryRefCard{transaction-write-16}{TW}{16}\\
	\queryRefCard{transaction-write-17}{TW}{17}\\
	\queryRefCard{transaction-write-18}{TW}{18}\\
	\queryRefCard{transaction-write-19}{TW}{19}\\
}

\noindent\begin{tabularx}{\queryCardWidth}{|>{\queryPropertyCell}p{\queryPropertyCellWidth}|X|}
	\hline
	query & Transaction / write / 9 \\ \hline
	title & Add invest between Company and Company \\ \hline

		pattern & \centering \includegraphics[scale=\yedscale,margin=0cm .2cm]{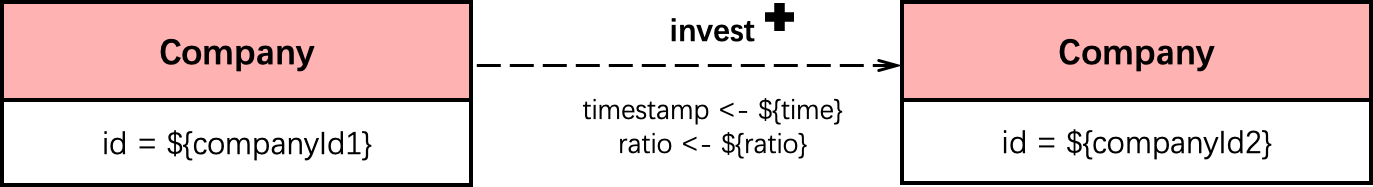} \tabularnewline \hline

	desc. & Add an \emph{invest} edge from a \emph{Company} to a \emph{Company}. \\ \hline

		params &
		\innerCardVSpace{\begin{tabularx}{\attributeCardWidth}{|>{\paramNumberCell}C{\attributeNumberWidth}|>{\varNameCell}M|>{\typeCell}m{\typeWidth}|Y|} \hline
		$\mathsf{1}$ & \$companyId1 & ID &  \\ \hline
		$\mathsf{2}$ & \$companyId2 & ID &  \\ \hline
		$\mathsf{3}$ & \$time & DateTime &  \\ \hline
		$\mathsf{4}$ & \$ratio & 64-bit Float &  \\ \hline
		\end{tabularx}}\innerCardVSpace \\ \hline
	
%
	
%
	%
	%
	%
	%
\end{tabularx}
\queryCardVSpace

\let\emph\oldemph
\let\textbf\oldtextbf

\renewcommand{\currentQueryCard}{0}
\renewcommand*{\arraystretch}{1.1}

\subsection*{Transaction / write / 10}
\label{sec:transaction-write-10}

\let\oldemph\emph
\renewcommand{\emph}[1]{{\footnotesize \sf #1}}
\let\oldtextbf\textbf
\renewcommand{\textbf}[1]{{\it #1}}\renewcommand{\currentQueryCard}{10}
\marginpar{
	\raggedleft
	\scriptsize
	
	\queryRefCard{transaction-write-01}{TW}{1}\\
	\queryRefCard{transaction-write-02}{TW}{2}\\
	\queryRefCard{transaction-write-03}{TW}{3}\\
	\queryRefCard{transaction-write-04}{TW}{4}\\
	\queryRefCard{transaction-write-05}{TW}{5}\\
	\queryRefCard{transaction-write-06}{TW}{6}\\
	\queryRefCard{transaction-write-07}{TW}{7}\\
	\queryRefCard{transaction-write-08}{TW}{8}\\
	\queryRefCard{transaction-write-09}{TW}{9}\\
	\queryRefCard{transaction-write-10}{TW}{10}\\
	\queryRefCard{transaction-write-11}{TW}{11}\\
	\queryRefCard{transaction-write-12}{TW}{12}\\
	\queryRefCard{transaction-write-13}{TW}{13}\\
	\queryRefCard{transaction-write-14}{TW}{14}\\
	\queryRefCard{transaction-write-15}{TW}{15}\\
	\queryRefCard{transaction-write-16}{TW}{16}\\
	\queryRefCard{transaction-write-17}{TW}{17}\\
	\queryRefCard{transaction-write-18}{TW}{18}\\
	\queryRefCard{transaction-write-19}{TW}{19}\\
}

\noindent\begin{tabularx}{\queryCardWidth}{|>{\queryPropertyCell}p{\queryPropertyCellWidth}|X|}
	\hline
	query & Transaction / write / 10 \\ \hline
	title & Add guarantee between Persons \\ \hline

		pattern & \centering \includegraphics[scale=\yedscale,margin=0cm .2cm]{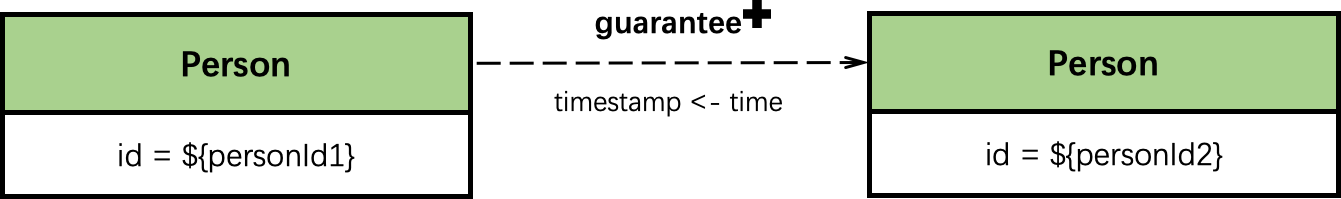} \tabularnewline \hline

	desc. & Add a \emph{guarantee} edge from a \emph{Person} to another
\emph{Person}. \\ \hline

		params &
		\innerCardVSpace{\begin{tabularx}{\attributeCardWidth}{|>{\paramNumberCell}C{\attributeNumberWidth}|>{\varNameCell}M|>{\typeCell}m{\typeWidth}|Y|} \hline
		$\mathsf{1}$ & \$personId1 & ID &  \\ \hline
		$\mathsf{2}$ & \$personId2 & ID &  \\ \hline
		$\mathsf{3}$ & \$time & DateTime &  \\ \hline
		\end{tabularx}}\innerCardVSpace \\ \hline
	
%
	
%
	%
	%
	%
	%
\end{tabularx}
\queryCardVSpace

\let\emph\oldemph
\let\textbf\oldtextbf

\renewcommand{\currentQueryCard}{0}
\renewcommand*{\arraystretch}{1.1}

\subsection*{Transaction / write / 11}
\label{sec:transaction-write-11}

\let\oldemph\emph
\renewcommand{\emph}[1]{{\footnotesize \sf #1}}
\let\oldtextbf\textbf
\renewcommand{\textbf}[1]{{\it #1}}\renewcommand{\currentQueryCard}{11}
\marginpar{
	\raggedleft
	\scriptsize
	
	\queryRefCard{transaction-write-01}{TW}{1}\\
	\queryRefCard{transaction-write-02}{TW}{2}\\
	\queryRefCard{transaction-write-03}{TW}{3}\\
	\queryRefCard{transaction-write-04}{TW}{4}\\
	\queryRefCard{transaction-write-05}{TW}{5}\\
	\queryRefCard{transaction-write-06}{TW}{6}\\
	\queryRefCard{transaction-write-07}{TW}{7}\\
	\queryRefCard{transaction-write-08}{TW}{8}\\
	\queryRefCard{transaction-write-09}{TW}{9}\\
	\queryRefCard{transaction-write-10}{TW}{10}\\
	\queryRefCard{transaction-write-11}{TW}{11}\\
	\queryRefCard{transaction-write-12}{TW}{12}\\
	\queryRefCard{transaction-write-13}{TW}{13}\\
	\queryRefCard{transaction-write-14}{TW}{14}\\
	\queryRefCard{transaction-write-15}{TW}{15}\\
	\queryRefCard{transaction-write-16}{TW}{16}\\
	\queryRefCard{transaction-write-17}{TW}{17}\\
	\queryRefCard{transaction-write-18}{TW}{18}\\
	\queryRefCard{transaction-write-19}{TW}{19}\\
}

\noindent\begin{tabularx}{\queryCardWidth}{|>{\queryPropertyCell}p{\queryPropertyCellWidth}|X|}
	\hline
	query & Transaction / write / 11 \\ \hline
	title & Add guarantee between Companies \\ \hline

		pattern & \centering \includegraphics[scale=\yedscale,margin=0cm .2cm]{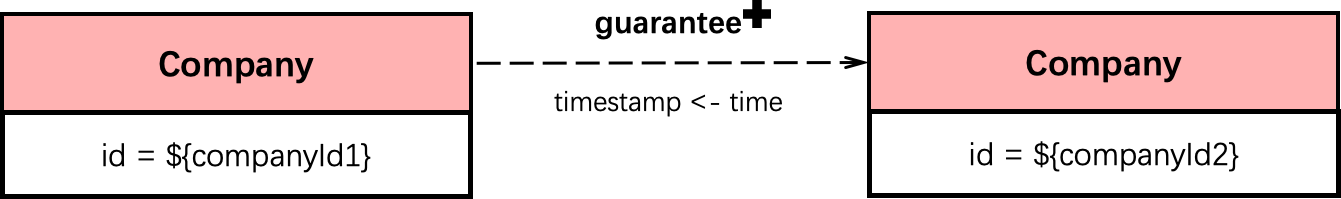} \tabularnewline \hline

	desc. & Add a \emph{guarantee} edge from a \emph{Company} to another
\emph{Company}. \\ \hline

		params &
		\innerCardVSpace{\begin{tabularx}{\attributeCardWidth}{|>{\paramNumberCell}C{\attributeNumberWidth}|>{\varNameCell}M|>{\typeCell}m{\typeWidth}|Y|} \hline
		$\mathsf{1}$ & \$companyId1 & ID &  \\ \hline
		$\mathsf{2}$ & \$companyId2 & ID &  \\ \hline
		$\mathsf{3}$ & \$time & DateTime &  \\ \hline
		\end{tabularx}}\innerCardVSpace \\ \hline
	
%
	
%
	%
	%
	%
	%
\end{tabularx}
\queryCardVSpace

\let\emph\oldemph
\let\textbf\oldtextbf

\renewcommand{\currentQueryCard}{0}
\renewcommand*{\arraystretch}{1.1}

\subsection*{Transaction / write / 12}
\label{sec:transaction-write-12}

\let\oldemph\emph
\renewcommand{\emph}[1]{{\footnotesize \sf #1}}
\let\oldtextbf\textbf
\renewcommand{\textbf}[1]{{\it #1}}\renewcommand{\currentQueryCard}{12}
\marginpar{
	\raggedleft
	\scriptsize
	
	\queryRefCard{transaction-write-01}{TW}{1}\\
	\queryRefCard{transaction-write-02}{TW}{2}\\
	\queryRefCard{transaction-write-03}{TW}{3}\\
	\queryRefCard{transaction-write-04}{TW}{4}\\
	\queryRefCard{transaction-write-05}{TW}{5}\\
	\queryRefCard{transaction-write-06}{TW}{6}\\
	\queryRefCard{transaction-write-07}{TW}{7}\\
	\queryRefCard{transaction-write-08}{TW}{8}\\
	\queryRefCard{transaction-write-09}{TW}{9}\\
	\queryRefCard{transaction-write-10}{TW}{10}\\
	\queryRefCard{transaction-write-11}{TW}{11}\\
	\queryRefCard{transaction-write-12}{TW}{12}\\
	\queryRefCard{transaction-write-13}{TW}{13}\\
	\queryRefCard{transaction-write-14}{TW}{14}\\
	\queryRefCard{transaction-write-15}{TW}{15}\\
	\queryRefCard{transaction-write-16}{TW}{16}\\
	\queryRefCard{transaction-write-17}{TW}{17}\\
	\queryRefCard{transaction-write-18}{TW}{18}\\
	\queryRefCard{transaction-write-19}{TW}{19}\\
}

\noindent\begin{tabularx}{\queryCardWidth}{|>{\queryPropertyCell}p{\queryPropertyCellWidth}|X|}
	\hline
	query & Transaction / write / 12 \\ \hline
	title & Add transfer between Accounts \\ \hline

		pattern & \centering \includegraphics[scale=\yedscale,margin=0cm .2cm]{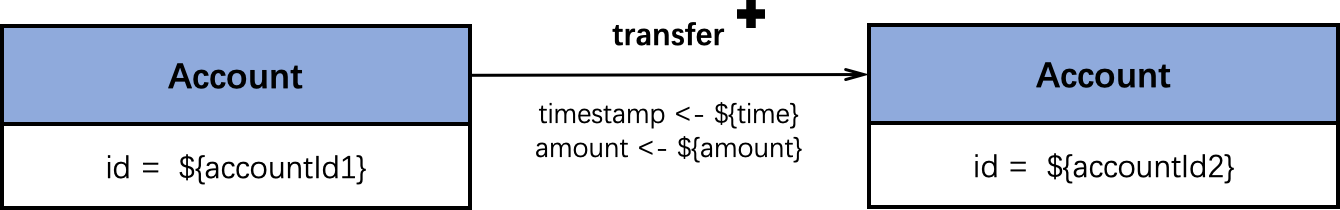} \tabularnewline \hline

	desc. & Add a \emph{transfer} edge from an \emph{Account} to another
\emph{Account}. \\ \hline

		params &
		\innerCardVSpace{\begin{tabularx}{\attributeCardWidth}{|>{\paramNumberCell}C{\attributeNumberWidth}|>{\varNameCell}M|>{\typeCell}m{\typeWidth}|Y|} \hline
		$\mathsf{1}$ & \$accountId1 & ID &  \\ \hline
		$\mathsf{2}$ & \$accountId2 & ID &  \\ \hline
		$\mathsf{3}$ & \$time & DateTime &  \\ \hline
		$\mathsf{4}$ & \$amount & 64-bit Float &  \\ \hline
		\end{tabularx}}\innerCardVSpace \\ \hline
	
%
	
%
	%
	%
	%
	%
\end{tabularx}
\queryCardVSpace

\let\emph\oldemph
\let\textbf\oldtextbf

\renewcommand{\currentQueryCard}{0}
\renewcommand*{\arraystretch}{1.1}

\subsection*{Transaction / write / 13}
\label{sec:transaction-write-13}

\let\oldemph\emph
\renewcommand{\emph}[1]{{\footnotesize \sf #1}}
\let\oldtextbf\textbf
\renewcommand{\textbf}[1]{{\it #1}}\renewcommand{\currentQueryCard}{13}
\marginpar{
	\raggedleft
	\scriptsize
	
	\queryRefCard{transaction-write-01}{TW}{1}\\
	\queryRefCard{transaction-write-02}{TW}{2}\\
	\queryRefCard{transaction-write-03}{TW}{3}\\
	\queryRefCard{transaction-write-04}{TW}{4}\\
	\queryRefCard{transaction-write-05}{TW}{5}\\
	\queryRefCard{transaction-write-06}{TW}{6}\\
	\queryRefCard{transaction-write-07}{TW}{7}\\
	\queryRefCard{transaction-write-08}{TW}{8}\\
	\queryRefCard{transaction-write-09}{TW}{9}\\
	\queryRefCard{transaction-write-10}{TW}{10}\\
	\queryRefCard{transaction-write-11}{TW}{11}\\
	\queryRefCard{transaction-write-12}{TW}{12}\\
	\queryRefCard{transaction-write-13}{TW}{13}\\
	\queryRefCard{transaction-write-14}{TW}{14}\\
	\queryRefCard{transaction-write-15}{TW}{15}\\
	\queryRefCard{transaction-write-16}{TW}{16}\\
	\queryRefCard{transaction-write-17}{TW}{17}\\
	\queryRefCard{transaction-write-18}{TW}{18}\\
	\queryRefCard{transaction-write-19}{TW}{19}\\
}

\noindent\begin{tabularx}{\queryCardWidth}{|>{\queryPropertyCell}p{\queryPropertyCellWidth}|X|}
	\hline
	query & Transaction / write / 13 \\ \hline
	title & Add withdraw between Accounts \\ \hline

		pattern & \centering \includegraphics[scale=\yedscale,margin=0cm .2cm]{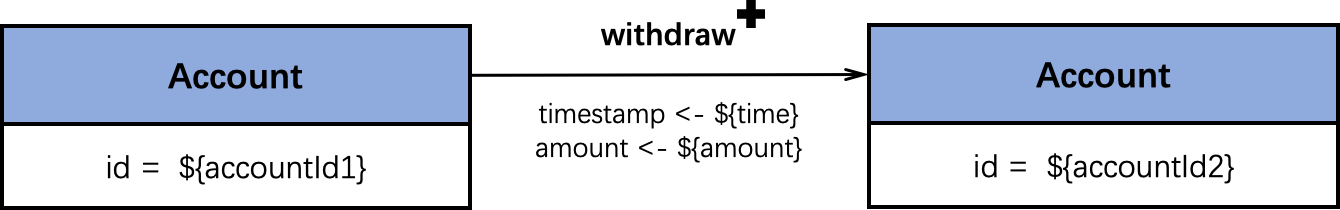} \tabularnewline \hline

	desc. & Add a \emph{withdraw} edge from an \emph{Account} to another
\emph{Account}. \\ \hline

		params &
		\innerCardVSpace{\begin{tabularx}{\attributeCardWidth}{|>{\paramNumberCell}C{\attributeNumberWidth}|>{\varNameCell}M|>{\typeCell}m{\typeWidth}|Y|} \hline
		$\mathsf{1}$ & \$accountId1 & ID &  \\ \hline
		$\mathsf{2}$ & \$accountId2 & ID &  \\ \hline
		$\mathsf{3}$ & \$time & DateTime &  \\ \hline
		$\mathsf{4}$ & \$amount & 64-bit Float &  \\ \hline
		\end{tabularx}}\innerCardVSpace \\ \hline
	
%
	
%
	%
	%
	%
	%
\end{tabularx}
\queryCardVSpace

\let\emph\oldemph
\let\textbf\oldtextbf

\renewcommand{\currentQueryCard}{0}
\renewcommand*{\arraystretch}{1.1}

\subsection*{Transaction / write / 14}
\label{sec:transaction-write-14}

\let\oldemph\emph
\renewcommand{\emph}[1]{{\footnotesize \sf #1}}
\let\oldtextbf\textbf
\renewcommand{\textbf}[1]{{\it #1}}\renewcommand{\currentQueryCard}{14}
\marginpar{
	\raggedleft
	\scriptsize
	
	\queryRefCard{transaction-write-01}{TW}{1}\\
	\queryRefCard{transaction-write-02}{TW}{2}\\
	\queryRefCard{transaction-write-03}{TW}{3}\\
	\queryRefCard{transaction-write-04}{TW}{4}\\
	\queryRefCard{transaction-write-05}{TW}{5}\\
	\queryRefCard{transaction-write-06}{TW}{6}\\
	\queryRefCard{transaction-write-07}{TW}{7}\\
	\queryRefCard{transaction-write-08}{TW}{8}\\
	\queryRefCard{transaction-write-09}{TW}{9}\\
	\queryRefCard{transaction-write-10}{TW}{10}\\
	\queryRefCard{transaction-write-11}{TW}{11}\\
	\queryRefCard{transaction-write-12}{TW}{12}\\
	\queryRefCard{transaction-write-13}{TW}{13}\\
	\queryRefCard{transaction-write-14}{TW}{14}\\
	\queryRefCard{transaction-write-15}{TW}{15}\\
	\queryRefCard{transaction-write-16}{TW}{16}\\
	\queryRefCard{transaction-write-17}{TW}{17}\\
	\queryRefCard{transaction-write-18}{TW}{18}\\
	\queryRefCard{transaction-write-19}{TW}{19}\\
}

\noindent\begin{tabularx}{\queryCardWidth}{|>{\queryPropertyCell}p{\queryPropertyCellWidth}|X|}
	\hline
	query & Transaction / write / 14 \\ \hline
	title & Add repay between Account and Loan \\ \hline

		pattern & \centering \includegraphics[scale=\yedscale,margin=0cm .2cm]{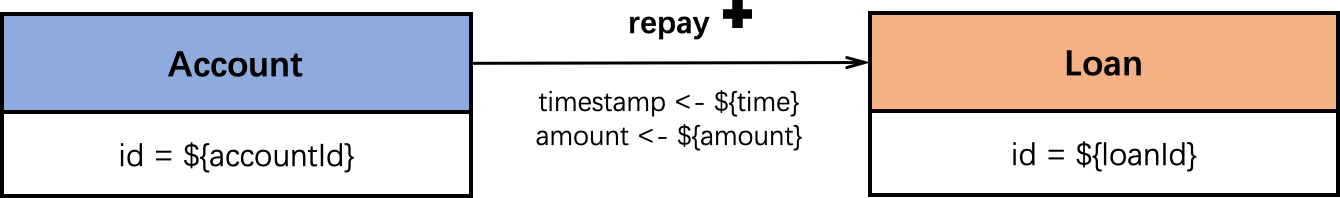} \tabularnewline \hline

	desc. & Add a \emph{repay} edge from an \emph{Account} to a \emph{Loan}. \\ \hline

		params &
		\innerCardVSpace{\begin{tabularx}{\attributeCardWidth}{|>{\paramNumberCell}C{\attributeNumberWidth}|>{\varNameCell}M|>{\typeCell}m{\typeWidth}|Y|} \hline
		$\mathsf{1}$ & \$accountId & ID &  \\ \hline
		$\mathsf{2}$ & \$loanId & ID &  \\ \hline
		$\mathsf{3}$ & \$time & DateTime &  \\ \hline
		$\mathsf{4}$ & \$amount & 64-bit Float &  \\ \hline
		\end{tabularx}}\innerCardVSpace \\ \hline
	
%
	
%
	%
	%
	%
	%
\end{tabularx}
\queryCardVSpace

\let\emph\oldemph
\let\textbf\oldtextbf

\renewcommand{\currentQueryCard}{0}
\renewcommand*{\arraystretch}{1.1}

\subsection*{Transaction / write / 15}
\label{sec:transaction-write-15}

\let\oldemph\emph
\renewcommand{\emph}[1]{{\footnotesize \sf #1}}
\let\oldtextbf\textbf
\renewcommand{\textbf}[1]{{\it #1}}\renewcommand{\currentQueryCard}{15}
\marginpar{
	\raggedleft
	\scriptsize
	
	\queryRefCard{transaction-write-01}{TW}{1}\\
	\queryRefCard{transaction-write-02}{TW}{2}\\
	\queryRefCard{transaction-write-03}{TW}{3}\\
	\queryRefCard{transaction-write-04}{TW}{4}\\
	\queryRefCard{transaction-write-05}{TW}{5}\\
	\queryRefCard{transaction-write-06}{TW}{6}\\
	\queryRefCard{transaction-write-07}{TW}{7}\\
	\queryRefCard{transaction-write-08}{TW}{8}\\
	\queryRefCard{transaction-write-09}{TW}{9}\\
	\queryRefCard{transaction-write-10}{TW}{10}\\
	\queryRefCard{transaction-write-11}{TW}{11}\\
	\queryRefCard{transaction-write-12}{TW}{12}\\
	\queryRefCard{transaction-write-13}{TW}{13}\\
	\queryRefCard{transaction-write-14}{TW}{14}\\
	\queryRefCard{transaction-write-15}{TW}{15}\\
	\queryRefCard{transaction-write-16}{TW}{16}\\
	\queryRefCard{transaction-write-17}{TW}{17}\\
	\queryRefCard{transaction-write-18}{TW}{18}\\
	\queryRefCard{transaction-write-19}{TW}{19}\\
}

\noindent\begin{tabularx}{\queryCardWidth}{|>{\queryPropertyCell}p{\queryPropertyCellWidth}|X|}
	\hline
	query & Transaction / write / 15 \\ \hline
	title & Add deposit between Loan and Account \\ \hline

		pattern & \centering \includegraphics[scale=\yedscale,margin=0cm .2cm]{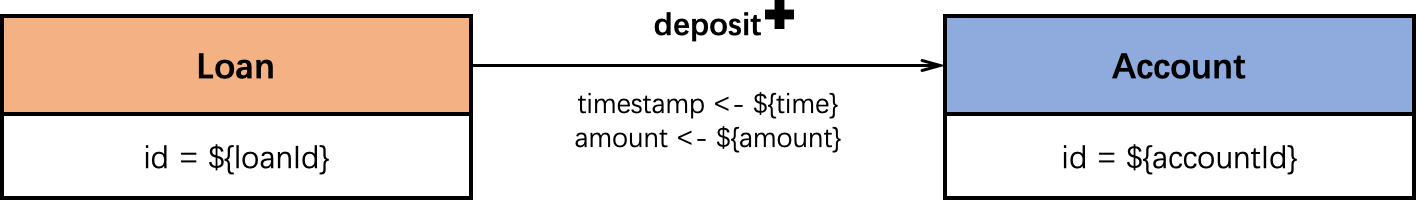} \tabularnewline \hline

	desc. & Add a \emph{deposit} edge from a \emph{Loan} to an \emph{Account}. \\ \hline

		params &
		\innerCardVSpace{\begin{tabularx}{\attributeCardWidth}{|>{\paramNumberCell}C{\attributeNumberWidth}|>{\varNameCell}M|>{\typeCell}m{\typeWidth}|Y|} \hline
		$\mathsf{1}$ & \$loanId & ID &  \\ \hline
		$\mathsf{2}$ & \$accountId & ID &  \\ \hline
		$\mathsf{3}$ & \$time & DateTime &  \\ \hline
		$\mathsf{4}$ & \$amount & 64-bit Float &  \\ \hline
		\end{tabularx}}\innerCardVSpace \\ \hline
	
%
	
%
	%
	%
	%
	%
\end{tabularx}
\queryCardVSpace

\let\emph\oldemph
\let\textbf\oldtextbf

\renewcommand{\currentQueryCard}{0}
\renewcommand*{\arraystretch}{1.1}

\subsection*{Transaction / write / 16}
\label{sec:transaction-write-16}

\let\oldemph\emph
\renewcommand{\emph}[1]{{\footnotesize \sf #1}}
\let\oldtextbf\textbf
\renewcommand{\textbf}[1]{{\it #1}}\renewcommand{\currentQueryCard}{16}
\marginpar{
	\raggedleft
	\scriptsize
	
	\queryRefCard{transaction-write-01}{TW}{1}\\
	\queryRefCard{transaction-write-02}{TW}{2}\\
	\queryRefCard{transaction-write-03}{TW}{3}\\
	\queryRefCard{transaction-write-04}{TW}{4}\\
	\queryRefCard{transaction-write-05}{TW}{5}\\
	\queryRefCard{transaction-write-06}{TW}{6}\\
	\queryRefCard{transaction-write-07}{TW}{7}\\
	\queryRefCard{transaction-write-08}{TW}{8}\\
	\queryRefCard{transaction-write-09}{TW}{9}\\
	\queryRefCard{transaction-write-10}{TW}{10}\\
	\queryRefCard{transaction-write-11}{TW}{11}\\
	\queryRefCard{transaction-write-12}{TW}{12}\\
	\queryRefCard{transaction-write-13}{TW}{13}\\
	\queryRefCard{transaction-write-14}{TW}{14}\\
	\queryRefCard{transaction-write-15}{TW}{15}\\
	\queryRefCard{transaction-write-16}{TW}{16}\\
	\queryRefCard{transaction-write-17}{TW}{17}\\
	\queryRefCard{transaction-write-18}{TW}{18}\\
	\queryRefCard{transaction-write-19}{TW}{19}\\
}

\noindent\begin{tabularx}{\queryCardWidth}{|>{\queryPropertyCell}p{\queryPropertyCellWidth}|X|}
	\hline
	query & Transaction / write / 16 \\ \hline
	title & Account signed in with Medium \\ \hline

		pattern & \centering \includegraphics[scale=\yedscale,margin=0cm .2cm]{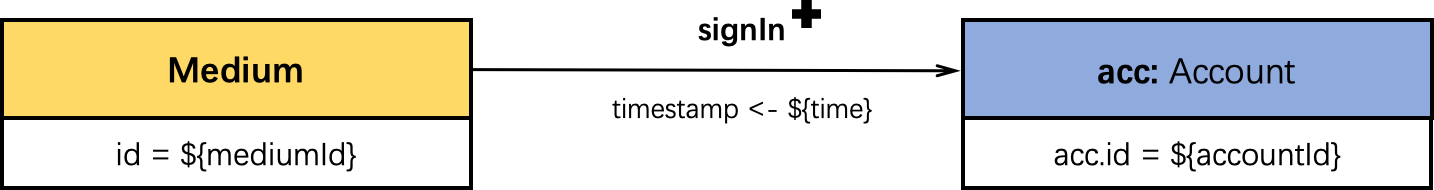} \tabularnewline \hline

	desc. & Add a \emph{signIn} edge from medium to an \emph{Account}. \\ \hline

		params &
		\innerCardVSpace{\begin{tabularx}{\attributeCardWidth}{|>{\paramNumberCell}C{\attributeNumberWidth}|>{\varNameCell}M|>{\typeCell}m{\typeWidth}|Y|} \hline
		$\mathsf{1}$ & \$mediumId & ID &  \\ \hline
		$\mathsf{2}$ & \$accountId & ID &  \\ \hline
		$\mathsf{3}$ & \$time & DateTime &  \\ \hline
		\end{tabularx}}\innerCardVSpace \\ \hline
	
%
	
%
	%
	%
	%
	%
\end{tabularx}
\queryCardVSpace

\let\emph\oldemph
\let\textbf\oldtextbf

\renewcommand{\currentQueryCard}{0}
\renewcommand*{\arraystretch}{1.1}

\subsection*{Transaction / write / 17}
\label{sec:transaction-write-17}

\let\oldemph\emph
\renewcommand{\emph}[1]{{\footnotesize \sf #1}}
\let\oldtextbf\textbf
\renewcommand{\textbf}[1]{{\it #1}}\renewcommand{\currentQueryCard}{17}
\marginpar{
	\raggedleft
	\scriptsize
	
	\queryRefCard{transaction-write-01}{TW}{1}\\
	\queryRefCard{transaction-write-02}{TW}{2}\\
	\queryRefCard{transaction-write-03}{TW}{3}\\
	\queryRefCard{transaction-write-04}{TW}{4}\\
	\queryRefCard{transaction-write-05}{TW}{5}\\
	\queryRefCard{transaction-write-06}{TW}{6}\\
	\queryRefCard{transaction-write-07}{TW}{7}\\
	\queryRefCard{transaction-write-08}{TW}{8}\\
	\queryRefCard{transaction-write-09}{TW}{9}\\
	\queryRefCard{transaction-write-10}{TW}{10}\\
	\queryRefCard{transaction-write-11}{TW}{11}\\
	\queryRefCard{transaction-write-12}{TW}{12}\\
	\queryRefCard{transaction-write-13}{TW}{13}\\
	\queryRefCard{transaction-write-14}{TW}{14}\\
	\queryRefCard{transaction-write-15}{TW}{15}\\
	\queryRefCard{transaction-write-16}{TW}{16}\\
	\queryRefCard{transaction-write-17}{TW}{17}\\
	\queryRefCard{transaction-write-18}{TW}{18}\\
	\queryRefCard{transaction-write-19}{TW}{19}\\
}

\noindent\begin{tabularx}{\queryCardWidth}{|>{\queryPropertyCell}p{\queryPropertyCellWidth}|X|}
	\hline
	query & Transaction / write / 17 \\ \hline
	title & Remove an Account \\ \hline

		pattern & \centering \includegraphics[scale=\yedscale,margin=0cm .2cm]{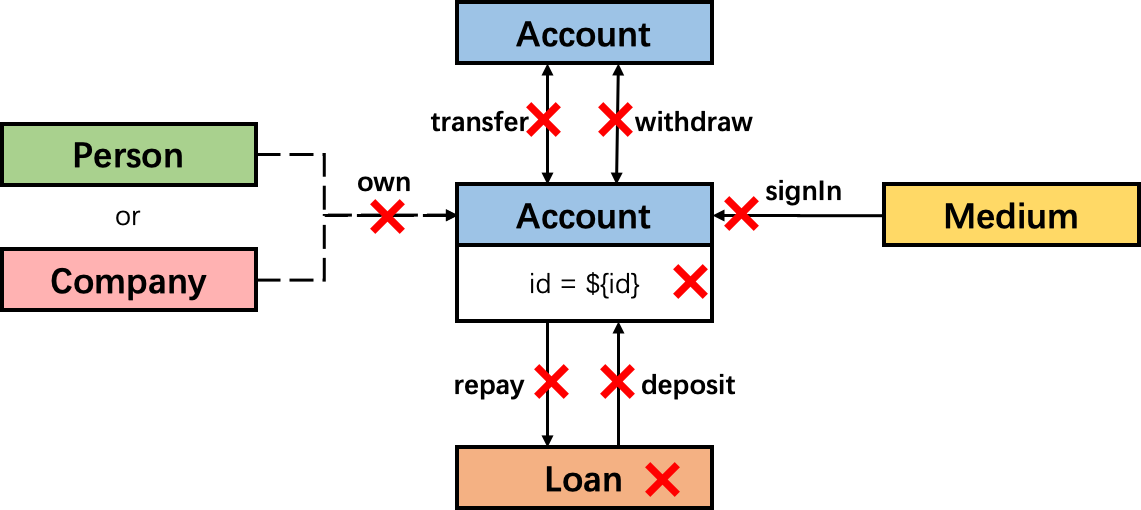} \tabularnewline \hline

	desc. & Given an id, remove the \emph{Account}, and remove the related edges
including \emph{own}, \emph{transfer}, \emph{withdraw}, \emph{repay},
\emph{deposit}, \emph{signIn}. Remove the connected \emph{Loan} vertex
in cascade. \\ \hline

		params &
		\innerCardVSpace{\begin{tabularx}{\attributeCardWidth}{|>{\paramNumberCell}C{\attributeNumberWidth}|>{\varNameCell}M|>{\typeCell}m{\typeWidth}|Y|} \hline
		$\mathsf{1}$ & \$accountId & ID &  \\ \hline
		\end{tabularx}}\innerCardVSpace \\ \hline
	
%
	
%
	%
	%
	%
	%
\end{tabularx}
\queryCardVSpace

\let\emph\oldemph
\let\textbf\oldtextbf

\renewcommand{\currentQueryCard}{0}
\renewcommand*{\arraystretch}{1.1}

\subsection*{Transaction / write / 18}
\label{sec:transaction-write-18}

\let\oldemph\emph
\renewcommand{\emph}[1]{{\footnotesize \sf #1}}
\let\oldtextbf\textbf
\renewcommand{\textbf}[1]{{\it #1}}\renewcommand{\currentQueryCard}{18}
\marginpar{
	\raggedleft
	\scriptsize
	
	\queryRefCard{transaction-write-01}{TW}{1}\\
	\queryRefCard{transaction-write-02}{TW}{2}\\
	\queryRefCard{transaction-write-03}{TW}{3}\\
	\queryRefCard{transaction-write-04}{TW}{4}\\
	\queryRefCard{transaction-write-05}{TW}{5}\\
	\queryRefCard{transaction-write-06}{TW}{6}\\
	\queryRefCard{transaction-write-07}{TW}{7}\\
	\queryRefCard{transaction-write-08}{TW}{8}\\
	\queryRefCard{transaction-write-09}{TW}{9}\\
	\queryRefCard{transaction-write-10}{TW}{10}\\
	\queryRefCard{transaction-write-11}{TW}{11}\\
	\queryRefCard{transaction-write-12}{TW}{12}\\
	\queryRefCard{transaction-write-13}{TW}{13}\\
	\queryRefCard{transaction-write-14}{TW}{14}\\
	\queryRefCard{transaction-write-15}{TW}{15}\\
	\queryRefCard{transaction-write-16}{TW}{16}\\
	\queryRefCard{transaction-write-17}{TW}{17}\\
	\queryRefCard{transaction-write-18}{TW}{18}\\
	\queryRefCard{transaction-write-19}{TW}{19}\\
}

\noindent\begin{tabularx}{\queryCardWidth}{|>{\queryPropertyCell}p{\queryPropertyCellWidth}|X|}
	\hline
	query & Transaction / write / 18 \\ \hline
	title & Block a Account of high risk \\ \hline

		pattern & \centering \includegraphics[scale=\yedscale,margin=0cm .2cm]{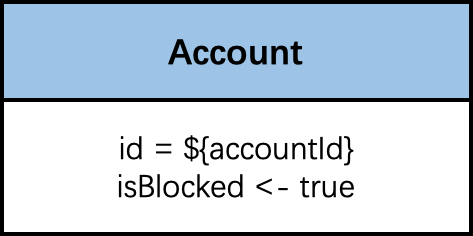} \tabularnewline \hline

	desc. & Set an \emph{Account}'s isBlocked to True. \\ \hline

		params &
		\innerCardVSpace{\begin{tabularx}{\attributeCardWidth}{|>{\paramNumberCell}C{\attributeNumberWidth}|>{\varNameCell}M|>{\typeCell}m{\typeWidth}|Y|} \hline
		$\mathsf{1}$ & \$accountId & ID &  \\ \hline
		\end{tabularx}}\innerCardVSpace \\ \hline
	
%
	
%
	%
	%
	%
	%
\end{tabularx}
\queryCardVSpace

\let\emph\oldemph
\let\textbf\oldtextbf

\renewcommand{\currentQueryCard}{0}
\renewcommand*{\arraystretch}{1.1}

\subsection*{Transaction / write / 19}
\label{sec:transaction-write-19}

\let\oldemph\emph
\renewcommand{\emph}[1]{{\footnotesize \sf #1}}
\let\oldtextbf\textbf
\renewcommand{\textbf}[1]{{\it #1}}\renewcommand{\currentQueryCard}{19}
\marginpar{
	\raggedleft
	\scriptsize
	
	\queryRefCard{transaction-write-01}{TW}{1}\\
	\queryRefCard{transaction-write-02}{TW}{2}\\
	\queryRefCard{transaction-write-03}{TW}{3}\\
	\queryRefCard{transaction-write-04}{TW}{4}\\
	\queryRefCard{transaction-write-05}{TW}{5}\\
	\queryRefCard{transaction-write-06}{TW}{6}\\
	\queryRefCard{transaction-write-07}{TW}{7}\\
	\queryRefCard{transaction-write-08}{TW}{8}\\
	\queryRefCard{transaction-write-09}{TW}{9}\\
	\queryRefCard{transaction-write-10}{TW}{10}\\
	\queryRefCard{transaction-write-11}{TW}{11}\\
	\queryRefCard{transaction-write-12}{TW}{12}\\
	\queryRefCard{transaction-write-13}{TW}{13}\\
	\queryRefCard{transaction-write-14}{TW}{14}\\
	\queryRefCard{transaction-write-15}{TW}{15}\\
	\queryRefCard{transaction-write-16}{TW}{16}\\
	\queryRefCard{transaction-write-17}{TW}{17}\\
	\queryRefCard{transaction-write-18}{TW}{18}\\
	\queryRefCard{transaction-write-19}{TW}{19}\\
}

\noindent\begin{tabularx}{\queryCardWidth}{|>{\queryPropertyCell}p{\queryPropertyCellWidth}|X|}
	\hline
	query & Transaction / write / 19 \\ \hline
	title & Block a Person of high risk \\ \hline

		pattern & \centering \includegraphics[scale=\yedscale,margin=0cm .2cm]{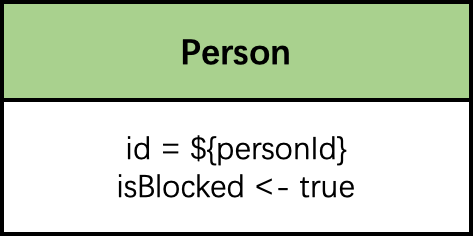} \tabularnewline \hline

	desc. & Set a \emph{Person}'s isBlocked to True. \\ \hline

		params &
		\innerCardVSpace{\begin{tabularx}{\attributeCardWidth}{|>{\paramNumberCell}C{\attributeNumberWidth}|>{\varNameCell}M|>{\typeCell}m{\typeWidth}|Y|} \hline
		$\mathsf{1}$ & \$personId & ID &  \\ \hline
		\end{tabularx}}\innerCardVSpace \\ \hline
	
%
	
%
	%
	%
	%
	%
\end{tabularx}
\queryCardVSpace

\let\emph\oldemph
\let\textbf\oldtextbf

\renewcommand{\currentQueryCard}{0}


\section{Read-Write Queries}
\label{sec:rw-queries}

\renewcommand*{\arraystretch}{1.1}

\subsection*{Transaction / read-write / 1}
\label{sec:transaction-read-write-01}

\let\oldemph\emph
\renewcommand{\emph}[1]{{\footnotesize \sf #1}}
\let\oldtextbf\textbf
\renewcommand{\textbf}[1]{{\it #1}}\renewcommand{\currentQueryCard}{1}
\marginpar{
	\raggedleft
	\scriptsize
	
	\queryRefCard{transaction-read-write-01}{TRW}{1}\\
	\queryRefCard{transaction-read-write-02}{TRW}{2}\\
	\queryRefCard{transaction-read-write-03}{TRW}{3}\\
}

\noindent\begin{tabularx}{\queryCardWidth}{|>{\queryPropertyCell}p{\queryPropertyCellWidth}|X|}
	\hline
	query & Transaction / read-write / 1 \\ \hline
	title & Transfer under transfer cycle detection strategy \\ \hline

		pattern & \centering \includegraphics[scale=\yedscale,margin=0cm .2cm]{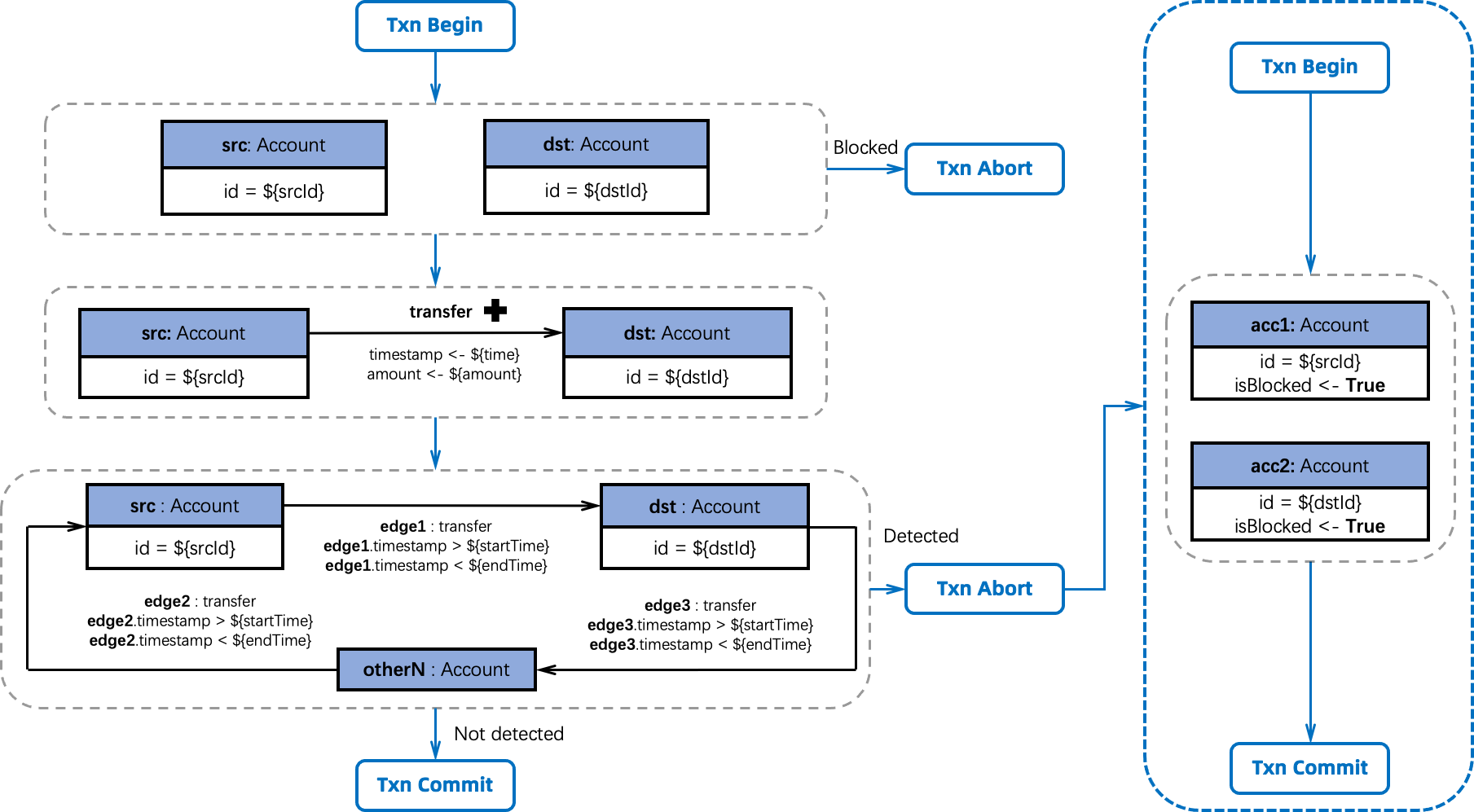} \tabularnewline \hline

		compose. & 
		This read-write query contains the reads and writes below,
		\begin{itemize}
			\item \hyperref[sec:transaction-simple-read-01] { Transaction / Simple Read / 1 }
			\item \hyperref[sec:transaction-write-12] { Transaction / Write / 12 }
			\item \hyperref[sec:transaction-complex-read-04] { Transaction / Complex Read / 4 }
			\item \hyperref[sec:transaction-write-18] { Transaction / Write / 18 }
			\end{itemize}
		\\ \hline

	desc. & The workflow of this read write query contains at least one transaction.
It works as:

\begin{itemize}
\tightlist
\item
  In the very beginning, read the blocked status of related accounts
  with given ids of two \emph{src} and \emph{dst} accounts. The
  transaction aborts if one of them is blocked. Move to the next step if
  none is blocked.
\item
  Add a transfer edge from \emph{src} to \emph{dst} inside a
  transaction. Given a specified time window between \emph{startTime}
  and \emph{endTime}, find the other accounts which received money from
  \emph{dst} and transferred money to \emph{src} in a specific time.
  Transaction aborts if a new transfer cycle is formed, otherwise the
  transaction commits.
\item
  If the last transaction aborts, mark the \emph{src} and \emph{dst}
  accounts as blocked in another transaction.
\end{itemize} \\ \hline

		params &
		\innerCardVSpace{\begin{tabularx}{\attributeCardWidth}{|>{\paramNumberCell}C{\attributeNumberWidth}|>{\varNameCell}M|>{\typeCell}m{\typeWidth}|Y|} \hline
		$\mathsf{1}$ & srcId & ID & id of the src Account \\ \hline
		$\mathsf{2}$ & dstId & ID & id of the dst Account \\ \hline
		$\mathsf{3}$ & time & DateTime & the timestamp of the transfer \\ \hline
		$\mathsf{4}$ & amount & 64-bit Float & the amount of the transfer \\ \hline
		$\mathsf{5}$ & startTime & DateTime & begin of the time window \\ \hline
		$\mathsf{6}$ & endTime & DateTime & end of the time window \\ \hline
		\end{tabularx}}\innerCardVSpace \\ \hline
	
%
	
%
	%
	%
	%
	%
\end{tabularx}
\queryCardVSpace

\let\emph\oldemph
\let\textbf\oldtextbf

\renewcommand{\currentQueryCard}{0}
\renewcommand*{\arraystretch}{1.1}

\subsection*{Transaction / read-write / 2}
\label{sec:transaction-read-write-02}

\let\oldemph\emph
\renewcommand{\emph}[1]{{\footnotesize \sf #1}}
\let\oldtextbf\textbf
\renewcommand{\textbf}[1]{{\it #1}}\renewcommand{\currentQueryCard}{2}
\marginpar{
	\raggedleft
	\scriptsize
	
	\queryRefCard{transaction-read-write-01}{TRW}{1}\\
	\queryRefCard{transaction-read-write-02}{TRW}{2}\\
	\queryRefCard{transaction-read-write-03}{TRW}{3}\\
}

\noindent\begin{tabularx}{\queryCardWidth}{|>{\queryPropertyCell}p{\queryPropertyCellWidth}|X|}
	\hline
	query & Transaction / read-write / 2 \\ \hline
	title & Transfer under in/out ratio strategy \\ \hline

		pattern & \centering \includegraphics[scale=\yedscale,margin=0cm .2cm]{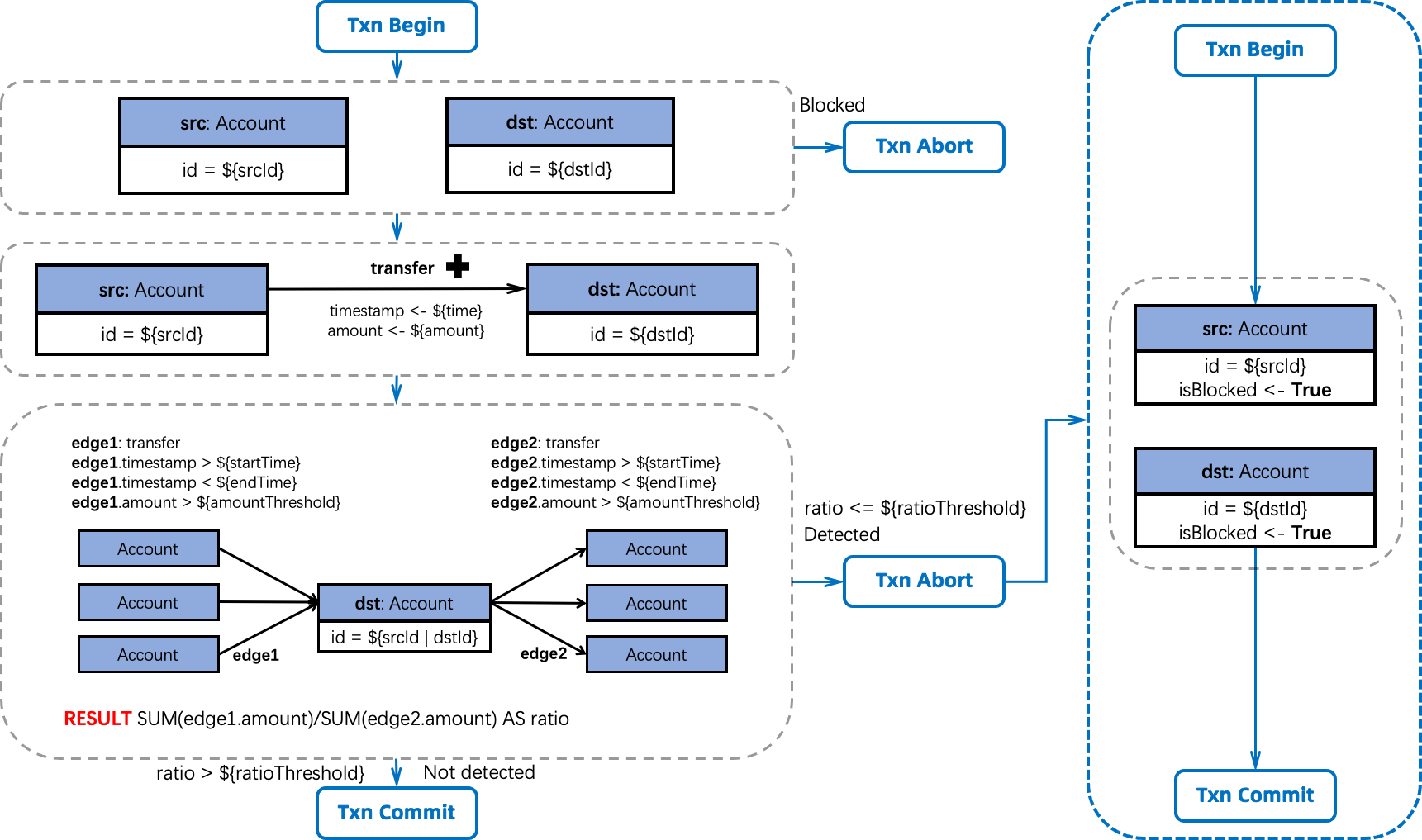} \tabularnewline \hline

		compose. & 
		This read-write query contains the reads and writes below,
		\begin{itemize}
			\item \hyperref[sec:transaction-simple-read-01] { Transaction / Simple Read / 1 }
			\item \hyperref[sec:transaction-write-12] { Transaction / Write / 12 }
			\item \hyperref[sec:transaction-complex-read-07] { Transaction / Complex Read / 7 }
			\item \hyperref[sec:transaction-write-18] { Transaction / Write / 18 }
			\end{itemize}
		\\ \hline

	desc. & The workflow of this read write query contains at least one transaction.
It works as:

\begin{itemize}
\tightlist
\item
  In the very beginning, read the blocked status of related accounts
  with given ids of two \emph{src} and \emph{dst} accounts. The
  transaction aborts if one of them is blocked. Move to the next step if
  none is blocked.
\item
  Add a transfer edge from \emph{src} to \emph{dst} inside a
  transaction. Given a specified time window between \emph{startTime}
  and \emph{endTime}, find all the \emph{transfer-in} and
  \emph{transfer-out} whose amount exceeds \emph{amountThreshold}.
  Transaction aborts if the ratio of transfers-in/transfers-out amount
  exceeds a given \emph{ratioThreshold}, both for the \emph{src} and
  \emph{dst} account. Otherwise the transaction commits.
\item
  If the last transaction aborts, mark the \emph{src} and \emph{dst}
  accounts as blocked in another transaction.
\end{itemize} \\ \hline

		params &
		\innerCardVSpace{\begin{tabularx}{\attributeCardWidth}{|>{\paramNumberCell}C{\attributeNumberWidth}|>{\varNameCell}M|>{\typeCell}m{\typeWidth}|Y|} \hline
		$\mathsf{1}$ & srcId & ID & id of the src Account \\ \hline
		$\mathsf{2}$ & dstId & ID & id of the dst Account \\ \hline
		$\mathsf{3}$ & time & DateTime & the timestamp of the transfer \\ \hline
		$\mathsf{4}$ & amount & 64-bit Float & the amount of the transfer \\ \hline
		$\mathsf{5}$ & amountThreshold & 64-bit Float & transfer amount threshold \\ \hline
		$\mathsf{6}$ & startTime & DateTime & begin of the time window \\ \hline
		$\mathsf{7}$ & endTime & DateTime & end of the time window \\ \hline
		$\mathsf{8}$ & ratioThreshold & 32-bit Float & ratio threshold of transfers-in over transfers-out \\ \hline
		$\mathsf{9}$ & truncationLimit & 32-bit Integer & maximum edges traversed at each step \\ \hline
		$\mathsf{10}$ & truncationOrder & Enum & the sort order before truncation at each step \\ \hline
		\end{tabularx}}\innerCardVSpace \\ \hline
	
%
	
%
	%
	%
	%
	%
\end{tabularx}
\queryCardVSpace

\let\emph\oldemph
\let\textbf\oldtextbf

\renewcommand{\currentQueryCard}{0}
\renewcommand*{\arraystretch}{1.1}

\subsection*{Transaction / read-write / 3}
\label{sec:transaction-read-write-03}

\let\oldemph\emph
\renewcommand{\emph}[1]{{\footnotesize \sf #1}}
\let\oldtextbf\textbf
\renewcommand{\textbf}[1]{{\it #1}}\renewcommand{\currentQueryCard}{3}
\marginpar{
	\raggedleft
	\scriptsize
	
	\queryRefCard{transaction-read-write-01}{TRW}{1}\\
	\queryRefCard{transaction-read-write-02}{TRW}{2}\\
	\queryRefCard{transaction-read-write-03}{TRW}{3}\\
}

\noindent\begin{tabularx}{\queryCardWidth}{|>{\queryPropertyCell}p{\queryPropertyCellWidth}|X|}
	\hline
	query & Transaction / read-write / 3 \\ \hline
	title & Guarantee under guarantee chain detection strategy \\ \hline

		pattern & \centering \includegraphics[scale=\yedscale,margin=0cm .2cm]{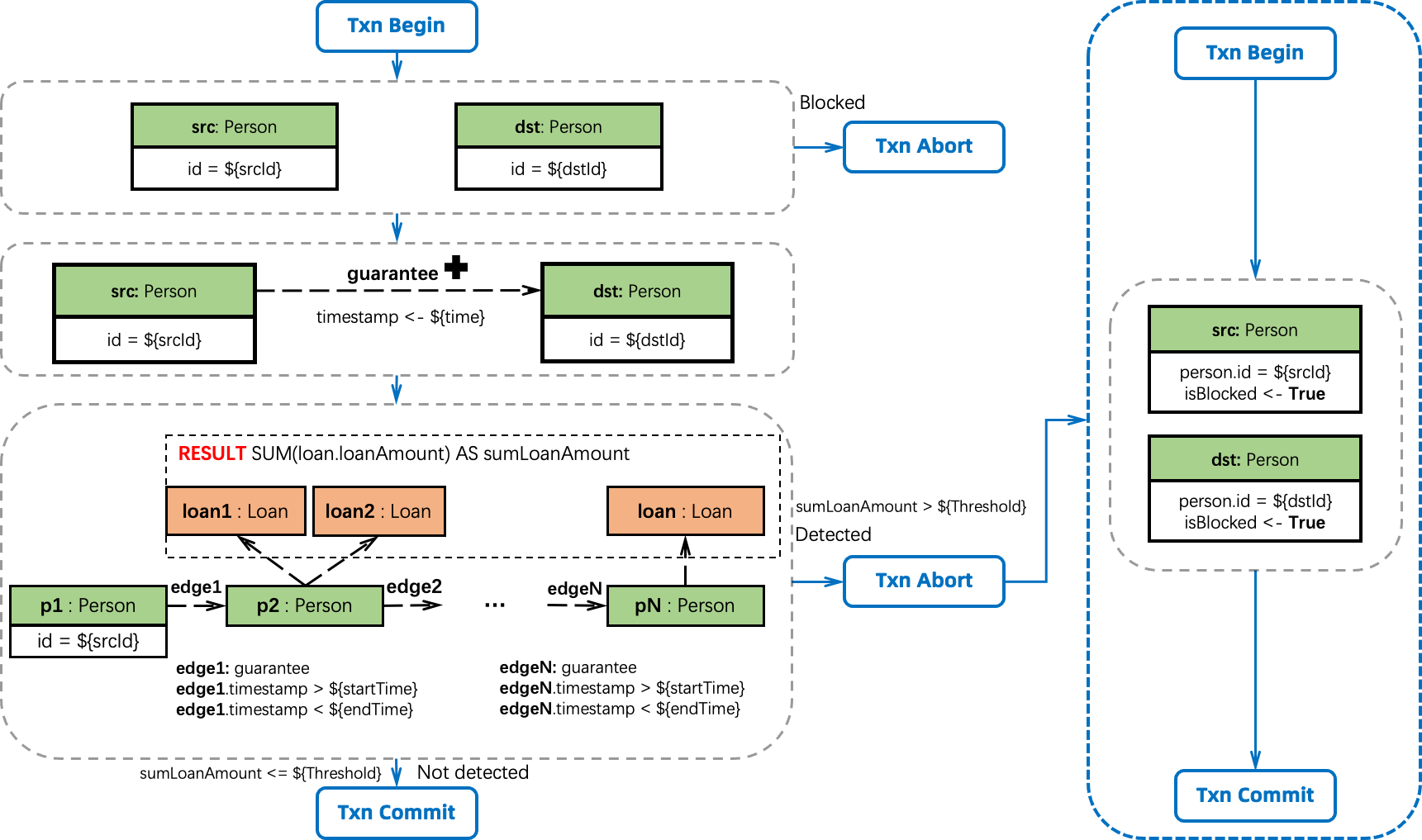} \tabularnewline \hline

		compose. & 
		This read-write query contains the reads and writes below,
		\begin{itemize}
			\item \hyperref[sec:transaction-simple-read-01] { Transaction / Simple Read / 1 }
			\item \hyperref[sec:transaction-write-10] { Transaction / Write / 10 }
			\item \hyperref[sec:transaction-complex-read-11] { Transaction / Complex Read / 11 }
			\item \hyperref[sec:transaction-write-19] { Transaction / Write / 19 }
			\end{itemize}
		\\ \hline

	desc. & The workflow of this read write query contains at least one transaction.
It works as:

\begin{itemize}
\tightlist
\item
  In the very beginning, read the blocked status of related persons with
  given ids of two \emph{src} and \emph{dst} persons. The transaction
  aborts if one of them is blocked. Move to the next step if none is
  blocked.
\item
  Add a guarantee edge between the \emph{src} and \emph{dst} persons
  inside a transaction. Given a specified time window between
  \emph{startTime} and \emph{endTime}, find all the persons in the
  guarantee chain of until end and their loans applied. Detect if a
  guarantee chain pattern formed, only for the \emph{src} person.
  Calculate if the amount sum of the related loans in the chain exceeds
  a given threshold. Transaction aborts if the sum exceeds the
  threshold. Otherwise the transaction commits.
\item
  If the last transaction aborts, mark the \emph{src} and \emph{dst}
  persons as blocked in another transaction.
\end{itemize} \\ \hline

		params &
		\innerCardVSpace{\begin{tabularx}{\attributeCardWidth}{|>{\paramNumberCell}C{\attributeNumberWidth}|>{\varNameCell}M|>{\typeCell}m{\typeWidth}|Y|} \hline
		$\mathsf{1}$ & srcId & ID & id of the src Person \\ \hline
		$\mathsf{2}$ & dstId & ID & id of the dst Person \\ \hline
		$\mathsf{3}$ & time & DateTime & the timestamp of the guarantee \\ \hline
		$\mathsf{4}$ & threshold & 64-bit Float & loan amount threshold in the guarantee chain \\ \hline
		$\mathsf{5}$ & startTime & DateTime & begin of the time window \\ \hline
		$\mathsf{6}$ & endTime & DateTime & end of the time window \\ \hline
		$\mathsf{7}$ & truncationLimit & 32-bit Integer & maximum edges traversed at each step \\ \hline
		$\mathsf{8}$ & truncationOrder & Enum & the sort order before truncation at each step \\ \hline
		\end{tabularx}}\innerCardVSpace \\ \hline
	
%
	
%
	%
	%
	%
	%
\end{tabularx}
\queryCardVSpace

\let\emph\oldemph
\let\textbf\oldtextbf

\renewcommand{\currentQueryCard}{0}


\chapter{Analytics Workload}
\label{sec:analytics-workload}

This workload is future work that will be released in the following version of \ldbcfinbench.


\chapter{ACID Test}
\label{sec:acid-test}

\newcommand{\bl}[1]{\textcolor{blue}{#1}}
\newcommand{\rd}[1]{\textcolor{red}{#1}}
\newcommand{\gn}[1]{\textcolor{green}{#1}}
\newcommand{\gy}[1]{\textcolor{grey}{\textit{#1}}}

\newcommand{\level}[1]{\textsf{#1}}
\newcommand{\anomaly}[1]{\rd{#1}}
\newcommand{\anolong}[1]{\emph{\rd{#1}}}
\newcommand{\tx}[1]{#1}

\newcommand{\cmark}{\ding{51}}
\newcommand{\xmark}{\ding{55}}

\begin{quote}
  \textit{This chapter is based on the chapter on "ACID tests" in the
    \ldbcsnb(LDBC SNB specification).The main difference between
    this section and \ldbcsnb\xspace is the schema design. The
    framework and reference implementations of the ACID test suite
    are available at \url{https://github.com/ldbc/ldbc_finbench_acid}.
  }
\end{quote}

Verifying ACID compliance is an important step in the benchmarking process for
enabling fair comparison between systems. The performance benefits of operating
with weaker safety guarantees are well established~\cite{DBLP:conf/ds/GrayLPT76}
but this can come at the cost of application correctness. To enable apples vs.
apples performance comparisons between systems it is expected they uphold the
ACID properties. Currently, LDBC provides no mechanism for validating ACID
compliance within the FinBench Transaction workflow.

This chapter presents the design of an implementation-agnostic ACID-compliance
test suite for the Transaction workload\footnote{We acknowledge verifying
  ACID compliance with a finite set of tests is not possible. However, the goal is
  not an exhaustive quality assurance test of a system's safety properties but
  rather to demonstrate that ACID guarantees are supported.}. Our guiding design
principle was to be agnostic of system-level implementation details, relying
solely on client observations to determine the occurrence of non-transactional
behavior. Thus all systems can be subjected to the same tests and fair
comparisons between FinBench Transaction performance results can be drawn. Tests
are described in the context of a graph database employing the property graph data
model~\cite{DBLP:journals/csur/AnglesABHRV17}. Reference implementations are
given in Cypher~\cite{DBLP:conf/sigmod/FrancisGGLLMPRS18}, the \emph{de facto}
standard graph query language.

Particular emphasis is given to testing isolation, covering ~10 known anomalies.
A conscious decision was made to keep tests relatively lightweight, as to not
add significant overhead to the benchmarking process.

\section{Background}

\begin{figure}[htbp]
  \centering
  \includegraphics[scale=0.5]{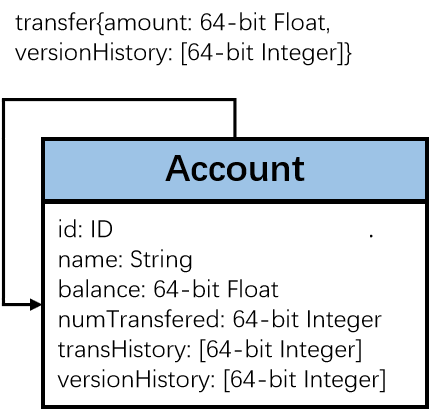}
  \caption{Graph schema for the ACID test queries}
  \label{figure:core-schema}
\end{figure}

The tests presented in this chapter are defined on a small core of LDBC FinBench
schema given in \autoref{figure:core-schema}.

\begin{figure}[h]
  \centering
  \scriptsize
\begin{tikzpicture}[xscale=0.77,yscale=0.47,trim left=0.8cm]
\node[text width=4cm,align=center] (RU) at (3,1) {
\level{Read Uncommitted}  \\
\anomaly{G0}
};

\node[text width=4cm,align=center] (RC) at (3,3) {
\level{Read Committed}  \\
\anomaly{+ G1\{a-c\}}
};

\node[text width=4cm,align=center] (ICI) at (9,1) {
\level{Item Cut Isolation}  \\
\anomaly{IMP}
};

\node[text width=4cm,align=center] (PCI) at (9,4) {
\level{Predicate Cut Isolation}  \\
\anomaly{+ PMP}
};

\node[text width=1.5cm,align=center] (MAV) at (6,5) {
\level{Monotonic Atomic View}  \\
\anomaly{+ OTV}
};

\node[text width=4cm,align=center] (CS) at (3,6.5) {
\level{Cursor Stability}  \\
\anomaly{+ G-Cursor(x), LU}
};

\node[text width=2.5cm,align=center] (RA) at (9,7) {
\level{Read Atomic}  \\
\anomaly{+ FR}
};

\node[text width=3.3cm,align=center] (SI) at (9,9) {
\level{Snapshot Isolation}  \\
\anomaly{+ LU}
};

\node[text width=2.2cm,align=center] (RR) at (3,9) {
\level{Repeatable Read}  \\
\anomaly{+ WS (G2-Item)}
}; 

\node[text width=4cm,align=center] (S) at (6,11) {
  \level{Serializability}
};

\draw [->,>=stealth] (RU) -- (RC);
\draw [->,>=stealth] (ICI) -- (PCI);
\draw [->,>=stealth] (RC) -- (MAV);
\draw (ICI) -- (MAV);
\draw [->,>=stealth] (MAV) -- (RR);
\draw [->,>=stealth] (PCI) -- (RA);
\draw [->,>=stealth] (MAV) -- (RA);
\draw [->,>=stealth] (RA) -- (RR);
\draw [->,>=stealth] (RA) -- (SI);
\draw [->,>=stealth] (SI) -- (S);
\draw [->,>=stealth] (RR) -- (S);
\draw [->,>=stealth] (CS) -- (RR);
\draw [->,>=stealth] (RC) -- (CS);

\end{tikzpicture}
\caption{Hierarchy of isolation levels as described in~\cite{DBLP:journals/tods/BailisFGHS16}. All anomalies are covered except \anomaly{G-Cursor(x)}.}
\label{figure:isolation}
\end{figure}
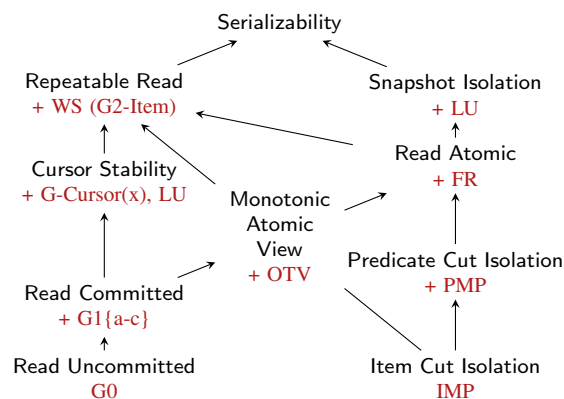

\section{Atomicity}

\emph{Atomicity} ensures that either all of a transaction's actions are
performed, or none are. Two atomicity tests have been designed.

  {\flushleft \textbf{Atomicity-C}} checks for every successful commit message a
client receives that any data items inserted or modified are subsequently visible.

  {\flushleft \textbf{Atomicity-RB}} checks for every aborted transaction that all
its modifications are not visible.

\paragraph{Test.}
\begin{enumerate*}[label={(\roman*)}]
  \item load a graph of \texttt{Account} vertices (\autoref{fig:ainitial}) each
  with a unique \texttt{id} and a set of \texttt{transHistory};
  \item a client executes a full graph scan counting the number of vertices, edges
  and transHistory (\autoref{fig:acheck}) using the result to initialize a
  counter \texttt{committed};
  \item $N$ transaction instances (\autoref{fig:ac}, \autoref{fig:arb}) of the
  required test are then executed, \texttt{committed} is incremented for
  each successful commit;
  \item repeat the full graph scan, storing the result in the variable
  \texttt{finalState};
  \item perform the anomaly check: \texttt{committed=finalState}.
\end{enumerate*}

The \textbf{Atomicity-C} transaction (\autoref{fig:ac}) randomly selects an
\texttt{Account}, creates a new \texttt{Account}, inserts a \texttt{transfer}
edge and appends a \texttt{newTrans} to \texttt{transHistory}. The \textbf{Atomicity-RB}
transaction (\autoref{fig:arb}) randomly selects an \texttt{Account}, appends a
\texttt{newTrans} and attempts to insert an \texttt{Account} only if it does
not exist. Note, for \textbf{Atomicity-RB} if the query API does not offer a
\texttt{ROLLBACK} statement constraints such as vertice uniqueness can be utilized
to trigger an abort.

\begin{figure}[htb]
  \centering

  \begin{lstlisting}[language=cypher,label=fig:ainitial,caption=Cypher query for creating initial data for the \tx{Atomicity} transactions.]
CREATE (:Account {id: 1, name: 'AliceAcc', transHistory: [100]}),
       (:Account {id: 2, name: 'BobAcc', transHistory: [50, 150]})
\end{lstlisting}

\end{figure}

\begin{figure}[htb]
  \centering
  \begin{minipage}{0.45\linewidth}
    \begin{lstlisting}[language=cypher,label=fig:ac,caption=\tx{Atomicity-C Tx.}]
<<BEGIN>>
MATCH (a1:Account {id: $account1Id})
CREATE (a1)-[t: transfer]->(a2:Account)
SET
  a1.transHistory = a1.transHistory + [$newTrans],
  a2.id = $account2Id,
  t.amount = $newTrans
<<COMMIT>>
\end{lstlisting}
  \end{minipage}
  \quad
  \begin{minipage}{0.52\linewidth}
    \begin{lstlisting}[language=cypher,label=fig:arb,caption=\tx{Atomicity-RB Tx.}]
<<BEGIN>>
MATCH (a1:Account {id: $account1Id})
SET a1.transHistory = a1.transHistory + [$newTrans]
<<IF>> MATCH (a2: Account {id: $account2Id}) exists
<<THEN>> <<ABORT>> <<ELSE>>
CREATE (a2:Account {id: $account2Id})
<<END>>
<<COMMIT>>
\end{lstlisting}
  \end{minipage}
\end{figure}

\begin{figure}[htb]
  \centering
  \begin{lstlisting}[language=cypher,label=fig:acheck,caption=\tx{Atomicity-C/Atomicity-RB:} counting entities in the graph.]
  MATCH (a:Account)
  RETURN count(a) AS numAccounts, count(a.name) AS numNames, sum(size(a.transHistory)) AS numTransferred
\end{lstlisting}
\end{figure}

\section{Isolation}
\label{sec:isolation}

The gold standard isolation level is \level{Serializability}, which offers
protection against all possible \emph{anomalies} that can occur from the
concurrent execution of transactions. Anomalies are occurrences of
non-serializable behavior. Providing \level{Serializability} can be detrimental
to performance~\cite{DBLP:conf/ds/GrayLPT76}. Thus systems offer numerous weak
isolation levels such as \level{Read Committed} and \level{Snapshot Isolation}
that allow a higher degree of concurrency at the cost of potential
non-serializable behavior. As such, isolation levels are defined in terms of
the anomalies they prevent~\cite{DBLP:conf/ds/GrayLPT76,DBLP:journals/pvldb/BailisDFGHS13}.
\autoref{figure:isolation} relates isolation levels to the anomalies they proscribe.

To allow fair comparison systems must disclose the isolation level used
during benchmark execution. The purpose of these isolation tests is to verify
that the claimed isolation level matches the expected behavior. To this end,
tests have been developed for each anomaly presented
in~\cite{DBLP:journals/tods/BailisFGHS16}. Formal definitions for each anomaly
are reproduced from~\cite{adya1999weak,DBLP:journals/tods/BailisFGHS16} using
their system model which is described below. General design considerations are
discussed before each test is described.

\subsection{System Model}
\label{sec:system-model}

Transactions consist of an ordered sequence of read and write operations to an
arbitrary set of data items, book-ended by a \texttt{BEGIN} operation and a
\texttt{COMMIT} or an \texttt{ABORT} operation. In a graph database data items
are vertices, edges and properties. The set of items a transaction reads from and
writes to is termed its \emph{item read set} and \emph{item write set}. Each
write creates a \emph{version} of an item, which is assigned a unique timestamp
taken from a totally ordered set (\eg natural numbers) version $i$ of item $x$
is denoted $x_i$. All data items have an initial \emph{unborn} version $\bot$
produced by an initial transaction $\tx{T_{\bot}}$. The unborn version is
located at the start of each item's version order. Execution of transactions
on a database is represented by a \emph{history}, H, consisting of
\begin{enumerate*}[label={(\roman*)}]
  \item an ordered sequence of read and write operations of each transaction,
  \item ordered data item versions read and written and
  \item commit or abort operations.
\end{enumerate*}~\cite{DBLP:journals/tods/BailisFGHS16}

There are three types of dependencies between transactions, which capture the
ways in which transactions can \emph{directly} conflict. \emph{Read dependencies}
capture the scenario where a transaction reads another transaction's write.
\emph{Antidependencies} capture the scenario where a transaction overwrites the
version another transaction reads. \emph{Write dependencies} capture the
scenario where a transaction overwrites the version another transaction writes.
Their definitions are as follows:

\begin{description}
  \item[Read-Depends]
    Transaction $\tx{T_j}$ \emph{directly read-depends} (\textsf{wr}) on
    $\tx{T_i}$ if $\tx{T_i}$ writes some version $x_i$ and $\tx{T_j}$ reads $x_i$.
  \item[Anti-Depends]
    Transaction $\tx{T_j}$ \emph{directly anti-depends} (\textsf{rw}) on
    $\tx{T_i}$ if $\tx{T_i}$ reads some version $x_k$ and $\tx{T_j}$ writes
    $x$'s next version after $x_k$ in the version order.
  \item[Write-Depends]
    Transaction $\tx{T_j}$ \emph{directly write-depends} (\textsf{ww}) on
    $\tx{T_i}$ if $\tx{T_i}$ writes some version $x_i$ and $\tx{T_j}$ writes
    $x$'s next version after $x_i$ in the version order.
\end{description}

Using these definitions, from a history $H$ a \emph{direct serialization graph}
$\textit{DSG}(H)$ is constructed. Each vertice in the $\textit{DSG}$ corresponds to
a committed transaction and edges correspond to the types of direct conflicts
between transactions. Anomalies can then be defined by stating properties about
the $\textit{DSG}$.

The above \emph{item-based} model can be extended to handle
\emph{predicate-based} operations~\cite{adya1999weak}. Database operations are
frequently performed on a set of items provided a certain condition called the
\emph{predicate}, $P$ holds. When a transaction executes a read or write based
on a predicate $P$, the database selects a version for each item to which $P$
applies, this is called the version set of the predicate-based denoted as
$\textit{Vset}(P)$. A transaction $\tx{T_j}$ changes the matches of a
predicate-based read $r_i(P_i)$ if $\tx{T_i}$ overwrites a version
in $\textit{Vset}(P_i)$.

\subsection{General Design}
\label{sec:design-cons}

Isolation tests begin by loading a \emph{test graph} into the database.
Configurable numbers of \emph{write clients} and \emph{read clients} then
execute a sequence of transactions on the database for some configurable time
period. After execution, results from read clients are collected and an
\emph{anomaly check} is performed. In some tests, an additional full graph scan
is performed after the execution period in order to collect information required
for the anomaly check.

The guiding principle behind test design was the preservation of data items
version history -- the key ingredient needed in the system model formalization
which is often not readily available to clients, if preserved at all. Several
anomalies are closely related, therefore, tests had to be constructed such that
other anomalies could not interfere with or mask the detection of the targeted
anomaly. Test descriptions provide
\begin{enumerate*}[label={(\roman*)}]
  \item informal and formal anomaly definitions,
  \item the required test graph,
  \item description of transaction profiles write and read clients execute, and
  \item reasoning for why the test works.
\end{enumerate*}

\subsection{Dirty Write}
\label{sec:dirty-write}

Informally, a \anolong{Dirty Write} (Adya's \anomaly{G0}~\cite{adya1999weak})
occurs when updates by conflicting transactions are interleaved. For example,
say $\tx{T_i}$ and $\tx{T_j}$ both modify items $\{x,y\}$. If version $x_i$
precedes version $x_j$ and $y_j$ precedes version $y_i$, a \anomaly{G0} anomaly
has occurred. Preventing \anomaly{G0} is especially important in a graph
database in order to maintain \emph{Reciprocal Consistency}~\cite{Waudby2020}.

\paragraph{Definition.}
A history $H$ exhibits phenomenon \anomaly{G0} if $\textit{DSG}(H)$ contains a
directed cycle consisting entirely of write-dependency edges.

\paragraph{Test.}
Load a test graph containing pairs of \texttt{Account} vertices connected by a
\texttt{transfer} edge. Assign each \texttt{Account} a unique \texttt{id} and
each \texttt{Account} and \texttt{transfer} edge a \texttt{versionHistory}
property of type list (initially empty). During the execution period, write
clients execute a sequence of \tx{G0 $T_\mathrm{W}$} instances, \autoref{fig:dw1}.
This transaction appends its ID to the \texttt{versionHistory} property for each
entity (2 \texttt{Accounts} and 1 \texttt{transfer} edge) in the \texttt{Account}
pair it matches. Note, transaction IDs are assumed to be globally unique. After
execution, a read client issues a \tx{G0 $T_\mathrm{R}$} for each
\texttt{Account} pair in the graph, \autoref{fig:dw2}. Retrieving the
\texttt{versionHistory} for each entity in an \texttt{Account} pair.

\paragraph{Anomaly check.}
For each \texttt{Account} pair in the test graph:
\begin{enumerate*}[label={(\roman*)}]
  \item prune each \texttt{versionHistory} list to remove any version numbers
  that do not appear in all lists; needed to account for interference from
  \anolong{Lost Update} anomalies (\autoref{sec:lost-update}),
  \item compare the contents of each entities' \texttt{versionHistory} list
  element-wise,
  \item if lists do not agree, a \anomaly{G0} anomaly has occurred.

\end{enumerate*}

\paragraph{Why it works.}
Each successful \tx{G0 $T_\mathrm{W}$} creates a new version of an
\texttt{Account} pair. Appending the transaction ID preserves the version
history of each entity in the \texttt{Account} pair. In a system that prevents
\anomaly{G0}, each entity of the \texttt{Account} pair should experience the
\emph{same} updates, in the \emph{same} order. Hence, each position in the
\texttt{versionHistory} lists should be equivalent. The additional pruning step
is needed as \anolong{Lost Updates} overwrites a version, effectively erasing it
from the history of a data item.

\begin{figure}[htb]
  \centering
  \begin{minipage}{0.53\linewidth}
    \begin{lstlisting}[language=cypher,label=fig:dw1,caption=\tx{Dirty Write (G0) $T_\mathrm{W}$}.]
MATCH
  (a1:Account {id: $account1Id})
  -[t:transfer]->(a2:Account {id: $account2Id})
SET a1.versionHistory = a1.versionHistory + [$tid]
SET a2.versionHistory = a2.versionHistory + [$tid]
SET t.versionHistory  = t.versionHistory  + [$tId]
\end{lstlisting}
  \end{minipage}
  \quad
  \begin{minipage}{0.431\linewidth}
    \begin{lstlisting}[language=cypher,label=fig:dw2,caption=\tx{Dirty Write (G0) $T_\mathrm{R}$}.]
MATCH (a1:Account {id: $account1Id})
-[t:transfer]->(a2:Account {id: $account2Id})
RETURN
  a1.versionHistory AS a1VersionHistory,
  t.versionHistory  AS tVersionHistory,
  a2.versionHistory AS a2VersionHistory
\end{lstlisting}
  \end{minipage}
\end{figure}

\subsection{Dirty Reads}
\label{sec:dirty-reads}

\subsection*{Aborted Reads}

Informally, an \anolong{Aborted Read} (\anomaly{G1a}) anomaly occurs when a
transaction reads the updates of a transaction that later aborts.

\paragraph{Definition.}
A history $H$ exhibits phenomenon \anomaly{G1a} if $H$ contains an aborted
transaction $\tx{T_a}$ and a committed transaction $\tx{T_c}$ such that
$\tx{T_c}$ reads a version written by $\tx{T_a}$.

\paragraph{Test.}
Load a test graph containing only \texttt{Account} vertices into the database.
Assign each \texttt{Account} a unique \texttt{id} and \texttt{balance}
initialized to 99 (or any odd number). During execution, write clients execute
a sequence of \tx{G1a $T_\mathrm{W}$} instances, \autoref{fig:ar1}. Selecting a
random \texttt{Account} \texttt{id} to populate each instance. This transaction
attempts to set \texttt{balance=200} (or any even number) but always aborts.
Concurrently, read clients execute a sequence of \tx{G1a $T_\mathrm{R}$}
instances, \autoref{fig:ar2}. This transaction retrieves the \texttt{balance}
property of an \texttt{Account}. Read clients store results, which are collected
after execution has finished.

\paragraph{Anomaly check.}
Each read should return \texttt{balance=99} (or any odd number). Otherwise, a
\anomaly{G1a} anomaly has occurred.

\paragraph{Why it works.}
Each transaction that attempts to set \texttt{balance} to an even number
\emph{always} aborts. Therefore, if a transaction reads \texttt{balance} to be
an even number, it must have read the write of an aborted transaction.

\begin{figure}[htb]
  \centering
  \begin{minipage}{0.45\linewidth}
    \begin{lstlisting}[language=cypher,label=fig:ar1,caption=\tx{Aborted Read (G1a) $T_\mathrm{W}$}.]
MATCH (a:Account {id: $accountId})
SET a.balance = 200
<<SLEEP($sleepTime)>>
<<ABORT>>
\end{lstlisting}

    \begin{lstlisting}[language=cypher,label=fig:ar2,caption=\tx{Aborted Read (G1a) $T_\mathrm{R}$}.]
MATCH (a:Account {id: $accountId})
RETURN a.balance as aBalance
\end{lstlisting}
  \end{minipage}
  \quad
  \begin{minipage}{0.45\linewidth}
    \begin{lstlisting}[language=cypher,label=fig:ir1,caption=\tx{Interm. Read (G1b) $T_\mathrm{W}$}.]
MATCH (a:Account {id: $accountId})
SET a.balance = $even
<<SLEEP($sleepTime)>>
SET a.balance = $odd
\end{lstlisting}

    \begin{lstlisting}[language=cypher,label=fig:ir2,caption=\tx{Interm. Read (G1b) $T_\mathrm{R}$}.]
MATCH (a:Account {id: $accountId})
RETURN a.balance as aBalance
\end{lstlisting}
  \end{minipage}
\end{figure}

\subsection*{Intermediate Reads}

Informally, an \anolong{Intermediate Read} (Adya's \anomaly{G1b}~\cite{adya1999weak})
anomaly occurs when a transaction reads the intermediate modifications of other
transactions.

\paragraph{Definition.}
A history $H$ exhibits phenomenon \anomaly{G1b} if $H$ contains a committed
transaction $\tx{T_i}$ that reads a version of an object $x_m$ written by
transaction $\tx{T_j}$, and $\tx{T_j}$ also wrote a version $x_n$ such that
$m < n$ in $x$'s version order.

\paragraph{Test.}
Load a test graph containing only \texttt{Account} vertices into the database.
Assign each \texttt{Account} a unique \texttt{id} and \texttt{balance}
initialized to 99 (or any odd number). During execution, write clients execute a
sequence of \tx{G1b $T_\mathrm{W}$} instances, \autoref{fig:ir1}. This
transaction sets \texttt{balance} to an even number, then an odd number before
committing. Concurrently read-clients execute a sequence of
\tx{G1b $T_\mathrm{R}$} instances, \autoref{fig:ir2}. Retrieving
\texttt{balance} property of an \texttt{Account}. Read clients store results
which are collected after execution has finished.

\paragraph{Anomaly check.}
Each read of \texttt{balance} should be an odd number.
Otherwise, a \anomaly{G1b} anomaly has occurred.

\paragraph{Why it works.}
The final balance modified by an \tx{G1b $T_\mathrm{W}$} instance is
\emph{never} an even number. Therefore, if a transaction reads \texttt{balance}
to be an even number it must have read an intermediate balance.

\subsection*{Circular Information Flow}

Informally, a \anolong{Circular Information Flow} (Adya's \anomaly{G1c}~\cite{adya1999weak})
anomaly occurs when two transactions affect each other; \ie both transactions
write data the other reads. For example, transaction $\tx{T_i}$ reads a
write by transaction $\tx{T_j}$ and transaction $\tx{T_j}$ reads a write by $\tx{T_i}$.

\paragraph{Definition.}
A history $H$ exhibits phenomenon \anomaly{G1c} if $\textit{DSG}(H)$ contains a
directed cycle that consists entirely of read-dependency and write-dependency edges.

\paragraph{Test.}
Load a test graph containing only \texttt{Account} vertices into the database.
Assign each \texttt{Account} a unique \texttt{id} and \texttt{balance}
initialized to 0. Read-write clients are required for this test, executing a
sequence of \tx{G1c $T_\mathrm{RW}$}, \autoref{fig:cif1}. This transaction
selects two different \texttt{Account} vertices, setting the \texttt{balance} of
one \texttt{Account} to the transaction ID and retrieving the \texttt{balance}
from the other. Note, transaction IDs are assumed to be globally unique.
Transaction results are stored in format \texttt{(txn.id, balanceRead)} and
collected after execution.

\paragraph{Anomaly check.}
For each result, check the result of the transaction the \texttt{balanceRead}
corresponds to, did not read the transaction of that result. Otherwise, a
\anomaly{G1c} anomaly has occurred.

\paragraph{Why it works.}
Consider the result set:
\texttt{\{($T_\mathrm{1}$, $T_\mathrm{2}$), ($T_\mathrm{2}$, $T_\mathrm{3}$),
  ($T_\mathrm{3}$, $T_\mathrm{2}$)\}}. $T_\mathrm{1}$ reads the balance written by
$T_\mathrm{2}$ and $T_\mathrm{2}$ reads the balance written by $T_\mathrm{3}$.
Here information flow is unidirectional from $T_\mathrm{1}$ to $T_\mathrm{2}$.
However, $T_\mathrm{2}$ reads the balance written by $T_\mathrm{3}$ and
$T_\mathrm{2}$ reads the balance written by $T_\mathrm{3}$. Here information flow
is circular from $T_\mathrm{2}$ to $T_\mathrm{3}$ and $T_\mathrm{3}$ to $T_\mathrm{2}$.
Thus a \anomaly{G1c} anomaly has been detected.

\begin{figure}[htb]
  \begin{lstlisting}[language=cypher,label=fig:cif1,caption=\tx{G1c $T_\mathrm{RW}$}.]
MATCH (a1:Account {id: $account1Id}) SET a1.balance = $transactionId
MATCH (a2:Account {id: $account2Id}) RETURN a2.balance AS account2Balance
\end{lstlisting}
\end{figure}

\subsection{Cut Anomalies}

\subsection*{Item-Many-Preceders}
\label{sec:cut-anomalies}

Informally, an \anolong{Item-Many-Preceders} (\anomaly{IMP})
anomaly~\cite{DBLP:journals/pvldb/BailisDFGHS13} occurs if a transaction observes
multiple versions of the same item (\eg transaction $\tx{T_i}$ reads versions
$x_1$ and $x_2$). In a graph database, this can be multiple reads of a vertice, edge,
property or label. Local transactions (involving a single data item) occur
frequently in graph databases, \eg in \emph{``Find properties of entities''}
\queryRefCard{transaction-simple-read-01}{TSR}{1}.

\paragraph{Definition.}
A history $H$ exhibits \anomaly{IMP} if $\textit{DSG}(H)$ contains a transaction
$\tx{T_i}$ such that $\tx{T_i}$ directly \emph{item-read-depends} on $x$ by more
than one other transaction.

\paragraph{Test.}
Load a test graph containing \texttt{Account} vertices. Assign each \texttt{Account}
a unique \texttt{id} and \texttt{balance} initialized to 1. During execution
write clients execute a sequence of \tx{IMP $T_\mathrm{W}$} instances, \autoref{fig:ic1}.
Selecting a random \texttt{id} and setting a new balance (globally unique) of the
\texttt{Account}. Concurrently read clients execute a sequence of \tx{IMP $T_\mathrm{R}$}
instances, \autoref{fig:ic2}. Performing multiple reads of the same \texttt{Account};
We can inject some wait time between reads to make conditions more favorable
for detecting an anomaly. Both reads within an \tx{IMP $T_\mathrm{R}$}
transaction are returned, stored and collected after execution.

\paragraph{Anomaly check.}
Each \tx{IMP $T_\mathrm{R}$} result set \texttt{(firstRead, secondRead)} should
contain the \emph{same} \texttt{Account} balance. If not, an \anomaly{IMP}
anomaly has occurred.

\paragraph{Why it works.}
By performing successive reads within the same transaction this test checks that
a system ensures consistent reads of the same data item. If the read balance
changes then a concurrent transaction has modified the data item and the reading
transaction is not protected from this change.

\begin{figure}[htb]
  \centering
  \begin{minipage}{0.35\linewidth}
    \begin{lstlisting}[language=cypher,label=fig:ic1,caption=\tx{IMP $T_\mathrm{W}$}.]
MATCH (a:Account {id: $accountId})
SET a.balance = $newBalance
\end{lstlisting}
    \begin{lstlisting}[language=cypher,label=fig:ic2,caption=\tx{IMP $T_\mathrm{R}$}.]
MATCH (a1:Account {id: $accountId})
WITH a1.balance AS firstRead
<<SLEEP($sleepTime)>>
MATCH (a2:Account {id: $accountId})
RETURN firstRead, 
  a2.balance AS secondRead
\end{lstlisting}
  \end{minipage}
  \quad
  \begin{minipage}{0.61\linewidth}
    \begin{lstlisting}[language=cypher,label=fig:pc1,caption=\tx{PMP $T_\mathrm{W}$}.]
MATCH (a1:Account {id: $account1Id}), (a2:Account {id: $account2Id})
CREATE (a1)-[:transfer]->(a2)
\end{lstlisting}
    \begin{lstlisting}[language=cypher,label=fig:pc2,caption=\tx{PMP $T_\mathrm{R}$}.]
MATCH (a2:Account {id: $accountId})<-[:transfer]-(a1:Account)
WITH count(a1) AS firstRead
<<SLEEP($sleepTime)>>
MATCH (a4:Account {id: $accountId})<-[:transfer]-(a3:Account)
RETURN firstRead, 
  count(a3) AS secondRead
\end{lstlisting}
  \end{minipage}
\end{figure}

\subsection*{Predicate-Many-Preceders}

Informally, a \anolong{Predicate-Many-Preceders} (\anomaly{PMP})
anomaly~\cite{DBLP:journals/pvldb/BailisDFGHS13} occurs if a transaction observes
different versions resulting from the same predicate read
(\eg $\tx{T_i}$ reads
$\textit{Vset}(P_i) =  \{x_1\}$ and
$\textit{Vset}(P_i) = \{x_1,y_2\}$).
Pattern matching is a common predicate read operation in a graph database.

\paragraph{Definition.}
A history $H$ exhibits the phenomenon \anomaly{PMP} if, for all predicate-based reads $r_i(P_i : \textit{Vset}(P_i))$ and $r_j(P_j : \textit{Vset}(P_j))$ in $\tx{T_k}$ such that the logical ranges of $P_i$ and $P_j$ overlap (call it $P_o$), the set of transactions that change the matches of $P_o$ for $r_i$ and $r_j$ differ.

\paragraph{Test.}
Load a test graph containing \texttt{Account} vertices. Assign each \texttt{Account}
a unique \texttt{id}. During execution write clients execute a sequence of
\tx{PMP $T_\mathrm{W}$} instances, inserting a \texttt{transfer} edge between a
randomly selected pair of \texttt{Accounts}, shown in \autoref{fig:pc1}.
Concurrently read clients execute a sequence of \tx{PMP $T_\mathrm{R}$}
instances, \autoref{fig:pc2}. Performing multiple reads of the pattern
\texttt{(a2:Account)<-[:transfer]-(a1:Account)} and counting the number of
\texttt{transfer} edges; successive reads can be separated by some artificially
injected wait time to make conditions more favorable for detecting an anomaly.
Both predicates reads within a \tx{PMP $T_\mathrm{R}$} transaction are returned,
stored and collected after test execution.

\paragraph{Anomaly check.}
For each \tx{PMP $T_\mathrm{R}$} transaction result set
\texttt{(firstRead, secondRead)}, the \texttt{firstRead} should be equal to
\texttt{secondRead}. Otherwise, a \anomaly{PMP} anomaly has occurred.

\paragraph{Why it works.}
By performing successive predicate reads and counting the number of
\texttt{transfer} edges within the same transaction this test checks that a
system ensures consistent reads of the same predicate. If the number of
\texttt{transfer} edges changes then a concurrent transaction has inserted a new
\texttt{transfer} edge and the reading transaction is not protected from this
change.

\subsection{Observed Transaction Vanishes}
\label{sec:observ-trans-vanish}

Informally, an \anolong{Observed Transaction Vanishes} (\anomaly{OTV})
anomaly~\cite{DBLP:journals/pvldb/BailisDFGHS13} occurs when a transaction
observes part of another transaction's updates but not all of them
(\eg $\tx{T_1}$ writes $x_1$ and $y_1$ and $\tx{T_2}$ reads $x_1$ and $y_\bot$).
Before formally defining \anomaly{OTV} the \emph{Unfolded Serialization Graph (USG)}
must be introduced~\cite{adya1999weak}. The $\textit{USG}$ is specified for an
individual transaction, $\tx{T_i}$ and a history, $H$ and is denoted by
$\textit{USG}(H,\tx{T_i})$. In a \emph{USG} the $\tx{T_i}$ vertice is split into
multiple vertices, one for each action read $r_i(\cdot)$ or  write $w_i(\cdot)$
within the transaction. The dependency edges are now incident on the relevant
event of $\tx{T_i}$. Additionally, actions within $\tx{T_i}$ are connected by
an \emph{order edge} \eg if $T_i$ reads object $y_j$ then immediately writes on
object $x$ an order edge exists from $w_i(x_i)$ to $r_i(y_j)$.

\paragraph{Definition.}
A history $H$ exhibits phenomenon \anomaly{OTV} if $\textit{USG}(H,T_i)$
contains a directed cycle consisting of
\begin{enumerate*}[label={(\roman*)}]
  \item exactly one read dependency edge induced by data item $x$ from
  $\tx{T_j}$ to $\tx{T_i}$ and
  \item a set of edges induced by data item $y$ containing at least one anti
  dependency edge from $\tx{T_i}$ to $\tx{T_j}$.
\end{enumerate*}
Additionally, $\tx{T_i}$'s read from $y$ precedes its read from $x$.

\paragraph{Test.}
Load a test graph containing a set of cycles of length 4 of \texttt{Accounts}
connected by \texttt{transfer} edges. Assign each \texttt{Account} an \texttt{id},
and \texttt{balance} property (initialized to 1). Note, \texttt{id} must be
unique across vertices. During execution write clients select an \texttt{id} and
executes a sequence of \tx{OTV $T_\mathrm{W}$} instances, \autoref{fig:otvfr1}.
This transaction effectively creates a new version of a given cycle. Concurrently
read-clients execute a sequence of \tx{OTV $T_\mathrm{R}$} instances, \autoref{fig:otvfr2}.
Matching a given cycle and performing multiple reads. Both reads within an
\tx{OTV $T_\mathrm{R}$} are returned, stored and collected after execution.

\paragraph{Anomaly check.}
For each \tx{OTV $T_\mathrm{R}$} result set \texttt{(firstRead,secondRead)},
the maximum \texttt{balance} in the \texttt{firstRead} should be less than or
equal to the minimum \texttt{balance} in the \texttt{secondRead}. Otherwise, an
\anomaly{OTV} anomaly has occurred.

\paragraph{Why it works.}
\tx{OTV $T_\mathrm{W}$} installs a new version of a cycle by updating the
\texttt{balance} property of each \texttt{Account}. Therefore when matching a
cycle once a transaction has observed some \texttt{balance} it should
\emph{at least} observe this same balance for every remaining entity in the cycle.
Unfortunately, this cannot be deduced from a single read of the cycle as results
from matching cycles often do not preserve the order in which graph entities
were read. This is solved by making multiple reads of the cycle. The maximum
\texttt{balance} of the \texttt{firstRead} determines the minimum
\texttt{balance} of \texttt{secondRead}. If this condition is violated then a
transaction has observed the effects of a transaction in the \texttt{firstRead}
then subsequently failed to observe it in the \texttt{secondRead} -- the
observed transaction has vanished!

\begin{figure}[htb]
  \centering
  \begin{minipage}{0.33\linewidth}
    \begin{lstlisting}[language=cypher,label=fig:otvfr1,caption=\tx{OTV/FR $T_\mathrm{W}$}.]
MATCH path = 
  (n:Account {id: $accountId})
  -[:transfer*..4]->(n)
UNWIND nodes(path)[0..4] AS a
SET a.balance = a.balance + 1
\end{lstlisting}
  \end{minipage}
  \quad
  \begin{minipage}{0.60\linewidth}
    \begin{lstlisting}[language=cypher,label=fig:otvfr2,caption=\tx{OTV/FR $T_\mathrm{R}$}.]
MATCH p1 = (a1:Account {id: $accountId})-[:transfer*..4]->(a1)
RETURN extract(a IN nodes(p1) | a.balance) AS firstRead
<<SLEEP($sleepTime)>>
MATCH p2 = (a2:Account {id: $accountId})-[:transfer*..4]->(a2)
RETURN extract(a IN nodes(p2) | a.balance) AS secondRead
\end{lstlisting}
  \end{minipage}
\end{figure}

\subsection{Fractured Read}
\label{sec:fractured-reads}

\begin{quote}
  \textit{This section is the same as \ldbcsnb, except the schema design.}
\end{quote}

Informally, a \anolong{Fractured Read} (\anomaly{FR})
anomaly~\cite{DBLP:journals/tods/BailisFGHS16} occurs when a transaction reads
\emph{across} transaction boundaries. For example, if $\tx{T_1}$ writes $x_1$
and $y_1$ and $\tx{T_3}$ writes $x_3$. If $\tx{T_2}$ reads $x_1$ and $y_1$, then
repeats its read of $x$ and reads $x_3$ a fractured read has occurred.

\paragraph{Definition.}
A transaction $\tx{T_j}$ exhibits phenomenon \anomaly{FR} if transaction
$\tx{T_i}$ writes versions $x_a$ and $y_b$ (in any order, where $x$ and $y$ may
or may not be distinct items), $\tx{T_j}$ reads version $x_a$ and version $y_c$,
and $c < b$.

\paragraph{Test.}
Same as the \anomaly{OTV} test.

\paragraph{Anomaly check.}
For each \tx{FR $T_\mathrm{R}$}  (\autoref{fig:otvfr2}) result set
\texttt{(firstRead, secondRead)}, all \texttt{balance} across both balance sets
should be equal. Otherwise, an \anomaly{FR} anomaly has occurred.

\paragraph{Why it works.}
\tx{FR $T_\mathrm{W}$} writes a new version of a cycle by updating the
\texttt{balance} properties on each \texttt{Account}. When \tx{FR $T_\mathrm{R}$}
observes a \texttt{balance} every subsequent read in that cycle should read the
\emph{same} \texttt{balance} as \tx{FR $T_\mathrm{W}$} (\autoref{fig:otvfr1})
installs the same \texttt{balance} for all \texttt{Account} vertices in the cycle.
Thus, if it observes a different \texttt{balance} it has observed the effect of
a different transaction and has read across transaction boundaries.

\subsection{Lost Update}
\label{sec:lost-update}

Informally, a \anolong{Lost Update} (\anomaly{LU})
anomaly~\cite{DBLP:journals/tods/BailisFGHS16} occurs when two transactions
concurrently attempt to make conditional modifications to the same data item(s).

\paragraph{Definition.}
A history $H$ exhibits phenomenon \anomaly{LU} if $\textit{DSG}(H)$ contains a
directed cycle having one or more anti-dependency edges and all edges are induced
by the same data item $x$.

\paragraph{Test.}
Load a test graph containing \texttt{Account} vertices. Assign each \texttt{Account}
a unique \texttt{id} and a property \texttt{numTransferred} (initialized to 0).
During execution write clients execute a sequence of \tx{LU $T_\mathrm{W}$}
instances, \autoref{fig:lu1}. Choosing a random \texttt{Account} and incrementing
its \texttt{numTransferred} property. Clients store local counters
(\texttt{expNumTransferred}) for each \texttt{Account}, which is incremented
each time an \texttt{Account} is selected \emph{and} the \tx{LU $T_\mathrm{W}$}
instance successfully commits. After the execution period, the
\texttt{numTransferred} is retrieved for each \texttt{Account} using
\tx{LU $T_\mathrm{R}$} in \autoref{fig:lu2} and \texttt{expNumTransferred} are
pooled from write clients for each \texttt{Account}.

\paragraph{Anomaly check.}
For each \texttt{Account} its \texttt{numTransferred} property should be equal
to the (global) \texttt{expNumTransferred} for that \texttt{Account}.

\paragraph{Why it works.}
Clients know how many successful \tx{LU $T_\mathrm{W}$} instances were issued
for a given \texttt{Account}. The observable \texttt{numTransferred} should
reflect this ground truth, otherwise, a \anomaly{LU} anomaly must have occurred.

\begin{figure}[htb]
  \centering
  \begin{minipage}{0.41\linewidth}
    \begin{lstlisting}[language=cypher,label=fig:lu1,caption=\tx{Lost Update $T_\mathrm{W}$}.]
MATCH (a1:Account {id: $account1Id})
CREATE (a1)-[:transfer]->(a2:Account {id: $account2Id})
SET a1.numTransferred = a1.numTransferred + 1
RETURN a1.numTransferred
\end{lstlisting}
  \end{minipage}
  \quad
  \begin{minipage}{0.52\linewidth}
    \begin{lstlisting}[language=cypher,label=fig:lu2,caption=\tx{Lost Update $T_\mathrm{R}$}.]
MATCH (a:Account {id: $accountId})
OPTIONAL MATCH (a)-[t:transfer]->()
WITH a, count(t) AS numTransferEdges
RETURN numTransferEdges,
  a.numTransferred AS numTransferred
\end{lstlisting}
  \end{minipage}
\end{figure}

\subsection{Write Skew}
\label{sec:write-skew}

\begin{quote}
  \textit{This section is similar to \ldbcsnb, except the schema design and
    constraint: \texttt{a1.id \% 2 = 1} in \tx{WS $T_\mathrm{R}$}, \autoref{fig:ws2}.
  }
\end{quote}

Informally, \anolong{Write Skew} (\anomaly{WS}) occurs when two transactions
simultaneously attempted to make \emph{disjoint} conditional modifications to
the same data item(s). It is referred to as \anomaly{G2-Item}
in~\cite{adya1999weak,DBLP:journals/tods/FeketeLOOS05}.

\paragraph{Definition.}
A history $H$ exhibits \anomaly{WS} if $\textit{DSG}(H)$ contains a directed
cycle having one or more anti-dependency edges.

\paragraph{Test.}
Load a test graph containing $n$ pairs of \texttt{Account} vertices
\texttt{(a1, a2)} for $k = 0, \ldots, n-1$, where the $k$th pair gets IDs
\texttt{a1.id = 2*k+1} and \texttt{a2.id = 2*k+2}, and balances
\texttt{a1.balance = 70} and \texttt{a2.balance = 80}. There is a constraint:
\texttt{a1.balance + a2.balance > 0}. During execution write clients execute a
sequence of \tx{WS $T_\mathrm{W}$} instances, \autoref{fig:ws1}. Selecting a
random \texttt{Account} pair and decrementing the \texttt{value} property of
one \texttt{Account} provided doing so would not violate the constraint. After
execution the database is scanned using \tx{WS $T_\mathrm{R}$}, \autoref{fig:ws2}.

\paragraph{Anomaly check.}
For each \texttt{Account} pair the constraint should hold true, otherwise, a
\anomaly{WS} anomaly has occurred.

\paragraph{Why it works.}
Under no \level{Serializable} execution of WS $T_\mathrm{W}$ instances would the
constraint \texttt{a1.balance + a2.balance > 0} be violated. Therefore, if
\tx{WS $T_\mathrm{R}$} returns a violation of this constraint it is clear a
\anomaly{WS} anomaly has occurred.

\begin{figure}[htb]
  \centering
  \begin{minipage}{0.55\linewidth}
    \begin{lstlisting}[language=cypher,label=fig:ws1,caption=\tx{WS $T_\mathrm{W}$}.]
MATCH (a1:Account {id: $account1Id}), 
      (a2:Account {id: $account2Id})
<<IF a1.balance + a2.balance < 100)>> <<THEN>> <<ABORT>> <<END>>
<<SLEEP($sleepTime)>>
account = <<pick randomly between account1Id, account2Id>>
MATCH (a:Account {id: $account})
SET a.balance = a.balance - 100
<<COMMIT>>
\end{lstlisting}
  \end{minipage}
  \quad
  \begin{minipage}{0.33\linewidth}
    \begin{lstlisting}[language=cypher,label=fig:ws2,caption=\tx{WS $T_\mathrm{R}$}.]
MATCH (a1:Account), 
      (a2:Account {id: a1.id+1})
WHERE a1.balance + a2.balance <= 0 
      and a1.id % 2 = 1
RETURN a1.id AS a1id, 
       a1.balance AS a1balance, 
       a2.id AS a2id, 
       a2.balance AS a2balance
\end{lstlisting}
  \end{minipage}
\end{figure}

\newpage

\section{Consistency and Durability Tests}
\label{sec:cd}

While this chapter mainly focused on \emph{atomicity} and \emph{isolation} from
the ACID properties, we provide a short overview of consistency and durability.

  {\bf Durability} is a hard requirement for FinBench Transaction and checking it
is part of the auditing process. The durability test requires the execution of
the LDBC FinBench transaction workload and uses the LDBC FinBench driver. Note,
the database and the driver must be configured in the same way as would be used
in the performance run. The durability test is executed as follows:

\begin{enumerate}[label={(\roman*)}]
  \item Execute the LDBC FinBench transaction workload;
  \item After 2 hours of execution, terminate all database processes
        \texttt{ungracefully}. This can be done by shutting down the entire
        machines or killing processes forcefully. Note, the ungraceful shutdown
        on different machines may differ:
        \begin{enumerate}[label={(\alph*)}]
          \item \emph{Amazon Web Services}: Using the \texttt{AWS CLI} to force
                stop the instance:
                \texttt{aws ec2 stop-instances --instance-ids \{ID\} --force};
          \item \emph{Alibaba Cloud}: Stopping the instance by \texttt{Force Stop}
                option on the \texttt{ECS Console};
          \item \emph{Bare Metal}: Force stop the machine by \texttt{poweroff -f}.
                Note, \texttt{shutdown -h now or shutdown -r now} are graceful;
          \item \emph{Others}: Depends on discussion.
        \end{enumerate}
  \item Restart the database system, retrieve the last entities (vertices or edges)
        updated by the last update operations before the crash from the driver
        logs;
  \item Issue read queries to get the value of the last entities. If the returned
        data matches the committed data according to the logs, the system
        passes the durability test.
\end{enumerate}

{\bf Consistency} is defined in terms of constraints: the database remains
consistent under updates; i.e. no constraint is violated. Relational database
systems usually support primary- and foreign-key constraints, as well as domain
constraints on column values and sometimes also support simple within-row
constraints. Graph database systems have a diversity of interfaces and generally
do not support constraints, beyond sometimes domain and primary key constraints
(in case indices are supported). However, we do note that we anticipate that
graph database systems will evolve to support constraints in the future. Beyond
equivalents of the relational ones, property graph systems might introduce
graph-specific constraints, such as (partial) compliance to a schema formulated
on top of property graphs, rules that guide the presence of labels or structural
graph constraints such as connectedness of the graph, absence of cycles, or
arbitrary well-formedness constraints~\cite{DBLP:journals/sosym/SemerathBHSV17}.
Here we provide an example of a consistency test (the consistency test also
requires the execution of the LDBC FinBench transaction workload and uses the
LDBC FinBench driver):

\begin{enumerate}[label={(\roman*)}]
  \item Add some precomputed properties (similar to materialized views) for vertex
        or edge. i.e. add property \emph{balance} for \emph{account}, which
        maintains the balance of the given account according to the associated
        transactions, and at the same time, the update queries need to be modified
        to maintain the balance. You can also design other constraints(i.e. vertice
        uniqueness);
  \item Execute the LDBC FinBench transaction workload;
  \item After 1 hour of execution, pause the execution of the workload;
        Issue read queries to check if the constraints are consistent after
        updating;
  \item Resume the execution of the workload. After another 1 hour of execution,
        terminate all database processes ungracefully;
  \item Restart the database system, Issue read queries to check if the
        constraints are consistent after recovery;
  \item If both of the above checks pass, the system passes the
        consistency test.
\end{enumerate}

\chapter{Auditing Rules}
\label{sec:auditing-rules}

\emph{This chapter contains the auditing policies for the LDBC Benchmarks. The initial draft of the auditing policies was published in the EU project deliverable D6.3.3 ``LDBC Benchmark Auditing Policies''.}


This chapter is divided into the following parts:
\begin{itemize}
    \item Motivation of benchmark result auditing
    \item General discussion of auditable aspects of benchmarks
    \item Specific checklists and running rules for \ldbcfinbench workloads
\end{itemize}

Many definitions and general considerations are shared between the benchmarks, hence it is justified to present the
principles first and to refer to these in the context of the benchmark-specific rules. The auditing process, including
the auditor certification exams, the possibility of challenging audited results, \etc, are defined in the LDBC
Byelaws~\cite{ldbc_byelaws}. Please refer to the latest Byelaws document when conducting audits.


\section{Rationale and General Principles}


The purpose of benchmark auditing is to improve the \emph{credibility} and \emph{reproducibility} of benchmark claims by involving a set of detailed execution rules and third-party verification of compliance with these.

Rules may exist separately from auditing but auditing is not meaningful unless the rules are adequately precise.
Aspects like auditor training and qualification cannot be addressed separately from a discussion of the matters the
auditor is supposed to verify. Thus, the credibility of the entire process hinges on a clear and shared understanding
of what a benchmark is expected to demonstrate and on the auditor being capable of understanding the process
and verifying that the benchmark execution is fair and does not abuse the rules or pervert the objectives of
the benchmark.

Due to the open-ended nature of technology and the agenda of furthering innovation via measurement, it is
not feasible or desirable to over-specify the limits of benchmark implementation. Hence, there will always remain
judgment calls for borderline cases. In this respect auditing and the LDBC are not separate. It is expected that
issues of compliance, as well as maintenance of rules, will come before the LDBC as benchmark claims are
made.


\section{Auditing Rules Overview}

\subsection{Auditor Training, Certification, and Selection}
\subsubsection{Auditor Training}
Auditor training consists of familiarization with the benchmark and existing implementations thereof. This involves the auditor candidate running the reference implementations of the benchmark to see what is normal behavior and practice in the workload. The training and practice may involve communication with the benchmark task force for clarifying the intent and details of the benchmark rules. This produces feedback for the task force for further specification of the rules.

\subsubsection{Auditor Certification}
The auditor certification and qualification are done in the form of an examination administered by the task force responsible for the benchmark being audited. The examination may be carried out by teleconference. The task force will subsequently vote on accepting each auditor, by a simple majority. An auditor is certified for a particular benchmark by the task force maintaining the benchmark in question.

\subsubsection{Auditor Selection}
In the default auditor selection, the task force responsible for the benchmark being audited appoints a third-party, impartial auditor. \emph{If needed, a Conflict of Interest Statement will be signed and provided.} The task force may in special cases appoint itself as auditor of a particular result. This is not,
however, the preferred course of action but may be done if no suitable third-party auditor is available.

\subsection{Auditing Process Stages}
\subsubsection{Getting Ready for a Benchmark Audit}
A benchmark result can be audited if it is a \emph{complete implementation} of an LDBC benchmark workload. This includes implementing all operations correctly, using official data sets, using the official LDBC driver (if available), and complying with the auditing rules of the workload (\eg workloads may have different rules regarding query languages, the allowance of materialized views, \etc).
Workloads may specify further requirements such as ACID compliance (checked using the LDBC FinBench ACID test suite).

\subsubsection{Performing a Benchmark Audit}
A benchmark result is to be audited by an LDBC-appointed auditor or the LDBC task force managing the benchmark. An LDBC audit may be performed by remote login and does not require the auditor's physical presence on site. The test sponsor shall grant the auditor any access necessary for validating the benchmark run. This will typically include administrator access to the SUT hardware.

\subsubsection{Benchmark-Specific Checklist}
Each benchmark specifies a checklist to be verified by the auditor. The benchmark run shall be performed by the auditor. The auditor shall make copies of relevant configuration files and test results for future checking and insertion into the full disclosure report.

\subsubsection{Producing the FDR}
The FDR is produced by the auditor or auditors, with any required input from the test sponsor. Each non-default configuration parameter needs to be included in the FDR and justification needs to be provided why the given parameter was changed.
The auditor produces an attestation letter that verifies the authenticity of the presented results. This letter is to be included in the FDR as an addendum. The attestation letter has no specific format requirements but shall state that the auditor has established compliance with a specified version of the benchmark specification.

\subsubsection{Publishing the FDR}
The FDR and any benchmark-specific summaries thereof shall be published on the LDBC website, \url{https://ldbcouncil.org/}.

\subsection{Challenge Procedure}

A benchmark result may be \emph{challenged} for non-compliance with LDBC rules. The benchmark task force responsible for the maintenance of the benchmark will rule on matters of compliance. A result found to be non-compliant will be withdrawn from the list of official LDBC benchmark results.



\section{Auditable Properties of Systems and Benchmark Implementations}

\subsection{Validation of Query Results}
\label{sec:validation}
A benchmark should be published with a deterministically reproducible validation data set. Validation queries applied to the validation data set will deterministically produce a set of correct answers. This is used in the first stage of the benchmark run to test for the correctness of A SUT or benchmark implementation. This validation stage is not timed.

\paragraph{Inputs for validation}
The validation takes the form of a set of data generator parameters, a set of test queries that at least include one instance of each of the workload query templates and the expected results.

\paragraph{Approximate results and error margin}
In certain cases, the results may be approximate. This may happen in cases of non-unique result ordering keys, imprecise numeric data types, random behaviors in certain graph analytics algorithms etc. Therefore, a validation set shall specify the degree of allowable error: For example, for counts, the value must be exact, for sums, averages and the like, at least 8 significant digits are needed, for statistical measures like graph centralities, the result must be within 1\% of the reference result. Each benchmark shall specify its expectation in an unambiguously verifiable manner.

\subsection{ACID Compliance}
\label{sec:acid-compliance}

As part of the auditing process for the Transaction workload, the auditors ascertain that the SUT satisfies the ACID properties,
\ie it provides atomic transactions, complies with its claimed isolation level, and ensures durability in case of failures.
This section outlines the transactional behaviors of SUTs which are checked in the course of auditing A SUT in a given benchmark.

A benchmark specifies transactional semantics that may be required for different parts of the workload. The requirements will typically be different for the initial bulk load of data and for the workload itself. Different sections of the workload may further be subject to different transactionality requirements.

No finite series of tests can prove that the ACID properties are fully supported. Passing the specified tests is a necessary, but not sufficient, condition for meeting the ACID requirements. However, for fairness of reporting, only the tests specified here are required and must appear in the FDR for a benchmark. (This is taken exactly from the \mbox{TPC-C} specification~\cite{tpcc}.)

The properties for ACID compliance are defined as follows:

\paragraph{Atomicity}
Either all the effects of the transaction are in effect after the transaction or none of the effects
is in effect. This is by definition only verifiable after a transaction has finished.

\paragraph{Consistency}
ADS such as secondary indices will be consistent among themselves as well as with the table or other PDS, if any. Such a consistency (compliance to all constraints, if these are declared in the schema, \eg primary key constraint, foreign key constraints and cardinality constraints) may be verified
after the commit or rollback of a transaction. If a single thread of control runs within a transaction, then
subsequent operations are expected to see a consistent state across all data indices of a table
or similar object. Multiple threads which may share a transaction context are not required to observe a
consistent state at all times during the execution of the transaction. Consistency will however always be
verifiable after the commit or rollback of any transaction, regardless of the number of threads that have
either implicitly or explicitly participated in the transaction. Any intra-transaction parallelism introduced
by the SUT will preserve transactional semantics statement-by-statement. If explicit, application created
sessions share a transaction context, then this definition of consistency does not hold: for example, if
two threads insert into the same table at the same time in the same transaction context, these may or may
not see a consistent image of (E)ADS for the parts affected by the other thread. All things will be
consistent after the commit or rollback, however, regardless of the number of threads, implicit or explicit
that have participated in the transaction.

\paragraph{Isolation}
Isolation is defined as the set of phenomena that may (or may not) be observed by operations running within a single transaction context. The levels of isolation are defined as follows:

\begin{description}
    \item[Read uncommitted] No guarantees apply.
    \item[Read committed] A transaction will never read a value that has at no point in time been part of a
        committed state.
    \item[Repeatable read] If a transaction reads a value several times during its execution, then it will see
        the original state with its modifications so far applied to it. If the transaction itself consists of
        multiple reading and updating threads then the ambiguities that may arise are beyond the scope of transaction isolation.
    \item[Serializable] The transactions see values that correspond to a fully serial execution of
        all client transactions. This is like a repeatable read except that if the transaction reads something, and
        repeats the read, it is guaranteed that no new values will appear for the same search condition on a
        subsequent read in the same transaction context. For example, a row that was seen not to exist when
        first checked will not be seen by a subsequent read. Likewise, counts of items will not be seen to
        change.
\end{description}

\paragraph{Durability}
Durability means that once the SUT has confirmed a successful commit, the committed state
will survive any instantaneous failure of the SUT (\eg a power failure, software crash, reboot or
the like). Durability is tied to atomicity in that if one part of the changes made by a transaction survives then
all parts must survive. 

\subsection{Data Format and Preprocessing}
\label{sec:auditing-data-format}

When producing the data sets, implementers are allowed to use custom formatting options (\eg use or omission of quotes, separator character, datetime format, \etc).
It is also allowed to convert the output of the DataGen into a format (\eg Parquet) that is loadable by the test-specific implementation of the data importer.
Additional preprocessing steps are also allowed, including adjustments to the CSV files (\eg with shell scripts), splitting and concatenating files, compressing and decompressing files, \etc
However, the preprocessing step shall not include a precomputation of (partial) query results.

\subsection{Query Languages}
\label{sec:query-languages}

In typical RDBMS benchmarks, online transaction processing (OLTP) benchmarks are allowed to be implemented via stored procedures, effectively amounting to explicit query plans.
Meanwhile, online analytical processing (OLAP) benchmarks prohibit the use of using general-purpose programming languages (\eg C, C\texttt{++}, Java) for query implementations and only allow domain-specific query languages.

In the graph processing space, there is currently (as of 2022) no standard query language and the systems are considerably more heterogeneous.
Therefore, the LDBC situation regarding declarative is not as simple as that of for example the \mbox{TPC-H} (where queries should be specified in SQL with the additional constraint of omitting any hints for OLAP workloads) and individual FinBench workloads specify their policy of either requiring a domain-specific query language or allowing the implementation of the queries in a general-purpose programming language.

In the case of domain-specific languages, systems are allowed to implement a FinBench query as a sequence of multiple queries.
A typical example of this is the following sequence:
(1)~create a projected graph,
(2)~run query,
(3)~drop projected graph.
However, it is not allowed to use sub-queries in an unrealistic and contrived manner, \ie the goal of overcoming optimization issues, \eg hard-coding a certain join order in a declarative query language.
It is the responsibility of the auditor to determine whether a sequence of queries can be considered realistic w.r.t.\ how a user would formulate their queries in the language provided by the system.

\subsubsection{Rules for Imperative Implementations Using a General-Purpose Programming Language}
An implementation where the queries are written in a general-purpose programming language (including imperative and ``API-based'' implementations) may choose between semantically equivalent implementations of an operation based on the query parameters. This simulates the behavior of a query optimizer in the presence of literal values in the query. If an implementation does this, all the code must be disclosed as part of the FDR and the decision must be based on values extracted from the database, not on hard-coded threshold values in the implementation.

The auditor must be able to reliably assess the compliance of implementation to guidelines specifying these matters. The actual specification remains benchmark-dependent. Borderline cases may be brought to the task force responsible for arbitration.

\subsubsection{Disclosure of Query Implementations in the FDR}
Benchmarks allowing imperative expression of workload should require full disclosure of all query implementation code.

\subsection{Materialization}

The mix of read and update operations in a workload will determine to which degree precomputation of results is beneficial. The auditor must check that materialized results are kept consistent at the end of each transaction.

\subsection{System Configuration and System Pricing}
\label{sec:system-config}


A benchmark execution shall produce a full disclosure report which specifies the hardware and software of the SUT, the benchmark implementation version and any specifics that are detailed in the benchmark specification. This clause gives a general minimum for disclosure for the SUT.

\subsubsection{Details of Machines Driving and Running the Workload}
A SUT may consist of one or more pieces of physical hardware. A SUT may include virtual or bare-metal machines in a cloud service.
For each distinct configuration, the FDR shall disclose the number of units of the type as well as the following:

\begin{enumerate}
    \item The used cloud provider (including the region where machines reside, if applicable).
    \item Common name of the item, \eg Dell PowerEdge xxxx or i3.2xlarge instance.
    \item Type and number of CPUs, cores \& threads per CPU, clock frequency, cache size.
    \item Amount of memory, type of memory and memory frequency, \eg 64GB DDR3 1333MHz.
    \item Disk controller or motherboard type if the disk controller is on the motherboard.
    \item For each distinct type of secondary storage device, the number and specification of the device, \eg 4xSeagate Constellation 2TB SATA 6Gbit/s.
    \item Number and type of network controllers, \eg 1x Mellanox QDR InfiniBand HCA, PCIE 2.0, 2x1GbE on motherboard. If the benchmark execution is entirely contained on a single machine, it must be stated, and the description of network controllers can be omitted.
    \item Number and type of network switches. If multiple switches are used, the wiring between the switches should be disclosed.
          Only the network switches and interfaces that participate in the run need to be reported. If the benchmark execution is entirely contained on a single machine, it must be stated, and the description of network switches can be omitted.
    \item Date of availability of the system as a whole, \ie the latest date of availability of any part.
\end{enumerate}

\subsubsection{System Pricing}
The price of the hardware in question must be disclosed. For cloud setups, the price of a dedicated instance for 3 years must be disclosed. The price should reflect the single quantity list price that any buyer could expect when purchasing one system with the given specification. The price may be either an item-by-item price or a package price if the system is sold as a package.
Reported prices should adhere to the TPC Pricing Specification 2.7.0~\cite{pricing,tpc-pricing}.
It is particularly important to ensure that the maintenance contract guarantees 24/7 support and 4~hour response time for problem recognition.

\subsubsection{Details of Software Components in the System}
The SUT software must be described at least as follows:
\begin{enumerate}
    \item The units of the SUT software are typically the DBMS and operating system.
    \item Name and version of each separately priced piece of the SUT software.
    \item If the price of the SUT software is tied to the platform or the count of concurrent users, these parameters must be disclosed.
    \item Price of the SUT software.
    \item Date of availability.
\end{enumerate}
Reported prices should adhere to the TPC Pricing Specification 2.5.0~\cite{pricing,tpc-pricing}.

The configuration of the SUT must be reported to include the following:
\begin{enumerate}
    \item The used LDBC specification, driver and data generator version.
    \item Complete configuration files of the DBMS, including any general server configuration files, any configuration scripts run on the DBMS for setting up the benchmark run etc.
    \item Complete schema of the DBMS, including eventual specification of storage layout.
    \item Any OS configuration parameters if other than default, \eg \verb+vm.swappiness+, \verb+vm.max_map_count+ in Linux.
    \item Complete source code of any server-side logic, \eg stored procedures, triggers.
    \item Complete source code of driver-side benchmark implementation.
    \item Description of the benchmark environment, including software versions, OS kernel version, DBMS version as well as versions of other major software components used for running the benchmark (Docker, Java Virtual Machine, Python, etc.).
    \item The SUT's highest configurable isolation level and the isolation level used for running the benchmark.
\end{enumerate}

\subsubsection{Audit of System Configuration}
The auditor must ascertain that a reported run has indeed taken place on the SUT in the disclosed configuration.
The full disclosure shall contain any relevant parameters of the benchmark execution itself, including:
\begin{enumerate}
    \item Parameters, switches, configuration file for data generation.
    \item Complete text of any data loading script or program.
    \item Parameters, switches, configuration files for any test driver. If the test driver is not an LDBC supplied open source package or is a modification of such, then the complete text or diff against a specific LDBC package must be disclosed.
    \item Test driver output files shall be part of the disclosure. In general, these must at least detail the following:
          \begin{enumerate}[label=\roman*)]
              \item Time and duration of data load and the timed portion of the benchmark execution.
              \item Count of each workload item (\eg query, transaction) successfully executed within the measurement window.
              \item Min/average/max execution time of each workload item, the specific benchmark shall specify additional details.
          \end{enumerate}
\end{enumerate}

Given this information, the number of concurrent database sessions at each point in the execution must be clearly stated. In the case of a cluster database, the possible spreading of connections across multiple server processes must be disclosed.

All parameters included in this section must be reported in the full disclosure report to guarantee that the benchmark run can be reproduced exactly in the future. Similarly, the test sponsor will inform the auditor of the scale factor to test. Finally, a clean test system with enough space to store the initial data set, the update streams, substitution parameters and anything that is part of the input and output as well as the benchmark run must be provided.

\subsection{Benchmark Specifics}

Similarly to TPC benchmarks, the LDBC benchmarks prohibit so-called benchmark specials (\ie extra software modules implemented in the core DBMS logic just to make a selected benchmark run faster are disallowed). Furthermore, upon request of the auditor, the test sponsor must provide all the source codes relevant to the benchmark.


\section{Auditing Rules for the Transaction Workload}

\begin{figure}[h]
    \centering
    \includegraphics[scale=\yedscale]{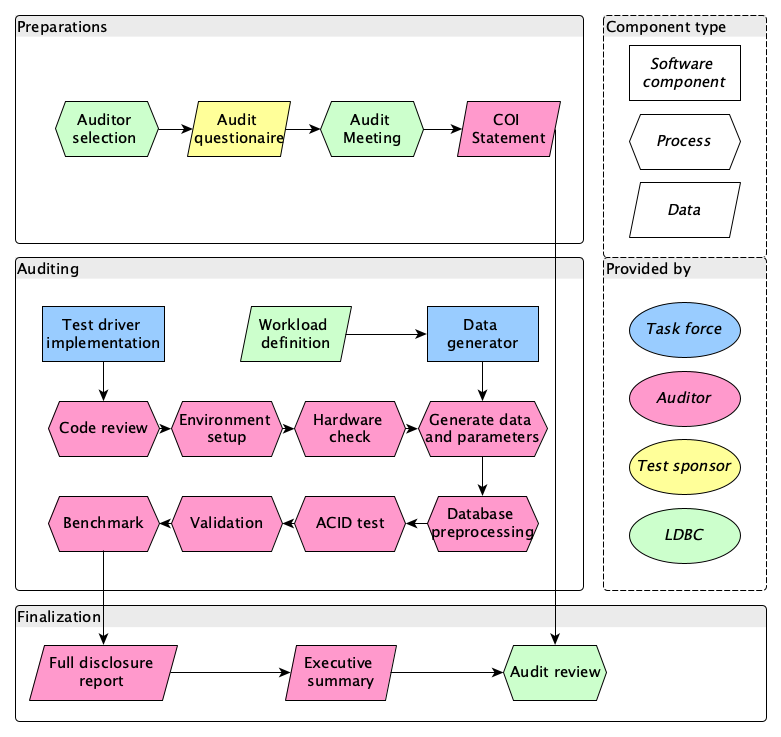}
    \caption{Benchmark execution and auditing workflow. For non-audited runs, the implementers perform the steps of the auditor.}
    \label{fig:audit-workflow}
\end{figure}

This section specifies a checklist (in the form of individual sections) that a benchmark audit shall cover in case of the FinBench Transaction workload. An overview of the benchmark audit workflow is shown in \autoref{fig:audit-workflow}. The three major phases of the audit are preparing the input data and validation query results (captured by \emph{Preparations} in the figure), validating the correctness of query results returned by the SUT using the validation scale factor and running the benchmark with all the prescribed workloads (\emph{Benchmarking}), and creating the FDR (\emph{Finalization}). The color codes capture the responsibilities of performing a step or providing some data in the workflow.

A key objective of the auditing guidelines for the Transaction workload is to \emph{allow a broad range of systems} to implement the benchmark.
Therefore, they do not impose constraints on the data model
(graph, relational, triple, \etc representations are allowed)
or on the query language
(both declarative and imperative languages are allowed).

\subsection{Scaling Factors}
\label{sec:transaction-workload-scaling}

The scale factor of a FinBench data set is the size of the data set in GiB of CSV (comma-separated values) files.
The size of a data set is characterized by scale factors: SF0.1, SF1, SF3 \etc (see \autoref{sec:scale-factors}).
All data sets contain data for three years of financial activities.

The \emph{validation run} shall be performed on the SF1 data set (see \autoref{sec:transaction-workload-validation-data-set}). Note that the auditor may perform additional validation runs of the benchmark implementation using smaller data sets (\eg SF1) and issue queries.

Audited \emph{benchmark runs} of the Transaction workload shall use SF10. The rationale behind this decision is to ensure that there is a sufficient number of update operations available to guarantee 2.5~hours of continuous execution (see \autoref{sec:transaction-workload-measurement-window}).

\subsection{Data Model}

FinBench may be implemented with different data models (\eg relational, RDF, and different graph data models). The reference schema is provided in the specification using a UML-like notation.

\subsection{Precomputation}

Precomputation of query results (both interim and end results) is allowed. However, systems must ensure that precomputed results (\eg materialized views) are kept consistent upon updates.

\subsection{Benchmark Software Components}
\label{sec:finbench-software-components}
LDBC provides a test driver, data generator, and summary reporting scripts. Benchmark implementations shall use a stable
version of the test driver. The SUT's database software should be a stable version that is available publicly or can be
purchased at the time of the release of the audit. Please see \autoref{sec:software-components} for more details.

\subsubsection{Adaptation of the Test Driver to a DBMS}
\label{sec:test-driver}
A qualifying run must use a test driver that adapts the provided test driver to interface with the SUT. Such an implementation, if needed, must be provided by the test sponsor. The parameter generation, result recording, and workload scheduling parts of the test driver should not be changed. The auditor must be given access to the test driver source code used in the reported run.

The test driver produces the following artifacts for each execution as a by-product of the run: Start and end timestamps in wall clock time, recorded with microsecond precision. The identifier of the operation and any substitution parameters.

\subsubsection{Summary of Benchmark Results}
\label{sec:performance-metrics}
A separate test summary tool provided with the test driver analyses the test driver log(s) after a measurement window is completed.

The tool produces for each of the distinct queries and transactions the following summary:
\begin{itemize}
    \item Run time of query in wall clock time.
    \item Count of executions.
    \item Minimum/mean/percentiles/maximum execution time.
    \item Standard deviation from the average execution time.
\end{itemize}
The tool produces for the complete run the following summary:
\begin{itemize}
    \item Operations per second for a given SF (throughput). This is the primary metric of this workload.
    \item The total execution time in wall clock time.
    \item The total number of completed operations.
\end{itemize}

\subsection{Implementation Language and Data Access Transparency}

The queries and updates may be implemented in a domain-specific query language or as procedural code written in a general-purpose programming language (\eg using the API of the database).

\subsubsection{Implementations Using a Domain-Specific Query Language}
\label{sec:finbench-domain-specific-query-language}

If a domain-specific query language is used, \eg SPARQL, SQL, Cypher, or Gremlin, then explicit query plans are prohibited in all read-only queries.%
\footnote{If the queries are not declarative clearly, the auditor must ensure that they do not specify explicit query plans by investigating their source code and experimenting with the query planner of the system (\eg using SQL's \texttt{EXPLAIN} command).}
The update transactions may still consist of multiple statements, effectively amounting to explicit plans.

Explicit query plans include but are not limited to:
\begin{itemize}
    \item Directives or hints specifying a join order or join type
    \item Directives or hints specifying an access path, \eg which index to use
    \item Directives or hints specifying an expected cardinality, selectivity, fanout or any other information that pertains to the expected number of results or cost of all or part of the query.
\end{itemize}

\begin{quote}
    \emph{Rationale behind the applied restrictions.} The updates are effectively OLTP and, therefore, the customary freedoms apply, including the use of stored procedures, however subject to access transparency. Declarative queries in a benchmark implementation should be such that they could plausibly be written by an application developer. Therefore, their formulation should not contain system-specific aspects that an application developer would be unlikely to know. In other words, making a benchmark implementation should not require uncommon sophistication on behalf of the developer. This is a regular practice in analytical benchmarks, \eg \mbox{TPC-H}.
\end{quote}

\subsubsection{Implementations Using a General-Purpose Programming Language}
\label{sec:finbench-general-purpose-programming-language}

Implementations using a general-purpose programming language for specifying the queries (including procedural, imperative, and API-based implementations) are expected to respect the rules described in \autoref{sec:query-languages}.
For these implementations, the rules in \autoref{sec:finbench-domain-specific-query-language} do not apply.

\subsection{Correctness of Benchmark Implementation}

\subsubsection{Validation data set}
\label{sec:transaction-workload-validation-data-set}
The scale factor 1 shall be used as a validation data set.

\subsubsection{ACID Compliance}
\label{sec:transaction-workload-acid-compliance}

The Transaction workload requires full ACID support (\autoref{sec:acid-compliance}) from the SUT.
This is tested using the LDBC ACID test suite.
For the specification of this test suite, see \autoref{sec:acid-test} and the related software repository at \url{https://github.com/ldbc/ldbc_finbench_acid}.

\paragraph{Expected level of isolation}
If a transaction reads the database with the intent to update, the DBMS must guarantee no dirty reads. In other words, this
corresponds to read committed isolation.

\paragraph{Durability and checkpoints}

A checkpoint is defined as the operation which causes data persisted in a transaction log to become durable outside the transaction log. Specifically, this means that A SUT restart after instantaneous failure following the completion of the checkpoint may not have recourse to transaction log entries written before the end of the checkpoint.

A checkpoint typically involves a synchronization barrier at which all data committed before the moment is required to be in durable storage that does not depend on the transaction log.
Not all DBMSs use a checkpoint mechanism for durability. For example, a system may rely on redundant storage of data for durability guarantees against the instantaneous failure of a single server.

The measurement window may contain a checkpoint. If the measurement window does not contain one, then the restart test will involve redoing all the updates in the window as part of the recovery test.

The timed window ends with an instantaneous failure of the SUT. Instantaneously killing all the SUT process(es) is adequate for simulating instantaneous failure. All these processes should be killed within one second of each other with an operating system action equivalent to the Unix \verb+kill -9+. If such is not available, then powering down each separate SUT component that has an independent power supply is also possible.

The restart test consists of restarting the SUT process(es) and finishes when the SUT is back online with all its functionality and the last successful update logged by the driver can be seen to be in effect in the database.

If the SUT hardware was powered down, the recovery period does not include the reboot and possible file system check time. The recovery time starts when the DBMS software is restarted.

\paragraph{Recovery}
The SUT is to be restarted after the measurement window and the auditor will verify that the SUT contains the entirety of the last update recorded by the test driver(s) as successfully committed. The driver or the implementation has to make this information available. The auditor may also check the \emph{audit log} of the SUT (if available) to confirm that the operations issued by the driver were saved.

Once an official run has been validated, the recovery capabilities of the system must be tested. The system and the driver must be configured in the same way as in during the benchmark execution. After a warm-up period, execution of the benchmark will be performed under the same terms as in the previous measured run.

\paragraph{Measuring recovery time}
At an arbitrary point close to 2 hours of wall clock time during the run, the machine will be shut down. Then, the auditor will restart the database system and will check that the last committed update (in the driver log file) is actually in the database. The auditor will measure the time taken by the system to recover from the failure. Also, all the information about how durability is ensured must be disclosed. If checkpoints are used, these must be performed for a period of 10 minutes at most.

\subsection{Benchmarking Workflow}
\label{sec:transaction-workload-benchmark-workflow}

A benchmark execution is divided into the following processes (these processes are also shown in \autoref{fig:audit-workflow}):

\begin{description}
    \item[Generate data] This includes running the data generator, placing the generated files in a staging area,
        configuring storage, setting up the SUT configuration and preparing any data partitions in the SUT. This may include
        preallocating database space but may not include loading any data or defining any schema having to do with the
        benchmark.
    \item[Preprocessing] If needed, the output from the data generator is to preprocess the data set (\autoref{sec:auditing-data-format}).
    \item[Create validation data] Using one of the reference implementations of the benchmark, the reference validation data is obtained in JSON format.
    \item[Data loading] The test sponsor must provide all the necessary documentation and scripts to load the data set
        into the database to test. This includes defining the database schema, if any, loading the initial database
        population, making this durably stored and gathering any optimizer statistics. The system under test must support
        the different data types needed by the benchmark for each of the attributes at their specified precision. No data
        can be filtered out, everything must be loaded. The test sponsor must provide a tool to perform arbitrary checks of
        the data or a shell to issue queries in a declarative language if the system supports it.
    \item[Run cross-validation] This step uses the data loader to populate the database, but the load is not timed. The
        validation data set is used to verify the correctness of the SUT. The auditor must load the provided data set and run the driver in validation mode, which will test that the queries provide the
        official results.  The benchmarking workflow will not go beyond this point unless the results match the expected
        output.
    \item[Warm-up] Benchmark runs are preceded by a warm-up which must be performed using the LDBC driver.
    \item[Run benchmark] The bulk load time is reported and is equal to the amount of elapsed wall clock time between
        starting the schema definition and receiving the confirmation message of the end of statistics gathering. The
        workflow runs begin after the bulk load is completed. If the run does not directly follow the bulk load, it must
        start at a point in the update stream that has not previously been played into the database. In other words, a run
        may only include update events whose timestamp is later than the latest message creation date in the database before
        the start of the run. The run starts when the first of the test drivers sends its first message to the SUT. If the
        SUT is running in the same process as the driver, the window starts when the driver starts. Also, make sure that
        the \verb|-rl/--results_log| is enabled. Make sure that all operations are enabled, and the frequencies are
        those for the selected scale factor (see the exact specification of the frequencies in
        \autoref{sec:sf-statistics}).
\end{description}

\subsubsection{Query Timing During Benchmark Run}
\label{sec:ontime-requirements}
A valid benchmark run must last at least 2 hours of wall clock time and at most 2 hours and 15 minutes.
In order to be valid, a benchmark run needs to meet the ``95\% on-time requirement''.
The \texttt{results\_log.csv} file contains the $\mathsf{actual\_start\_time}$ and the $\mathsf{scheduled\_start\_time}$ of each of the issued queries. To have a valid run, 95\% of the queries must meet the following condition:
\begin{equation*}
    \mathsf{actual\_start\_time} - \mathsf{scheduled\_start\_time} < 1\
    \mathrm{second}
\end{equation*}

If the execution of the benchmark is valid, the auditor must retrieve all the files from the directory specified by
\verb|--results_dir| which includes configuration settings used, results log and results summary. All of which must be
disclosed.

\subsubsection{Measurement Window}
\label{sec:transaction-workload-measurement-window}

\begin{figure}[h]
    \centering
    \includegraphics[width=.7\linewidth]{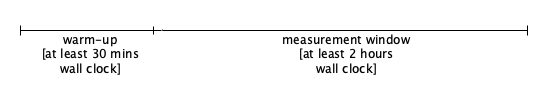}
    \caption{Warm-up and measurement window for the benchmark run.}
    \label{fig:measurement-window-selection}
\end{figure}

Benchmark runs execute the workload on the SUT in two phases (\autoref{fig:measurement-window-selection}). First, the
SUT must undergo a warm-up period that takes at least 30 minutes and at most 35 minutes. The goal of this is to put the
system in a steady state which reflects how it would behave in a normal operating environment. The performance of the
operations during warm-up is not considered. Next, the SUT is benchmarked during a two-hour measurement window.
Operation times are recorded and checked to ensure the ``95\% on-time requirement'' is satisfied.

The FinBench \DataGen produces 3~years worth data of which 3\% is used for updates
(\autoref{sec:transaction-workload-data-sets}), \ie approximately $3 \times 365 \times 0.03 = 32.85~\text{days} =
    788.4~\text{hours}$. To ensure that the 2.5~hours wall clock period has enough input data, the lower bound of TCR is
defined as 0.001 (if $2628$ hours of updates are played back at more than $1000\times$ speed, the benchmark framework
runs out of updates to execute). A system that can achieve a better compression (\ie lower TCR value) on a given scale
factor should use larger SFs for their benchmark runs -- otherwise their total runs will be less than 2.5~hours, making
them unsuitable for auditing.


\subsection{Full Disclosure Report}
\label{sec:transaction-workload-fdr}

Upon successful completion of the audit, an FDR is compiled. In addition to the general requirements, the full disclosure shall cover the following:

\begin{itemize}
    \item General terms: an executive summary and declaration of the credibility of the audit
    \item Conflict of Interest Statement between the auditor and the test sponsor, if needed.
    \item System description and pricing summary
    \item Data generation and data loading
    \item Test driver details
    \item Performance metrics
    \item Validation results
    \item ACID compliance
    \item List of supplementary materials
\end{itemize}

To ensure the reproducibility of the audited results, a supplementary package is attached to the full disclosure report. This package should contain:

\begin{itemize}
    \item A README file with instructions specifying how to set up the system and run the benchmark
    \item Configuration files of the database, including database-level configuration such as buffer size and schema descriptors (if necessary)
    \item Source code or binary of a generic driver that can be used to interact with the DBMS
    \item SUT-specific LDBC driver implementation (similarly to the projects in \url{https://github.com/ldbc/ldbc_finbench_transaction_impls})
    \item Script or instructions to compile the LDBC Java driver implementation
    \item Instructions on how to reach the server through CLI and/or web UI (if applicable), \eg the URL (including port number), username and password
    \item LDBC configuration files (\texttt{.properties}), including the \texttt{time\_compression\_ratio} values used in the audited runs
    \item Scripts to preprocess the input files (if necessary) and to load the data sets into the database
    \item Scripts to create validation data sets and to run the benchmark
    \item The implementations of the queries and the update operations, including their complete source code (\eg declarative queries specifications, stored procedures, \etc)
    \item Implementation of the ACID test suite
    \item Binary package of the DBMS (\eg \texttt{.deb} or \texttt{.rpm})
\end{itemize}


\chapter{Related Work}
\label{sec:related-work}

\begin{quote}
    \textit{A detailed list of LDBC publications is curated at~\url{https://ldbcouncil.org/publications}.}
\end{quote}



\ldbcfinbench is designed based on the \ldbcsnb and introduces the new features in financial scenarios.

\appendix

\chapter{Choke Points}
\label{sec:choke-points}

\newcommand{\tpcCard}[1]{\colorbox{lightgray}{\tt TPC-H #1}}
\newcommand{\tpcCPSection}[4][]{%
\subsection*{%
CP-#2: [#3] #4%
\ifthenelse{\equal{#1}{}}{}{\hfill \tpcCard{#1}}%
}%
\label{choke_point_#2}}

\newcommand{\snbCPSection}[4][]{%
\subsection*{%
CP-#2: [#3] #4%
{\hfill {\colorbox{lightgray}{\tt From SNB}}}%
}%
\label{choke_point_#2}}

\newcommand{\newCPSection}[4][]{%
\subsection*{%
CP-#2: [#3] #4%
{\hfill {\colorbox{lightgray}{\tt New in FinBench}}}%
}%
\label{choke_point_#2}}


\section*{Introduction}

An interesting benchmark should be designed with representative read-world
scenarios and also chokepoints embedded in the deeper technical level.
Chokepoints capture particularly challenging aspects of queries. The
correlations between chokepoints and read queries are displayed in
\autoref{tab:query_choke_point}. To help understand the following chokepoints,
there are some annotations.
\begin{itemize}
    \item The capital abbreviations are short for the aspects the chokepoints
          affect. \begin{itemize}
              \item \emph{QOPT}: Those aimed at testing aspects of the query
                    optimizer.
              \item \emph{QEXE}: Those aimed at testing aspects of the execution
                    engine.
              \item \emph{STORAGE}: Those aimed at testing aspects of the
                    storage system.
              \item \emph{LANG}: Those aimed at testing aspects of the
                    expression capability of DSL.
              \item \emph{UPD}: Those aimed at testing aspects of the update
                    operation performance.
          \end{itemize}
    \item The gray boxes in the top right corner annotate the source of the
          chokepoints. \begin{itemize}
              \item \colorbox{lightgray}{\emph{TPC-H}} means the chokepoint is
                    from the paper \emph{TPC-H
                    Analyzed}~\cite{DBLP:conf/tpctc/BonczNE13}. You can refer to
                    the paper for the chokepoint details.
              \item \colorbox{lightgray}{\emph{From SNB}} means the chokepoint
                    refers to the ones in LDBC SNB~\cite{ldbc_snb_docs}.
              \item \colorbox{lightgray}{\emph{New in FinBench}} means the
                    chokepoint is summarized newly from FinBench.
          \end{itemize}
\end{itemize}


{
\setlength{\tabcolsep}{.05em}
\begin{table}[htbp]
\scriptsize
\caption{Coverage of choke points by queries.}
\label{tab:query_choke_point}
\end{table}
}


\section{Aggregation Performance}

\tpcCPSection[1.2]{1.1}{QOPT}{Interesting orders}

This choke point tests the ability of the query optimizer to exploit the
interesting orders induced by some operators. Apart from clustered indices
providing key order, other operators also preserve or even induce tuple
orderings. Sort-based operators create new orderings, typically on the probe-side
of a hash join conserves its order, etc.


\IfFileExists{choke-point-query-mapping/cp-1-1}{\paragraph{Queries}
{\raggedright
\queryRefCard{transaction-complex-read-05}{TCR}{5}

}}{}

\tpcCPSection[1.1]{1.2}{QEXE}{High cardinality group-by performance}

This choke point tests the ability of the execution engine to parallelize
group-by operations with a large number of groups. Some queries require
performing large group-by operations. In such a case, if an aggregation produces
a significant number of groups, intra-query parallelization can be exploited as
each thread may make its own partial aggregation. Then, to produce the result,
these have to be re-aggregated. In order to avoid this, the tuples entering the
aggregation operator may be partitioned by a hash of the grouping key and be
sent to the appropriate partition. Each partition would have its own thread so
that only that thread would write the aggregation, hence avoiding costly
critical sections as well. A high cardinality distinct modifier in a query is a
special case of this choke point. It is amenable to the same solution with
intra-query parallelization and partitioning as the group-by. We further note
that scale-out systems have an extra incentive for partitioning since this will
distribute the CPU and memory pressure over multiple machines, yielding better
platform utilization and scalability.


\IfFileExists{choke-point-query-mapping/cp-1-2}{\paragraph{Queries}
{\raggedright
\queryRefCard{transaction-complex-read-07}{TCR}{7}

}}{}

\snbCPSection{1.3}{QOPT}{Top-k pushdown}

This choke point tests the ability of the query optimizer to perform
optimizations based on top-$k$ selections. Many times queries demand for
returning the top-$k$ elements based on some property. Engines can exploit that
once $k$ results are obtained, extra restrictions in a selection can be added
based on the properties of the $k$th element currently in the top-$k$, being
more restrictive as the query advances, instead of sorting all elements and
picking the highest $k$.


\IfFileExists{choke-point-query-mapping/cp-1-3}{}{}

\tpcCPSection[1.3]{1.4}{QEXE}{Low cardinality group-by performance}

This choke point tests the ability to efficiently perform group-by evaluation
when only a very limited set of groups is available.  This can require special
strategies for parallelization, \eg pre-aggregation when possible. This case
also allows using special strategies for grouping like using array lookup if the
domain of keys is small.


\IfFileExists{choke-point-query-mapping/cp-1-4}{}{}


\section{Join Performance}

\tpcCPSection[2.3]{2.1}{QOPT}{Rich join order optimization}

This choke point tests the ability of the query optimizer to find optimal join
orders. A graph can be traversed in different ways. In the relational model,
this is equivalent to different join orders. The execution time of these orders
may differ by orders of magnitude. Therefore, finding an efficient join
(traversal) order is important, which in general, requires enumeration of all
the possibilities. The enumeration is complicated by operators that are not
freely re-orderable like semi-, \mbox{anti-,} and outer-joins. Because of this
difficulty most join enumeration algorithms do not enumerate all possible plans,
and therefore can miss the optimal join order. Therefore, this choke point tests
the ability of the query optimizer to find optimal join (traversal) orders.


\IfFileExists{choke-point-query-mapping/cp-2-1}{}{}

\tpcCPSection[2.4]{2.2}{QOPT}{Late projection}

This choke point tests the ability of the query optimizer to delay the
projection of unneeded attributes until late in the execution. Queries where
certain columns are only needed late in the query. In such a situation, it is
better to omit them from initial table scans, as fetching them later by row-id
with a separate scan operator, which is joined to the intermediate query result,
can save temporal space, and therefore I/O. Late projection does have a
trade-off involving locality, since late in the plan the tuples may be in a
different order, and scattered I/O in terms of tuples/second is much more
expensive than sequential I/O. Late projection specifically makes sense in
queries where the late use of these columns happens at a moment where the amount
of tuples involved has been considerably reduced; for example after an
aggregation with only few unique group-by keys or a top-$k$ operator.


\IfFileExists{choke-point-query-mapping/cp-2-2}{}{}
\snbCPSection{2.3}{QOPT}{Join type selection}

This choke point tests the ability of the query optimizer to select the proper
join operator type, which implies accurate estimates of cardinalities. Depending
on the cardinalities of both sides of a join, a hash or an index-based join
operator is more appropriate. This is especially important with column stores,
where one usually has an index on everything. Deciding to use a hash join
requires a good estimation of cardinalities on both the probe and build sides.
In TPC-H, the use of hash join is almost a foregone conclusion in many cases,
since an implementation will usually not even define an index on foreign key
columns. There is a break even point between index and hash based plans,
depending on the cardinality on the probe and build sides.


\IfFileExists{choke-point-query-mapping/cp-2-3}{}{}

\tpcCPSection[2.2]{2.4}{QOPT}{Sparse foreign key joins}

This choke point tests the performance of join operators when the join is
sparse. Sometimes joins involve relations where only a small percentage of rows
in one of the tables is required to satisfy a join. When tables are larger,
typical join methods can be sub-optimal. Partitioning the sparse table, using
Hash Clustered indices or implementing Bloom-filter tests inside the join are
techniques to improve the performance in such
situations~\cite{DBLP:journals/csur/Graefe93}.


\IfFileExists{choke-point-query-mapping/cp-2-4}{}{}

\snbCPSection{2.5}{QEXE}{Worst-case optimal joins}

This choke point tests the query engine's ability to use multi-way, worst-case
optimal joins to evaluate cyclic queries which are required to efficiently
compute some dense subgraphs such as the triangle, the 4-cycle, and the diamond
(4-cycle with a cross-edge). The absence of multi-way joins (\eg in systems
which only support binary joins), implies that join performance will be provably
suboptimal for cyclic queries.


\IfFileExists{choke-point-query-mapping/cp-2-5}{}{}

\snbCPSection{2.6}{QEXE}{Factorized query execution}

Query results produced by many-to-many joins often have redundancies when
represented as tuples. Factorization~\cite{DBLP:journals/sigmod/OlteanuS16}
provides a more compact (relational) representation by eliminating repetitions,
while still allowing some operations (\eg aggregation) to be performed without
flattening the relation.


\IfFileExists{choke-point-query-mapping/cp-2-6}{}{}


\section{Data Access Locality}

\tpcCPSection[3.3]{3.1}{QOPT}{Detecting correlation}

This choke point tests the ability of the query optimizer to detect data
correlations and exploiting them. If a schema rewards creating clustered
indices, the question then is which of the date or data columns to use as key.
In fact it should not matter which column is used, as range-propagation between
correlated attributes of the same table is relatively easy. One way is through
the creation of multi-attribute histograms after detection of attribute
correlation. With MinMax indices, range-predicates on any column can be
translated into qualifying tuple position ranges. If an attribute value is
correlated with tuple position, this reduces the area to scan roughly equally to
predicate selectivity.


\IfFileExists{choke-point-query-mapping/cp-3-1}{}{}

\snbCPSection{3.2}{STORAGE}{Dimensional clustering}

This choke point tests suitability of the identifiers assigned to entities by
the storage system to better exploit data locality. A data model where each
entity has a unique synthetic identifier, \eg RDF or graph models, has some
choice in assigning a value to this identifier. The properties of the entity
being identified may affect this, \eg type (label), other dependent properties,
\eg geographic location, date, position in a hierarchy, etc., depending on the
application. Such identifier choice may create locality which in turn improves
efficiency of compression or index access.


\IfFileExists{choke-point-query-mapping/cp-3-2}{\paragraph{Queries}
{\raggedright
\queryRefCard{transaction-complex-read-01}{TCR}{1}
\queryRefCard{transaction-complex-read-02}{TCR}{2}
\queryRefCard{transaction-complex-read-03}{TCR}{3}
\queryRefCard{transaction-complex-read-04}{TCR}{4}
\queryRefCard{transaction-complex-read-05}{TCR}{5}
\queryRefCard{transaction-complex-read-06}{TCR}{6}
\queryRefCard{transaction-complex-read-07}{TCR}{7}
\queryRefCard{transaction-complex-read-08}{TCR}{8}
\queryRefCard{transaction-complex-read-09}{TCR}{9}
\queryRefCard{transaction-complex-read-10}{TCR}{10}
\queryRefCard{transaction-complex-read-11}{TCR}{11}
\queryRefCard{transaction-complex-read-12}{TCR}{12}

}}{}

\snbCPSection{3.3}{QEXE}{Scattered index access patterns}

This choke point tests the performance of indices when scattered accesses are
performed. The efficiency of index lookup is very different depending on the
locality of keys coming to the indexed access. Techniques like vectoring
non-local index accesses by simply missing the cache in parallel on multiple
lookups vectored on the same thread may have high impact. Also detecting absence
of locality should turn off any locality dependent optimizations if these are
costly when there is no locality. A graph neighborhood traversal is an example
of an operation with random access without predictable locality.


\IfFileExists{choke-point-query-mapping/cp-3-3}{}{}

\newCPSection{3.4}{STORAGE}{Temporal access locality and performance}

When filtering edge in navigational pattern on a high-degree vertex, the
performance of queries with temporal window filters can be improved when the
edges are sorted by timestamp in the embedded storage. This placement optimizes
the data access locality for timestamps avoiding scanning.


\IfFileExists{choke-point-query-mapping/cp-3-4}{\paragraph{Queries}
{\raggedright
\queryRefCard{transaction-complex-read-01}{TCR}{1}
\queryRefCard{transaction-complex-read-02}{TCR}{2}
\queryRefCard{transaction-complex-read-03}{TCR}{3}
\queryRefCard{transaction-complex-read-04}{TCR}{4}
\queryRefCard{transaction-complex-read-05}{TCR}{5}
\queryRefCard{transaction-complex-read-06}{TCR}{6}
\queryRefCard{transaction-complex-read-07}{TCR}{7}
\queryRefCard{transaction-complex-read-08}{TCR}{8}
\queryRefCard{transaction-complex-read-09}{TCR}{9}
\queryRefCard{transaction-complex-read-10}{TCR}{10}
\queryRefCard{transaction-complex-read-11}{TCR}{11}
\queryRefCard{transaction-complex-read-12}{TCR}{12}

}}{}


\section{Expression Calculation}

\tpcCPSection[4.2a]{4.1}{QOPT}{Common subexpression elimination}

This choke point tests the ability of the query optimizer to detect common
sub-expressions and reuse their results. A basic technique helpful in multiple
queries is common subexpression elimination (CSE). CSE should recognize also
that \lstinline{avg} aggregates can be derived afterwards by dividing a
\lstinline{sum} by the \lstinline{count} when those have been computed.


\IfFileExists{choke-point-query-mapping/cp-4-1}{}{}

\tpcCPSection[4.2d]{4.2}{QOPT}{Complex boolean expression joins and selections}

This choke point tests the ability of the query optimizer to reorder the
execution of boolean expressions to improve the performance. Some boolean
expressions are complex, with possibilities for alternative optimal evaluation
orders. For instance, the optimizer may reorder conjunctions to test first those
conditions with larger selectivity~\cite{DBLP:conf/vldb/Moerkotte98}.


\IfFileExists{choke-point-query-mapping/cp-4-2}{}{}
\snbCPSection{4.3}{QEXE}{Low overhead expressions interpretation}

This choke point tests the ability of efficiently evaluating simple expressions
on a large number of values. A typical example could be simple arithmetic
expressions, mathematical functions like floor and absolute or date functions
like extracting a year.


\IfFileExists{choke-point-query-mapping/cp-4-3}{}{}



\section{Correlated Sub-Queries}

\tpcCPSection[5.1]{5.1}{QOPT}{Flattening sub-queries}

This choke point tests the ability of the query optimizer to flatten execution
plans when there are correlated sub-queries. Many queries have correlated
sub-queries and their query plans can be flattened, such that the correlated
sub-query is handled using an equi-join, outer-join or anti-join. In TPC-H Q21,
for instance, there is an \lstinline{EXISTS} clause (for orders with more than
one supplier) and a \lstinline{NOT EXISTS} clauses (looking for an item that was
received too late). To execute this query well, systems need to flatten both
sub-queries, the first into an equi-join plan, the second into an anti-join
plan. Therefore, the execution layer of the database system will benefit from
implementing these extended join variants.

The ill effects of repetitive tuple-at-a-time sub-query execution can also be
mitigated if execution systems by using vectorized, or blockwise query
execution, allowing to run sub-queries with thousands of input parameters
instead of one. The ability to look up many keys in an index in one API call
creates the opportunity to benefit from physical locality, if lookup keys
exhibit some clustering.


\IfFileExists{choke-point-query-mapping/cp-5-1}{}{}

\tpcCPSection[5.3]{5.2}{QEXE}{Overlap between outer and sub-query}

This choke point tests the ability of the execution engine to reuse results when
there is an overlap between the outer query and the sub-query. In some queries,
the correlated sub-query and the outer query have the same joins and selections.
In this case, a non-tree, rather DAG-shaped~\cite{DBLP:conf/btw/NeumannM09}
query plan would allow to execute the common parts just once, providing the
intermediate result stream to both the outer query and correlated sub-query,
which higher up in the query plan are joined together (using normal query
decorrelation rewrites). As such, the benchmark rewards systems where the
optimizer can detect this and the execution engine supports an operator that can
buffer intermediate results and provide them to multiple parent operators.


\IfFileExists{choke-point-query-mapping/cp-5-2}{}{}

\tpcCPSection[5.2]{5.3}{QEXE}{Intra-query result reuse}

This choke point tests the ability of the execution engine to reuse sub-query
results when two sub-queries are mostly identical. Some queries have almost
identical sub-queries, where some of their internal results can be reused in
both sides of the execution plan, thus avoiding to repeat computations.


\IfFileExists{choke-point-query-mapping/cp-5-3}{}{}


\section{Parallelism and Concurrency}

\tpcCPSection[6.3]{6.1}{QEXE}{Inter-query result reuse}

This choke point tests the ability of the query execution engine to reuse
results from different queries. Sometimes with a high number of streams a
significant amount of identical queries emerge in the resulting workload. The
reason is that certain parameters, as generated by the workload generator, have
only a limited amount of parameters bindings. This weakness opens up the
possibility of using a query result cache, to eliminate the repetitive part of
the workload. A further opportunity that detects even more overlap is the work
on recycling, which does not only cache final query results, but also
intermediate query results of a ``high worth''. Here, worth is a combination of
partial-query result size, partial-query evaluation cost, and observed (or
estimated) frequency of the partial-query in the workload.


\IfFileExists{choke-point-query-mapping/cp-6-1}{}{}

\newCPSection{6.2}{QEXE}{Intra-query parallelization on hub vertex}

When traversing on hub vertex, the number of edges is beyond estimation based on
the degree distribution of the graph. This chokepoint tests the query optimizer
to automate the intra-query parallelization when traversing on hub vertex to
speed up.


\IfFileExists{choke-point-query-mapping/cp-6-2}{\paragraph{Queries}
{\raggedright
\queryRefCard{transaction-complex-read-01}{TCR}{1}
\queryRefCard{transaction-complex-read-02}{TCR}{2}
\queryRefCard{transaction-complex-read-03}{TCR}{3}
\queryRefCard{transaction-complex-read-04}{TCR}{4}
\queryRefCard{transaction-complex-read-05}{TCR}{5}
\queryRefCard{transaction-complex-read-06}{TCR}{6}
\queryRefCard{transaction-complex-read-07}{TCR}{7}
\queryRefCard{transaction-complex-read-08}{TCR}{8}
\queryRefCard{transaction-complex-read-09}{TCR}{9}
\queryRefCard{transaction-complex-read-10}{TCR}{10}
\queryRefCard{transaction-complex-read-11}{TCR}{11}
\queryRefCard{transaction-complex-read-12}{TCR}{12}

}}{}

\newCPSection{6.3}{QEXE}{Write operation contention and conflicts}

Read-write query is expected to execute inside a transaction. The transaction
like a possible write down to storage (I/O) after a long time read starting with
a write operation in memory. This means long time write transactions that hold
write locks longer than expected. This may result in contention and conflicts
between write operations to the same datum.


\IfFileExists{choke-point-query-mapping/cp-6-3}{}{}


\section{Graph Specifics}

\snbCPSection{7.1}{QEXE}{Incremental path computation}

This choke point tests the ability of the execution engine to reuse work across
graph traversals. For example, when computing paths within a range of distances,
it is often possible to incrementally compute longer paths by reusing paths of
shorter distances that were already computed.


\IfFileExists{choke-point-query-mapping/cp-7-1}{\paragraph{Queries}
{\raggedright
\queryRefCard{transaction-complex-read-01}{TCR}{1}
\queryRefCard{transaction-complex-read-02}{TCR}{2}
\queryRefCard{transaction-complex-read-05}{TCR}{5}
\queryRefCard{transaction-complex-read-08}{TCR}{8}
\queryRefCard{transaction-complex-read-12}{TCR}{12}

}}{}

\snbCPSection{7.2}{QOPT}{Cardinality estimation of transitive paths}

This choke point tests the ability of the query optimizer to properly estimate
the cardinality of intermediate results when executing transitive paths. A
transitive path may occur in a ``fact table'' or a ``dimension table'' position.
A transitive path may cover a tree or a graph, \eg descendants in a geographical
hierarchy \vs graph neighborhood or transitive closure in a many-to-many
connected social network. In order to decide proper join order and type, the
cardinality of the expansion of the transitive path needs to be correctly
estimated. This could for example take the form of executing on a sample of the
data in the cost model or of gathering special statistics, \eg the depth and
fan-out of a tree. In the case of hierarchical dimensions, \eg geographic
locations or other hierarchical classifications, detecting the cardinality of
the transitive path will allow one to go to a star schema plan with scan of a
fact table with a selective hash join. Such a plan will be on the other hand
very bad for example if the hash table is much larger than the ``fact table''
being scanned.


\IfFileExists{choke-point-query-mapping/cp-7-2}{}{}

\snbCPSection{7.3}{QEXE}{Execution of a transitive step}

This choke point tests the ability of the query execution engine to efficiently
execute transitive steps. Graph workloads may have transitive operations, for
example finding the shortest path between vertices. This involves repeated execution
of a short lookup, often on many values at the same time, while usually having
an end condition, \eg the target vertice being reached or having reached the border
of a search going in the opposite direction. For the best efficiency, these
operations can be merged or tightly coupled to the index operations themselves.
Also, parallelization may be possible but may need to deal with a global state,
\eg set of visited vertices. There are many possible tradeoffs between generality
and performance.


\IfFileExists{choke-point-query-mapping/cp-7-3}{}{}

\snbCPSection{7.4}{QEXE}{Efficient evaluation of termination criteria for transitive queries}

This tests the ability of a system to express termination criteria for
transitive queries so that not the whole transitive relation has to be evaluated
as well as efficient testing for termination.


\IfFileExists{choke-point-query-mapping/cp-7-4}{\paragraph{Queries}
{\raggedright
\queryRefCard{transaction-complex-read-01}{TCR}{1}
\queryRefCard{transaction-complex-read-02}{TCR}{2}
\queryRefCard{transaction-complex-read-05}{TCR}{5}
\queryRefCard{transaction-complex-read-11}{TCR}{11}

}}{}

\snbCPSection{7.5}{QEXE}{Unweighted shortest paths}

A common problem in graph queries is determining the distance between a vertice and
a set of vertices. To compute the distance values, systems may employ BFS or a
single-source shortest path algorithm with uniform weights. To compute the
distance between two given vertices, systems can use bidirectional search
algorithms.


\IfFileExists{choke-point-query-mapping/cp-7-5}{}{}

\snbCPSection{7.6}{QEXE}{Weighted shortest paths}

Computing single-source shortest path is a fundamental problem in graph queries.
While there are well-known algorithms to compute it, \eg Dijkstra's algorithm or
the Bellman-Ford algorithm, system often use na\"ive approaches such as
enumerating all paths which makes these queries intractable.


\IfFileExists{choke-point-query-mapping/cp-7-6}{}{}

\snbCPSection{7.7}{QEXE}{Composition of graph queries}

In many cases, it is desirable to specify multiple graph queries, where the
first one defines an induced subgraph or an overlay graph on the original graph,
which is then passed two the next query, and so on. Expressing such computations
as a sequence of composable graph queries would be desirable from both
usability, optimization, and execution aspects. However, currently many graph
dabases lack support for composable graph queries.

The \mbox{G-CORE}~\cite{DBLP:conf/sigmod/AnglesABBFGLPPS18} design language
tackled problem this by introducing the \emph{path property graph} data model
(consisting of vertices, edges, and paths) and defining queries such that they
return a graph (while also providing means to return a tabular output).


\IfFileExists{choke-point-query-mapping/cp-7-7}{}{}

\snbCPSection{7.8}{QEXE}{Reachability between disconnected components}

For path finding queries, the result is often that the specified path does not
exist in the graph. For example, for a single-source single-destination search,
when one of the endpoints is in a small component (\eg the endpoint is an
isolated vertice), systems using a bidirectional search algorithm can quickly
determine that there is no path to be found.


\IfFileExists{choke-point-query-mapping/cp-7-8}{}{}

\newCPSection{7.9}{STORAGE}{Hub vertex storage balance}

Especially in distributed systems, hub vertices means bigger data unit, \eg
shard, which may need to split to balance the storage, load and inter-shard
communication.


\IfFileExists{choke-point-query-mapping/cp-7-9}{}{}

\newCPSection{7.10}{STORAGE}{Multiplicity support in Graph Model}

Edge multiplicity requires that systems support multiple edges between the same
vertex pair. Another dimension is required to annotate the edge id.


\IfFileExists{choke-point-query-mapping/cp-7-10}{}{}

\newCPSection{7.11}{QEXE}{Intermediate Result Propagation}

When calculating some final share or final ratio values, there is a common
pattern in computing that each value need to calculate with the value in last
hop which is similar to propagation, (\eg label propagation). To make the
computation more efficient, some intermediate results should be cached to reuse
in the next computing stage.


\IfFileExists{choke-point-query-mapping/cp-7-11}{\paragraph{Queries}
{\raggedright
\queryRefCard{transaction-complex-read-11}{TCR}{11}

}}{}


\section{Language Features}

\snbCPSection{8.1}{LANG}{Complex patterns}

\paragraph{Description.}

A natural requirement for graph query systems is to be able to express complex
graph patterns.

\paragraph{Transitive edges.} Transitive closure-style computations are common
in graph query systems, both with fixed bounds (\eg get vertices that can be
reached through at least 3 and at most 5 \textsf{knows} edges), and without
fixed bounds (\eg get all \textsf{Messages} that a \textsf{Comment} replies to).

\paragraph{Negative edge conditions.} Some queries define \emph{negative pattern
conditions}. For example, the condition that a certain \textsf{Message} does not
have a certain \textsf{Tag} is represented in the graph as the absence of a
\textsf{hasTag} edge between the two vertices. Thus, queries looking for cases
where this condition is satisfied check for negative patterns, also known as
negative application conditions (NACs) in graph transformation
literature~\cite{DBLP:journals/fuin/HabelHT96}.


\IfFileExists{choke-point-query-mapping/cp-8-1}{}{}

\snbCPSection{8.2}{LANG}{Complex aggregations}

\paragraph{Description.}

BI workloads are heavy on aggregation, including queries with \emph{subsequent
aggregations}, where the results of an aggregation serves as the input of
another aggregation. Expressing such operations requires some sort of query
composition or chaining (see also CP-8.4). It is also common to \emph{filter on
aggregation results} (similarly to the \lstinline[language=sql]{HAVING} keyword
of SQL).


\IfFileExists{choke-point-query-mapping/cp-8-2}{}{}

\snbCPSection{8.3}{LANG}{Ranking-style queries}

\paragraph{Description.}

Along with aggregations, BI workloads often use \emph{window functions}, which
perform aggregations without grouping input tuples to a single output tuple.  A
common use case for windowing is \emph{ranking}, \ie selecting the top element
with additional values in the tuple (vertices, edges or
attributes).\footnote{PostgreSQL defines the \lstinline[language=sql]{OVER}
keyword to use aggregation functions as window functions, and the
\lstinline[language=sql]{rank()} function to produce numerical ranks, see
\url{https://www.postgresql.org/docs/9.1/static/tutorial-window.html} for
details.}


\IfFileExists{choke-point-query-mapping/cp-8-3}{}{}

\snbCPSection{8.4}{LANG}{Query composition}

\paragraph{Description.}

Numerous use cases require \emph{composition} of queries, including the reuse of
query results (\eg vertices, edges) or using scalar subqueries (\eg selecting a
threshold value with a subquery and using it for subsequent filtering
operations).


\IfFileExists{choke-point-query-mapping/cp-8-4}{}{}

\snbCPSection{8.5}{LANG}{Dates and times}

\paragraph{Description.}

Handling dates and times is a fundamental requirement for production-ready
database systems. It is particularly important in the context of BI queries as
these often calculate aggregations on certain periods of time (\eg on entities
created during the course of a month).


\IfFileExists{choke-point-query-mapping/cp-8-5}{}{}

\snbCPSection{8.6}{LANG}{Handling paths}

\paragraph{Description.}

Handling paths as first-class citizens is one of the key distinguishing features
of graph database systems~\cite{DBLP:conf/sigmod/AnglesABBFGLPPS18}. Hence,
additionally to reachability-style checks, a language should be able to express
queries that operate on elements of a path, \eg calculate a score on each edge
of the path. Also, some use cases specify uniqueness constraints on
paths~\cite{DBLP:journals/csur/AnglesABHRV17}: \emph{arbitrary path},
\emph{shortest path}, \emph{no-repeated-node semantics} (also known as
\emph{simple paths}), and \emph{no-repeated-edge semantics} (also known as
\emph{trails}). Other variants are also used in rare cases, such as
\emph{maximal} (non-expandable) or \emph{minimal} (non-contractable) paths.

\paragraph{Note on terminology.}

The \emph{Glossary of graph theory terms} page of
Wikipedia\footnote{\url{https://en.wikipedia.org/wiki/Glossary_of_graph_theory_terms}}
defines \emph{paths} as follows: ``A path may either be a walk (a sequence of
vertices and edges, with both endpoints of an edge appearing adjacent to it in the
sequence) or a simple path (a walk with no repetitions of vertices or edges),
depending on the source.'' In this work, we use the first definition, which is
more common in modern graph database systems and is also followed in a recent
survey on graph query languages~\cite{DBLP:journals/csur/AnglesABHRV17}.


\IfFileExists{choke-point-query-mapping/cp-8-6}{}{}

\newCPSection{8.7}{LANG}{Concise temporal window expression}

Temporal window filtering is a common expression pattern when filtering edges in
navigational pattern. The common scenario is that the whole pattern is expected
bounded by the timestamp filter, including \emph{BEFORE}, \emph{AFTER} and
\emph{BETWEEN}. It is supported that adding timestamp filtering on each vertex
and edge in the pattern to express a temporal window, which is a verbose
expression. A more concise expression is desired. A possible solution is adding
keywords like \emph{RANGE\_SLICE}, \emph{LEFT\_SLICE} and \emph{RIGHT\_SLICE}
referring to an extension of \emph{Cypher}~\cite{tcypher}.


\IfFileExists{choke-point-query-mapping/cp-8-7}{\paragraph{Queries}
{\raggedright
\queryRefCard{transaction-complex-read-01}{TCR}{1}
\queryRefCard{transaction-complex-read-02}{TCR}{2}
\queryRefCard{transaction-complex-read-03}{TCR}{3}
\queryRefCard{transaction-complex-read-04}{TCR}{4}
\queryRefCard{transaction-complex-read-05}{TCR}{5}
\queryRefCard{transaction-complex-read-06}{TCR}{6}
\queryRefCard{transaction-complex-read-07}{TCR}{7}
\queryRefCard{transaction-complex-read-08}{TCR}{8}
\queryRefCard{transaction-complex-read-09}{TCR}{9}
\queryRefCard{transaction-complex-read-10}{TCR}{10}
\queryRefCard{transaction-complex-read-11}{TCR}{11}
\queryRefCard{transaction-complex-read-12}{TCR}{12}

}}{}

\newCPSection{8.8}{LANG}{Recursive path filtering pattern}

Sometimes when tracing a fund flow, such a pattern is expected that find a path
with recursive filters. For example, filters are expected to assume a path A
-[${e_1}$]-> B -[${e_2}$]-> ... -> X.

\begin{itemize}
    \item The timestamp order: ${e_1}$ < ${e_2}$ < … < ${e_i}$
    \item The amount order: ${e_1}$ > ${e_2}$ > … > ${e_i}$
    \item The time window: ${e_{i-1}}$ < ${e_i}$ < ${e_{i-1}}$ + $\vec{\Delta}$,
    $\vec{\Delta}$ is a given constant.
\end{itemize}

Such queries that require \emph{all timestamps in the transfer trace are in ascending order} or the \emph{upstream} edge are
difficult to explain in plain Cypher (or GQL or SQL/PGQ) because they require support for the category of queries
\emph{Regular expression with memory} as described in this paper\cite{10.1145/2274576.2274585}. Another possible solution is
adding keywords like \emph{SEQUENTIAL} and \emph{DELTA} referring to an extension of \emph{Cypher}~\cite{tcypher}. 


\IfFileExists{choke-point-query-mapping/cp-8-8}{\paragraph{Queries}
{\raggedright
\queryRefCard{transaction-complex-read-01}{TCR}{1}
\queryRefCard{transaction-complex-read-02}{TCR}{2}
\queryRefCard{transaction-complex-read-05}{TCR}{5}

}}{}

\newCPSection{8.9}{LANG}{Traversal limit pattern}

When traversing on hub vertex, the data amount touched may experience
exponential growth, which is a common challenge to systems. When the performance
is not enough to satisfy the queries on hub vertex, a language feature is needed
that the number of edges traversed out from the hub vertex can be limited. Such
keyword may be \emph{truncation\_limit}.


\IfFileExists{choke-point-query-mapping/cp-8-9}{}{}


\section{Update Operations}

\snbCPSection{9.1}{UPD}{Insert vertice}

This choke point tests the ability of the database to insert a vertice.


\IfFileExists{choke-point-query-mapping/cp-9-1}{}{}

\snbCPSection{9.2}{UPD}{Insert edge}

This choke point tests the ability of the database to insert an edge.


\IfFileExists{choke-point-query-mapping/cp-9-2}{}{}

\snbCPSection{9.3}{UPD}{Delete vertice}

This choke point tests the ability of the database to delete a vertice.


\IfFileExists{choke-point-query-mapping/cp-9-3}{}{}

\snbCPSection{9.4}{UPD}{Delete edge}

This choke point tests the ability of the database to delete an edge.


\IfFileExists{choke-point-query-mapping/cp-9-4}{}{}

\snbCPSection{9.5}{UPD}{Delete recursively}

This choke point tests the ability of the database to recursively perform a delete operation, \eg delete an entire message thread.


\IfFileExists{choke-point-query-mapping/cp-9-5}{}{}

\chapter{Scale Factor Statistics}
\label{sec:sf-statistics}

\section{Number of Entities for FinBench Transaction v0.1.0}

\begin{table}[htb]
    \setlength{\tabcolsep}{.10em}
    \centering
    {
        \normalsize
        \begin{tabular}{|>{\sffamily}c|>{\tt}l|r|r|r|r|r|r|r|r|r|r|r|r|r|}
            \hline
            \tableHeaderFirst{C} & \tableHeader{File} & \tableHeader{SF0.01} & \tableHeader{SF0.1} & \tableHeader{SF0.3} & \tableHeader{SF1}  & \tableHeader{SF3}  & \tableHeader{SF10}  \\ \hline
            \hline
            N                    & account            & \numprint{2633}      & \numprint{26347}    & \numprint{79199}    & \numprint{264075}  & \numprint{791769}  & \numprint{1980883}  \\
            N                    & company            & \numprint{2633}      & \numprint{4000}     & \numprint{12000}    & \numprint{40000}   & \numprint{120000}  & \numprint{300000}   \\
            E                    & companyApplyLoan   & \numprint{524}       & \numprint{5332}     & \numprint{15761}    & \numprint{52820}   & \numprint{158678}  & \numprint{397060}   \\
            E                    & companyGuarantee   & \numprint{248}       & \numprint{2315}     & \numprint{7123}     & \numprint{23870}   & \numprint{71716}   & \numprint{179526}   \\
            E                    & companyInvest      & \numprint{860}       & \numprint{8639}     & \numprint{25853}    & \numprint{86092}   & \numprint{259884}  & \numprint{650190}   \\
            E                    & companyOwnAccount  & \numprint{864}       & \numprint{8805}     & \numprint{26356}    & \numprint{88119}   & \numprint{264352}  & \numprint{660625}   \\
            E                    & deposit            & \numprint{5199}      & \numprint{51686}    & \numprint{153521}   & \numprint{512680}  & \numprint{1534595} & \numprint{3829905}  \\
            N                    & loan               & \numprint{1597}      & \numprint{16138}    & \numprint{47772}    & \numprint{159166}  & \numprint{476670}  & \numprint{1189072}  \\
            E                    & loanTransfer       & \numprint{4886}      & \numprint{49180}    & \numprint{145679}   & \numprint{484657}  & \numprint{1453874} & \numprint{3625556}  \\
            N                    & medium             & \numprint{1000}      & \numprint{10000}    & \numprint{30000}    & \numprint{100000}  & \numprint{300000}  & \numprint{2000000}  \\
            N                    & person             & \numprint{800}       & \numprint{8000}     & \numprint{24000}    & \numprint{80000}   & \numprint{240000}  & \numprint{600000}   \\
            E                    & personApplyLoan    & \numprint{1073}      & \numprint{10806}    & \numprint{32011}    & \numprint{106346}  & \numprint{317992}  & \numprint{792012}   \\
            E                    & personGuarantee    & \numprint{469}       & \numprint{4694}     & \numprint{14221}    & \numprint{47935}   & \numprint{144064}  & \numprint{359283}   \\
            E                    & personInvest       & \numprint{1650}      & \numprint{17296}    & \numprint{52002}    & \numprint{174064}  & \numprint{520584}  & \numprint{1300980}  \\
            E                    & personOwnAccount   & \numprint{1769}      & \numprint{17542}    & \numprint{52843}    & \numprint{175956}  & \numprint{527417}  & \numprint{1320258}  \\
            E                    & repay              & \numprint{5046}      & \numprint{50495}    & \numprint{149559}   & \numprint{497033}  & \numprint{1488916} & \numprint{3715487}  \\
            E                    & signIn             & \numprint{4384}      & \numprint{44540}    & \numprint{134532}   & \numprint{451362}  & \numprint{1350759} & \numprint{8996781}  \\
            E                    & transfer           & \numprint{14145}     & \numprint{138209}   & \numprint{411882}   & \numprint{1379527} & \numprint{4136803} & \numprint{11005032} \\
            E                    & withdraw           & \numprint{20557}     & \numprint{201119}   & \numprint{609548}   & \numprint{2011359} & \numprint{6013709} & \numprint{15056721} \\
            \hline
        \end{tabular}
    }
    \caption{The number of entities per SF and per file in the Transaction workload (produced by the LDBC FinBench
        DataGen). To derive these numbers, 100\% of the network was generated as an initial bulk data set with no update
        streams. Notation -- \textsf{C}: entity category, \textsf{N}: node, \textsf{E}: edge.}
    \label{tab:number-of-entities-transaction-v0.1.0}
\end{table}

\printbibliography[heading=bibintoc, title={References}]

\end{document}